
\overfullrule=0pt
\magnification=1100
\baselineskip=3ex
\raggedbottom
\font\eightpoint=cmr8
\font\fivepoint=cmr5
\font\ninepoint=cmr9
\headline={\hfill{\fivepoint VBEHLJPS -- 25/Nov/93}}
\def\v{{\overline v}}
\def\u{{\overline u}}
\def\g{{\overline g}}
\def\l{{\overline l}}
\def\h{{\overline h}}
\def\k{{\overline k}}
\def\R{{\bf R}}
\def\P{{\cal P}}
\def\Z{{\bf Z}}
\def\HF{{\rm HF}}
\def\mod{{\rm mod}}
\def\Det{{\rm Det}}
\def\sr{{\cal R}}
\def\H{{\cal H}}
\def\N{{\cal N}}
\def\W{{\cal W}}
\def\E{{\cal E}}
\def\Q{{\rm Q}}
\def\F{{\cal F}}
\def\G{{\cal G}}
\def\C{{\bf C}}
\def\1{{\bf 1}}
\def\U{{\cal U}}
\def\Tr{{\rm Tr}}
\def\lanbox{{$\, \vrule height 0.25cm width 0.25cm depth 0.01cm \,$}}
\def\uprho{\raise1pt\hbox{$\rho$}}
\def\mfr#1/#2{\hbox{${{#1} \over {#2}}$}}
\catcode`@=11
\def\eqalignii#1{\,\vcenter{\openup1\jot \m@th
\ialign{\strut\hfil$\displaystyle{##}$&
        $\displaystyle{{}##}$\hfil&
        $\displaystyle{{}##}$\hfil\crcr#1\crcr}}\,}
\catcode`@=12

\def\s{{\rm s}}
\def\bp{{\rm bp}}
\def\ph{{\rm ph}}
\def\ps{{\rm ps}}
\def\S{{\cal S}}
\def\N{{\cal N}}
\def\OS{{\widetilde{\cal S}}}
\def\phagger{{\phantom{\dagger}}}

\def\sr{{\cal R}}
\def\srr{{\cal R}}
\centerline{\bf GENERALIZED HARTREE-FOCK THEORY}
\centerline{\bf AND THE HUBBARD MODEL}
\vskip 1.5truecm
{\baselineskip=3ex
\centerline{Volker Bach$^{(1)}$,
Elliott H. Lieb$^{(2),(3)}$
and
Jan Philip Solovej$^{(3)}$
\vfootnote{}{\eightpoint \hskip -\parindent
This work was supported in part by Sonderforschungsbereich
288 of the Deutsche Forschungsgemeinschaft (VB) and U.S. National Science
Foundation grants PHY90-19433 A02 (VB and EHL) and DMS92-03829 (JPS).}
\vfootnote{}{\eightpoint \hskip -\parindent
\copyright 1993 by the authors. Reproduction
of this article by any means is permitted for non-commercial
purposes.}}}
\bigskip\bigskip
\noindent{\it
$^{(1)}$FB Mathematik, Technische Universit\"at
Berlin, Strasse des
17 Juni 136, D-10623 Berlin, Germany\hfil\break
$^{(2)}$Department of Physics, Jadwin Hall, Princeton
University, P.O. Box 708, Princeton, NJ, 08544\hfil\break
$^{(3)}$Department of Mathematics, Fine Hall, Princeton
University,
Princeton, NJ, 08544\hfil}
\bigskip\bigskip\bigskip
\centerline{\it Dedicated to Philippe Choquard on his sixty fifth
birthday}
\bigskip\bigskip\bigskip
{\narrower{\it Abstract:\/}
The familiar unrestricted Hartree-Fock variational principle is
generalized to include quasi-free states. As we show,
these are in one-to-one correspondence with the one-particle density
matrices and these, in turn provide a convenient formulation of
a generalized Hartree-Fock variational principle,
which includes the BCS theory as a special case. While this
generalization is not new, it is not well known and we
begin by elucidating it.
The Hubbard model, with its particle-hole symmetry, is well
suited to exploring this theory because BCS states for the attractive
model turn into usual HF states for the repulsive model.
We rigorously determine the true, unrestricted minimizers for
zero and for nonzero temperature in several cases, notably
the half-filled band. For the cases treated here, we can exactly
determine all broken and unbroken spatial and gauge symmetries
of the Hamiltonian.
\smallskip}
\vfill\eject

\vbox{{\narrower\baselineskip=3ex
\noindent{\bf Table of Contents}
\item{I.} {\sl Introduction} \dotfill 2
\item{II.}  {\sl Definition and properties of generalized HF
states} \dotfill 6
\itemitem{II.a} Definition of HF states\dotfill6
\itemitem{II.b} One-particle density matrices\dotfill9
\itemitem{II.c} The generalized HF functional\dotfill17
\item{III.} {\sl The generalized HF theory for the Hubbard model with
attractive interaction}\dotfill26
\itemitem{III.a} Definitions\dotfill26
\itemitem{III.b} Linearization of the pressure functional\dotfill29
\itemitem{III.c} Gap and zero temperature limit\dotfill34
\itemitem{III.d} Broken gauge symmetries\dotfill37
\itemitem{III.e} Spatial uniqueness of minimizers\dotfill47
\itemitem{III.f} Spatial symmetries\dotfill51
\itemitem{III.g} The translation invariant case\dotfill52
\item{IV.}  {\sl The generalized HF theory for the Hubbard model with
repulsive interaction} \dotfill 54
\itemitem{IV.a} Linearization of the pressure functional\dotfill54
\itemitem{IV.b} Constant density lemma for bipartite lattices at
half-filling\dotfill57
\itemitem{IV.c} Particle-hole symmetry\dotfill59
\itemitem{IV.d} Ferromagnetism at infinite repulsion\dotfill60
\item{V.}  {\sl Summary of HF theory of the Hubbard model}\dotfill 63
\itemitem{V.a} Introduction\dotfill63
\itemitem{V.b} Symmetries of the Hubbard Hamiltonian\dotfill64
\itemitem{} Tables\dotfill66
\item{} {\sl References}\dotfill67
\smallskip}}
\bigskip
\bigskip
\noindent
{\bf I.  INTRODUCTION}

The Hubbard model has long been recognized as an interesting imitation of
electron-electron interactions and of the correlations they induce.  It
also turns out, as we show here, that it is an interesting
testing ground for Hartree-Fock (HF) theory, and it is one of the very few
examples in many-body theory for which many properties of the {\it true},
energy minimizing HF ground state and pressure maximizing positive
temperature state can be precisely elucidated
without approximation, restriction or unjustified assumptions.
(To avoid misunderstandings, we should make it clear at the outset that
we consider only unrestricted HF theory.)

While studying the HF theory of the Hubbard model we were led to a critical
study of HF theory itself---namely, the proper context in which to view it,
as well as some of its very general features.  This is the content of
Sect.~II which we can summarize as follows.

The usual HF theory for an $N$-particle system starts with a Slater determinant
$\psi$ formed from $N$ orthonormal, one-particle functions
of space and spin; the energy is
then minimized with respect to the choice of these $N$ functions.  This
$\psi$ will generally break certain symmetries inherent in the original
problem --- typical examples being translation invariance and spin or spatial
angular momenta.  That being the case, it is not worse to permit a
violation of particle-number conservation as well --- if the energy can
thereby be lowered.
This is precisely what was done by Bardeen, Cooper and Schrieffer [BCS]
in their theory of superconductivity.
It is important, here to recall the simple fact that if a state
violates a symmetry of the Hamiltonian (e.g. angular momentum, particle
number) and if that state is then decomposed into its irreducible
components, at least one component will have an energy no greater than
(and often less than) the original state.

The proper Hilbert space, then, is Fock space, $\F$, the sum of all the
original $N$-particle spaces with $N = 0, 1, 2, \dots$.  The simplest
Hamiltonians on $\F$ are, of course, the quadratic ones, and they are all
diagonalizable by  Bogoliubov unitary transformations, which transforms
creation operators $c^\dagger$ into linear combinations of $c$'s and
$c^\dagger$'s.
In the context of fermionic theories, this transformation was discovered
at the same time by Valatin [VJ] and Bogoliubov (see [BN]).
The ground state of a quadratic Hamiltonian $H^Q$ is a
Bogoliubov transformation $\U$ applied to the vacuum (zero-particle state)
$\vert 0 \rangle$.  We call all such states of the form $\U \vert 0 \rangle$
a {\bf generalized HF state} because the original $N$-particle determinant
(which has the form $c^\dagger_1 c^\dagger_2 \dots c^\dagger_N \vert 0 \rangle$
and which we call a {\bf normal state}) is only the special case
corresponding to an $H^Q$ that contains terms of the form $c^\dagger c$ and
$N$ terms of the form $cc^\dagger$, but not the particle nonconserving
terms $cc$ or $c^\dagger c^\dagger$.  Such generalized HF states, $\U \vert
0 \rangle$, are also called {\bf quasi-free states}
because they satisfy the conclusion of Wick's Theorem.  Indeed, all
quasi-free states are of this form, as we show below.

A generalized HF ground state is thus the ground state of some quadratic
$H^Q$ in Fock space.  A positive temperature HF state is, likewise, the
usual grand canonical Gibbs state for such an $H^Q$.  In analogy with the
ground state, such a grand canonical Gibbs state is called {\bf normal}
when $H^Q$ is particle conserving.  For the ground state, the $H^Q$ is
determined so that the expectation value of the original many-body
Hamiltonian, $H$, of our system is as small as possible.  For positive
temperatures, $H^Q$ is chosen (in a temperature dependent way) to maximize
the pressure in the grand canonical ensemble.

The paired (BCS) state, which is so important in
superconductivity theory, is also of the form $\U \vert 0 \rangle$, a
Bogoliubov transformation of the vacuum.  Thus, normal HF theory and BCS
theory are but two aspects of the same general theory: find the best substitute
quadratic Hamiltonian or, equivalently, find the best Bogoliubov
transformation.  This relationship was certainly known
[BN], [dG], [BR], [DJK],
but our personal experience is that it is far from
being universally appreciated.  At first it seems surprising that
one-particle states can somehow evolve into pair states, and the
explanation is roughly the following:
It is always true that for
each mode $\alpha$, $\U c_{\alpha}^\dagger \U^\dagger
= d^\dagger + e$, where $d^\dagger$ (resp. $e$) is proportional to a
creation (resp. annihilation) operator, but it is possible that
$d^\dagger=0$ or $e=0$.  If $d^\dagger = 0$ then $\psi = \U
\vert 0 \rangle$ contains the mode $e^\dagger$, i.e., $e^\dagger
e\psi=\hbox{(pos. const)}\psi$.
If $e = 0$, $\psi$ just
contains a factor proportional to $|0\rangle$, i.e.,
$d\psi=0$. If $e$ and $d^\dagger$ are
both nonvanishing then
such modes must come in pairs (as we prove in Theorem~2.2) and
$\psi$ is found to contain a pairing factor $(1+d^\dagger e^\dagger)$
acting on $\vert 0 \rangle$.
Alternatively, it turns out that to every eigenvalue 1 of the
particle conserving part, $\gamma$, of the one-particle density matrix
(1-$pdm$), $\Gamma$, associated with the state $\psi = \U
\vert 0 \rangle$ there corresponds a simple one-particle state in $\psi$,
but to every eigenvalue of $\gamma$ between 0 and 1 there corresponds a
pair state in $\psi$.

The 1-$pdm$ plays an essential structural mathematical role in HF theory.
The set of quasi-free states (generalized or not) does not have a linear
or convex structure. A linear or convex combination of such states is
not necessarily a quasi-free state. However, a convex combination of
1-{\it pdms} is a 1-$pdm$, i.e., $\Gamma=\lambda\Gamma_1+(1-\lambda)\Gamma_2$
is a 1-$pdm$ if $0\leq\lambda\leq1$ and $\Gamma_1$ and $\Gamma_2$ are 1-{\it
pdms}.
This fact allows us to compensate for the missing convex structure of
quasi-free states. Thus given $\psi_{1,2}=\U_{1,2}\vert 0 \rangle$ with
$\U_{1,2}$ being two Bogoliubov transformations, we can form the 1-{\it pdms}
$\Gamma_1$ and $\Gamma_2$ and then form $\Gamma$ as above.
Finally, we can return to the level of the quasi-free states and thereby
define a quasi-free state that interpolates between the two original states.
Sect.~II contains a detailed description of quasi-free states, density
matrices and quadratic Hamiltonians.  We present this partly for the
reader's convenience, but also because we could not find quite what we need
in the literature on quasi-free states (which usually concentrates on
quasi-free states in terms of algebraic automorphisms rather than
operators) or in the excellent book by Blaizot and Ripka [BR] which does
not deal explicitly with the infinite dimensional Hilbert space $L^2
(\R^3)$, the space of square integrable functions on $\R^3$, needed for our
other theorems in Sect.~II on atomic HF theory.

The notation in Sect.~II is a bit complicated and one reason for this is
the necessity to introduce antiunitary transformations (because $c$'s
transform by antiunitaries if $c^\dagger$'s transform by unitaries).
Consequently, if one tries to write equations in a basis independent way,
one needs more than the usual notation of linear operator theory.  If one
fixes a basis, however, one can use the ordinary linear operator notation,
but then complex conjugates (denoted by superscript bars) and indices
appear in profusion.  We have opted for the second route.

The general Theorem 2.14 in Sect.~II about the usual ($N$-particle) HF theory
should have been well known but seemingly was not.  It applies to
repulsive two-body potentials (as in the real world of electrons with
Coulomb interaction) and states two things.  The first is
that the $N$ one-particle states
are precisely the energetically {\it lowest} eigenvectors of the HF operator.
(This fact was stated in [LS], and the proof was sketched in [LE4]).
While the $N$ HF orbitals are distinct
eigenvectors of the HF operator, it is not obvious, a priori, that they are
the lowest ones; indeed, this might not be true when the interactions are
attractive.  The second part is surprising, for its
conclusion runs counter to
what one might naively assume.
{\it There are never unfilled shells} (for any choice of $N$).  That is to
say, the degeneracy of the last level of the HF operator is {\it always}
precisely what is needed to accommodate the available number of electrons
--- not more than that!

Readers who are already familiar with the formalities of Sect.~II are
advised to skip over it and to turn to Sect.~III, which contains the
HF analysis of the attractive Hubbard model.  It beautifully illustrates the
relationship between BCS and normal states discussed
above.  It is well known that a particle-hole transformation (on the down
spins alone) converts the repulsive and the attractive models into each
other.  What does this do to generalized HF states?  The answer, simply, is
that a normal HF state may be turned into a BCS state (in which there is
pure pairing without isolated one-particle states).  Indeed, it turns out
that the repulsive Hubbard model at half-filling (i.e., the expected
particle number equals $\vert \Lambda \vert$, the number of sites in
$\Lambda$) always has a normal state
as its optimal state (for zero and for positive temperature).  The
attractive model at half-filling then has a BCS state (and, when the
lattice is bipartite, also a normal state of the same energy) as its ground
state.
This was well understood by Dichtel {\it et. al.\/} [DJK],
but they did not prove that their state was, indeed, the true
minimum energy HF state. We do so here as a special case of our results
in Sects.~III and IV.

At the outset we emphasize that translation invariance is not assumed.
By the word ``lattice'' we mean a collection of points connected by bonds
(or edges). Perhaps ``graph'' would be more accurate, but physicists are
accustomed
to the word lattice. If our lattice does have translation invariance, e.g. a
hypercubic lattice, we shall say so explicitly. We do, in fact, investigate
translation invariant cases and we do discuss the cases in which the
translation invariance is broken by the HF state. Thus, there is a special
column in the tables of Sect.~V for lattices that have the additional property
of translation invariance (or some other spatial symmetry).
In any case, our systems are always finite.

Among the things we can prove
about HF theory for a bipartite lattice is the existence of a phase
transition from a BCS state at low temperature to a normal state at high
temperature.

Sect.~IV contains the analysis of the repulsive Hubbard model; most of the
results here---but not all---are a transcription of the results in
Sect.~III via a particle-hole transformation.
One of the earliest HF studies of this model was by Penn [PD].

One question that is peculiar to unrestricted HF theory is
whether or not the orbitals
(which are well defined even for the generalized, particle nonconserving
theory) are simple products of spatial functions and spin functions (the
latter being one of two types, either spin up $\uparrow$ or spin down
$\downarrow$).  For the half-filled band and real hopping matrices we can
show this to be true for both the attractive and repulsive models;
this is one of our more complicated proofs, and it
involves a somewhat delicate convexity argument.

Another question concerns the uniqueness of the HF state for a finite
system.  Apart from possible global gauge transformations (those
describes as ``broken'' in Tables 1--3 in Sect.~V) uniqueness
does hold for a half-filled band as we prove in Theorems~3.12 and 4.5.

The main thing one wishes to know about the true HF state is whether or not
it is qualitatively correct.  From that point of view, a main question is
whether the HF state breaks the symmetries of the problem, and if it does
so in conformity with what one believes to be the case in the corresponding
exact quantum state.  For example, the repulsive Hubbard model on a
bipartite lattice has total spin equal to $\big\vert \vert A \vert - \vert
B \vert \big\vert$ (where $\vert A \vert$ and $\vert B \vert$ are the number
of sites in the two sublattices) in a finite system ground state [LE2]
(see also [LE3])
and is expected to have Ne\'el long range order in three or
more dimensions.  This Ne\'el state {\it partially} breaks the original
translation invariance of the Hamiltonian (if there is any to start with)
into a smaller group consisting of translations on each sublattice
separately.
This is exactly what we prove to be the case for the HF ground or Gibbs state.
Indeed, after a suitable rotation of the spin basis, we will find the spins
to point upward on the $A$-sublattice and downward on the $B$-sublattice
This validates the predictions of mean field calculations for the translation
invariant case in the physics literature
[DJK], [CM], [FE].
As far as we know, this self-consistent
antiferromagnetic $A$-$B$ spin structure was assumed to be valid in the
energy minimizing ground state,
but it was never proved that this was indeed true. In principle, some sort of
further symmetry breaking could occur. Our results show that this does not
happen.

Sect.~V summarizes what we can prove about the breaking of symmetry in
different cases.  Unfortunately, different combinations of conditions have
to be treated separately; the basic possible postulates are bipartite (or
non-bipartite) lattice, real (or complex) hopping matrix, repulsive or
attractive interaction, half-filled (or not half-filled) band, translation
invariance (or no invariance).  The symmetries to be investigated are spin
$SU(2)$, pseudospin $SU(2)$, $U(1)$ (particle conservation) $\Z_2$
(particle-hole symmetry) and translation invariance.  For the reader's
convenience, our conclusions are encapsulated in three tables.

The phrase ``symmetry breaking'' does require a precise definition and we
supply that in Sect.~V.  To us it means that the state fails to have the
symmetry that the Hamiltonian has.  Some authors, e.g., [BR] use
a more restrictive definition, namely that the state also fails to belong
to exactly one irreducible representation of the symmetry
group.  For example, the
completely magnetized ground state of a Heisenberg ferromagnet breaks
rotation symmetry in our sense, but not in the restrictive sense.  As a
matter of experience, however, this case is rare, i.e.,
usually the two senses agree in practice.  At least this is so in the cases
we can analyze completely here.
In any event, as explained in Sect.~IV, the restrictive definition
requires a linear structure and, therefore, is not appropriate
for HF theory.

{\it Acknowledgment:} We thank J.~Poelchau, H.T.~Yau and A.S.~Wightman
for helpful discussions.
\bigskip\bigskip
\noindent
\vbox{\noindent{\bf II.  DEFINITION AND PROPERTIES OF GENERALIZED HF STATES}
\bigskip
\bigskip

Since its introduction in 1930 the notion of a Hartree-Fock ground state
and positive
temperature state has evolved.  Our purpose in this section is to state
clearly several definitions of these states, demonstrate their equivalence
and prove some of their fundamental properties.  Despite years of attention
to the subject it is surprising that some of the basic properties have not
been clearly stated, much less proved.  Two of these are
in Theorem~2.12 (the variational principle [LE1]) and Theorem~2.14,
which states that there are {\it never} unfilled shells, regardless
of the particle number.}
\bigskip\noindent
{\bf II.a  DEFINITIONS OF HF STATES}

The original point of view was that a HF state is a single determinant of
one-particle orbitals, in the variables $z_1, z_2, \dots ,z_N$, where $z =
(x, \sigma)$ denotes a space-spin variable for one particle.  The inner
product of two functions is $\langle \varphi_1 \vert \varphi_2 \rangle =
\sum \limits_\sigma\int\overline{\varphi_1} \varphi_2 (x, \sigma) dx$.
The usual HF state is then
$$\psi_{\HF} = (N!)^{-1/2} \Det [\varphi_i (x_j)]_{1 \leq i,j \leq N} =:
(N!)^{-1/2} (\varphi_1 \wedge \dots \wedge \varphi_N) ,\eqno(2a.1)$$
in which $\Det [\langle \varphi_i \vert \varphi_j \rangle ] = 1$.

This is not general enough for our purposes because we also want to allow
for HF states in which the particle number is not conserved.  After all,
there are other quantities such as total spin, total momentum and total
angular momentum that are not necessarily sharp in a HF state, and there
is no reason why the particle number should not suffer a similar fate.
In any case, when we go to positive
temperature, the HF state
should not be expected to be a pure state if it is to have any physical
relevance.  For these reasons we are going to make definitions that go
beyond simple determinants.  There is nothing new about the definitions
given here but it is important for us to be very clear about them.

Abstractly we begin with a one-particle Hilbert space $\H$ (finite or
infinite dimensional, but always separable).  We define the fermionic
$N$-particle space to be the antisymmetric tensor product of
$\H^{(N)} := \overbrace{\H \wedge \dots \wedge \H}^{N \
{\rm times}}$ for all $N = 1,2,3 \dots$.  A {\bf simple vector} in $\H^{(N)}$
is of the form
$$f_1 \wedge f_2 \wedge \dots \wedge f_N := \sum \limits_{{\rm
permutations}} (-)^\pi \cdot f_{\pi (1)} \otimes f_{\pi (2)} \otimes \dots
\otimes f_{\pi (N)}, \eqno(2a.2)$$
where each $f_i$ is in $\H$.  A general vector in $\H^{(N)}$ is a (possibly
infinite) linear
combination of such simple vectors.  We note that the squared norm of the
above simple vector is\footnote{$^\dagger$}{\eightpoint
Dirac notation will be used.
$\langle f \vert g \rangle$ is the inner product of $f$ and $g$, which is
linear in $g$ and conjugate linear in $f$.  $\langle f \vert A \vert g
\rangle = \langle f \vert Ag \rangle$ is the inner product of $f$ with the
vector $Ag$, i.e., the operator $A$ acts to the right.  It is important to
keep this in mind when $A$ is not self-adjoint.}
$$\langle f_1 \wedge \dots \wedge f_N \big\vert f_1 \wedge \dots \wedge f_N
\rangle = N! \ \Det [\langle f_i \vert f_j \rangle ]_{1 \leq i,j \leq N}.
\eqno(2a.3)$$

With the identification $\H^{(1)} = \H$ and $\H^{(0)} = \C \cdot \vert 0
\rangle$ where $\vert 0 \rangle$ is the vacuum and $\langle 0 \vert 0
\rangle := 1$ (of course, $\vert 0 \rangle \not\in \H$), we can define the
Fock Hilbert space
$$\F := \H^{(0)} \oplus \H^{(1)} \oplus \H^{(2)} \oplus \dots . \eqno(2a.4)$$

To any vector $f \in \H$ we associate a creation operator $c^\dagger (f)$
and an annihilation operator $c(f)$, each acting on $\F$.  The creation
operator acts on simple vectors by
$$c^\dagger (f) (f_1 \wedge \dots \wedge f_N) = (N + 1)^{-1/2} (f \wedge f_1
\wedge \dots \wedge f_N). \eqno(2a.5)$$
This definition extends to $\F$ by linearity, and $c(f)$ is defined to be the
adjoint operator of $c^\dagger (f)$.  Note that $c(f) \vert 0 \rangle = 0$ and
that $c(\lambda f) = \overline \lambda c(f)$ for any $f \in \H$ and
any $\lambda \in \C$.
(Here and elsewhere complex conjugation is denoted by a bar.)  By this
construction the creation and annihilation operators fulfill the canonical
anticommutation relations (CAR):
$$\eqalignno{\{ c(f), c^\dagger (g) \} &:= c(f) c^\dagger (g) + c^\dagger
(g) c(f) = \langle f \vert g \rangle\cdot \1, \cr
\{ c^\dagger (f), c^\dagger (g) \} &= \{ c(f), c(g) \} = 0. \qquad&(2a.6)\cr}$$
Here, $\1$ is the identity operator on $\F$.
We remark that the Fock space $\F$ is determined by the vacuum
$\vert 0 \rangle$ and a complete set of operators $c^\dagger (f), c(f)$
that obey the CAR.  Indeed, simple vectors can be written as
$$f_1 \wedge \dots \wedge f_N = (N!)^{1/2} c^\dagger (f_1) c^\dagger
(f_2) \dots c^\dagger (f_N) \vert 0 \rangle. \eqno(2a.7)$$

A {\bf Bogoliubov transformation} of $\F$ is a unitary operator $\W$, on
$\F$, with the following special property:  For each vector $f \in \H$
$$d^\dagger (f) := \W c^\dagger (f) \W^\dagger = c^\dagger (g) + c(h) =
c^\dagger (\widehat u f) + c(\widehat v f). \eqno(2a.8)$$
Here $g$ and $h$ are vectors in $\H$ and, by the linearity of $\W$ we see
that there must exist a linear operator $\widehat u$ on $\H$ such that $g =
\widehat u f$ and
there must exist an antilinear operator $\widehat v$ on $\H$ such that $h =
\widehat v f$ (antilinear means that $\widehat v (\lambda a + b) = \overline
\lambda \widehat v (a) + \widehat v (b)$).
It is easy to check that the unitarity of $\W$, which in particular implies
the CAR for the operators $d^\dagger (f), d(f)$,
results in certain conditions on $\widehat u$ and $\widehat v$.
Because $\widehat v$ is antilinear these are not easy to state.
One way is to choose
an orthonormal basis $f_1, f_2, \dots$ in $\H$ and define the matrix
elements $v_{ij} = \langle f_i \vert \widehat v f_j \rangle$ and
$u_{ij} = \langle f_i \vert \widehat u f_j \rangle$.  In terms of the
linear operators $u$ and $v$ having the same matrix elements the condition
is that the linear operator
$$W = \pmatrix{u &v \cr \v
& \u\cr} ,\eqno(2a.9)$$
acting on $\H \oplus \H$, is a unitary matrix.  Here $\u,
\v$ denote the linear operators with the complex conjugate matrix
elements $\u_{ij}, \v_{ij}$.  An equivalent, basis independent
statement is this:  to every antilinear operator $\widehat v$ we can
always associate an antilinear operator $\widehat v^T$ such that $\langle f
\vert \widehat v g \rangle =
\overline{\langle \widehat v^T f \vert g \rangle}$ for every $f,g \in \H$.  The
condition becomes
$$\eqalign{\widehat u^\dagger \widehat u + \widehat v^T \widehat v &= \1
= \widehat u \widehat u^\dagger + \widehat v \widehat v^T \cr
\widehat v^T \widehat u + \widehat u^\dagger \widehat v &= 0 = \widehat u
\widehat v^T + \widehat v \widehat u^\dagger \cr}\eqno(2a.10) $$

In case $\H$ is finite dimensional the converse is also true, i.e., if a pair
$(\widehat u,\widehat v)$ satisfies (2a.10) then there is a
unique unitary operator $\W$ on $\F$
satisfying (2a.8).  If $\H$ is infinite dimensional it may not be possible
to find a $\W$, even if (2a.10) is true.  This phenomenon occurs, for
example, when $\widehat u = 0$.  A sufficient condition for the existence
of $\W$ is given in Theorem~2.2.

The second notion needed for generalized HF theory is a
quasi-free state.  In general a state, $\uprho$, on the set of bounded
operators that act on $\F$ is a (complex-valued) linear map (i.e., for all
operators $A$ and $B$, $\quad \uprho (\lambda A + B) = \lambda \uprho
(A) + \uprho
(B)$) satisfying the conditions $\uprho (\1) = 1$ and $\uprho (A^\dagger A)
\geq 0$ (which implies $\overline{\uprho (A)} = \uprho (A^\dagger)$).
The example that will concern us most is a {\bf pure state} $\uprho
(A) = \langle \psi \vert A \psi \rangle$ for some $\psi \in \F$.
Another important example is the {\bf Gibbs state} $\uprho (A) = Z^{-1} \Tr
[A \exp (-\beta H)]$ with $Z = \Tr [\exp (-\beta H)]$ for some Hamiltonian
$H$ on $\F$ with $Z < \infty$.

A state $\uprho$ is {\bf quasi-free} if all correlation functions can be
computed from Wick's Theorem, i.e., if the operators $e_1, e_2, \dots ,
e_{2N}$ are
each either a $c^\dagger$ or a $c$, then $\uprho (e_1 e_2 \dots e_{2N-1}) =
0$ and
$$\uprho (e_1 e_2 \dots e_{2N}) = \mathop{{\sum}^\prime}_\pi (-)^\pi \uprho
(e_{\pi(1)} e_{\pi(2)}) \dots \uprho (e_{\pi (2N-1)} e_{\pi(2N)})
\eqno(2a.11)$$
where $\mathop{{\sum}^\prime}_\pi$ is the sum over permutations $\pi$ which
satisfy $\pi (1) < \pi (3) < \dots < \pi (2N-1)$ and $\pi (2j - 1) <
\pi (2j)$ for all $1 \leq j \leq N$.  The right side of (2a.11) is also
known as the Pfaffian of the
triangular array $[\uprho (e_i e_j)]_{1 \leq i < j \leq
2N}$.  In particular, we have the important formula
$$\uprho (e_1 e_2 e_3 e_4) = \uprho (e_1 e_2) \uprho (e_3 e_4) - \uprho
(e_1 e_3) \uprho (e_2 e_4) + \uprho (e_1 e_4) \uprho (e_2 e_3).
\eqno(2a.12)$$
When this is applied later to the expectation value of the two-body
potential these terms will correspond to the direct, the exchange and the
pairing energies (see (2c.8)).

We remark that the quasi-free states are invariant under Bogoliubov
transformations, i.e., if $\uprho$ is quasi-free and $\W$ is a Bogoliubov
transformation, then the state $\uprho_\W (A) := \uprho (\W A \W^\dagger)$ is
quasi-free, too.

If $\uprho$ is a pure state, i.e., $\uprho (A) = \langle \psi \vert A \vert
\psi
\rangle$, and if the vector $\psi$ lies solely in some fixed $\H^{(N)}$
(including the possibility $N =0$) then $\uprho$ is a quasi-free state if
and only if $\psi$ is a normalized {\it simple} vector
(including the possibility
$\psi = \vert 0 \rangle$).  This state is {\it the usual $N$-particle Slater
determinant state \/} defined by taking expectation values with respect to the
vector $\psi_{\HF}$ given in (2a.1).

We can define the (unbounded) particle {\bf number operator} on $\F$ by the
formal sum
$$\N = \sum \limits^\infty_{N=0} N \Pi^{(N)},\eqno(2a.13)$$
where $\Pi^{(N)}$ is the projector onto the subspace $\H^{(N)} \subset \F$.
(Note that $\1 = \sum \nolimits_N \Pi^{(N)}$.)  A state $\uprho$ {\bf has
finite particle number} if
$$\uprho (\N) := \sum \limits^\infty_{N=0} N
\uprho (\Pi^{(N)})\eqno(2a.14)$$
is finite.  These are the states of primary physical interest.

{\it A} {\bf generalized Hartree-Fock state} {\it is defined to be any
quasi-free state having finite particle number.}

\bigskip
\bigskip\noindent
\vbox{\noindent{\bf II.b  ONE-PARTICLE DENSITY MATRICES}

Let $\uprho$ be a state and let $\{ f_1, f_2, \dots \}$ be an orthonormal
basis in $\H$.  We define the one-particle density matrix (1-$pdm)\
\Gamma$ to be the self-adjoint operator on $\H \oplus \H$ whose matrix
elements are}
$$\left\langle {h_1 \choose h_2} \biggl\vert \Gamma {g_1 \choose g_2}
\right\rangle = \uprho \left( \big[c^\dagger (g_1) + c
(\g_2)\big] \big[ c (h_1) + c^\dagger (\h_2) \big]
\right), \eqno(2b.1)$$
where $\g := \sum \limits_k \overline \mu_k f_k$ when $g = \sum
\limits_k \mu_k f_k$.  Note the ordering of the operators here.  The matrix
$\1 - \Gamma$ has a more ``natural'' appearance; using the CAR we find that
$$\left\langle {h_1 \choose h_2} \ \biggl\vert \ (\1 - \Gamma) {g_1 \choose
g_2} \right\rangle = \uprho \big( [c(h_1) + c^\dagger (\h_2)] [c^\dagger
(g_1) + c^\dagger (\g_2)] \big). \eqno(2b.2)$$

Note that the definition of $\Gamma$ as an operator on $\H \oplus \H$
depends on the choice of the basis $\{ f_1, f_2, \dots \}$.  If this basis
is changed then $\Gamma$ itself
changes if the antiunitary map $g \mapsto \g$ changes (which
may or may not occur).  The underlying reason that $\Gamma$ cannot be
defined in a basis independent way is that $\Gamma$ is not, intrinsically,
a linear map on $\H \oplus \H$; it also has an antilinear component.  This
basis dependence is the price we pay for avoiding the introduction of an
abstract antiunitary map.  The quantities we are going to compute later by
means of the 1-$pdm \ \Gamma$ will, however, be independent of the choice of
the basis $\{ f_1, f_2, \dots \}$, which we shall consider to be fixed
henceforth.

{\bf 2.1 LEMMA:}  {\it For any state $\uprho$ and any orthonormal basis
 $\{
f_1, f_2, \dots \}$
$$0 \leq \Gamma \leq \1 \eqno(2b.3)$$
holds as an operator on $\H \oplus \H$.}

{\it Proof:}  Let $\varphi = (f,g) \in \H \oplus \H$ be normalized.  Then
$1 = \Vert \varphi \Vert^2 = \Vert f \Vert^2 + \Vert g \Vert^2 = \Vert f
\Vert^2 + \Vert \g \Vert^2$ and, from (2b.1) and the CAR, it
follows that
$$\eqalignno{0 \leq \langle \varphi \vert \Gamma \varphi \rangle &= \uprho
\big[ (c^\dagger (f) + c (\g)) (c (f) + c^\dagger
(\g)) \big] \cr
&= (\Vert f \Vert^2 + \Vert \g \Vert^2) \uprho (\1) - \uprho
\big[ (c(f) + c^\dagger (\g)) (c^\dagger (f) + c
(\g))\big] \leq 1. \quad \hbox{\lanbox} \qquad&(2b.4)\cr}$$
It is convenient to view $\Gamma$ as a $2 \times 2$-matrix of operators on
$\H$.  Writing
$$\Gamma =: \pmatrix{\gamma &\alpha \cr\vrule height15ptwidth0pt \alpha^\dagger
&1- \overline \gamma \cr} \eqno(2b.5)$$
one easily finds, using $c^\dagger_k := c^\dagger (f_k)$, that
$$\eqalignno{\langle f_m \vert \gamma f_k \rangle &= \uprho (c^\dagger_k
c^{{\phantom{\dagger}}}_m), \cr
\langle f_m \vert \alpha^\dagger f_k \rangle &= \uprho (c^\dagger_k
c^\dagger_m), \qquad&(2b.6)\cr}$$
where the operator $\overline A$ is defined by
$$\langle f_m \vert \overline A f_k \rangle := \overline{\langle f_m
\vert A f_k \rangle}. \eqno(2b.7)$$
Note that
$$\gamma^\dagger = \gamma, \quad \alpha^T = - \alpha \eqno(2b.8)$$
(where $\alpha^T := \overline{\alpha^\dagger}$).

In view of (2a.11) and the density of polynomials in the algebra of
observables, the 1-$pdm \ \Gamma$ of a quasi-free state $\uprho$
uniquely determines $\uprho$.  More importantly, any $\Gamma$ is the
1-$pdm$ of a quasi-free state $\uprho$, as we shall show in Theorem~2.3
below.
We shall, however, restrict attention to finite particle states.
It is easily seen that $\N=\sum_kc_k^\dagger c_k$ and that,
for any state,
$$
        \uprho(\N)=\sum_k\uprho(c_k^\dagger c_k).
$$

First, we give the relationship between the 1-$pdm \ \Gamma$ of a
state $\uprho$ and the 1-$pdm \ \Gamma_\W$ of the transformed state
$\uprho_\W$, $\uprho_\W (A) := \uprho (\W A \W^\dagger)$, assuming $\W$ is a
Bogoliubov transformation.  Using (2a.9) we can write the Bogoliubov
transformation (2a.8) as
$${d_k \choose d^\dagger_k} = \sum \limits_i \pmatrix{
u^\dagger_{ki} &\v^\dagger_{ki} \cr\vrule height15ptwidth0pt v^\dagger_{ki}
&\overline
u^\dagger_{ki}
\cr} {c_i \choose c^\dagger_i} =: \sum \limits_i (W^\dagger)_{ki} {c_i
\choose c^\dagger_i}. \eqno(2b.9)$$
Note, as indicated above, that each $(W^\dagger)_{ki}$ is a $2 \times 2$
matrix.  We then find that
$$(\1 - \Gamma_\W)_{kl} = \pmatrix{\uprho (d_k d^\dagger_l) &\uprho (d_k
d_l) \cr\vrule height15ptwidth0pt \uprho (d^\dagger_k d_l) &\uprho (d^\dagger_k
d_l) \cr}
= \sum \limits_{i,j} (W^\dagger)_{ki} (\1 - \Gamma)_{ij} (W)_{jl}.
\eqno(2b.10)$$
Thus
$$\Gamma_\W = W^\dagger \Gamma W. \eqno(2b.11)$$

We now give a sufficient condition for the operator $W$ in (2a.9) to
represent a Bogoliubov transformation in $\F$.

{\bf 2.2  THEOREM:}  {\it Any unitary operator $W$ of the form (2a.9) that
satisfies the condition $\Tr
[vv^\dagger] < \infty$ always corresponds to a Bogoliubov transformation
$\W$ on $\F$ with
$$d^\dagger (f) := \W c^\dagger (f) \W^\dagger = c^\dagger (uf) + c(v \overline
f). \eqno(2b.12)$$

Moreover, $\W \vert 0 \rangle = \vert \psi \rangle$ is the state
$$\vert \psi \rangle = \prod \limits_i \Bigl\{ (1 - \zeta_i)^{1/2} +
\zeta^{1/2}_i c^\dagger (h_i) c^\dagger (k_i)\Bigr\} \prod
\limits^r_{i=1} c^\dagger (g_i) \vert 0 \rangle . \eqno(2b.13)$$
Here, $\zeta_1, \zeta_2, \dots$, denote the eigenvalues of $vv^\dagger$
in the interval (0,1) counted with half their multiplicity (these
eigenvalues are all evenly
degenerate), and $r$ is the multiplicity of 1 as
an eigenvalue of $vv^\dagger$.  The vectors $g_1, g_2, \dots , g_r, h_1,
k_1, h_2, k_2, \dots$ form an orthonormal family of eigenvectors of
$vv^\dagger$ with $g_1, g_2, \dots , g_r$ being the eigenvectors of
eigenvalue 1.  The pair of vectors $h_i$, and $k_i := (\zeta_i -
\zeta^2_i)^{-1/2} u\overline v^\dagger \h_i$ are eigenvectors of
$vv^\dagger$ of eigenvalue $\zeta_i$.}

{\it Remark:}  We refer the reader to [AH] where the first
statement of the
lemma, together with its converse (which is not needed here), is proved
([AH], Theorem 7).  However, we are not aware that the explicit formula
(2.22), which we do require, is readily accessible.

{\it Proof:}  The unitarity of the operator $W$ implies the following
conditions similar to (2a.10) for the linear operators $u$ and $v$:
$$\eqalign{u^\dagger u + \v^\dagger \v &= \1 =
uu^\dagger + vv^\dagger \cr
\v^\dagger \u + u^\dagger v &= 0 = u\v^\dagger +
v \u^\dagger. \cr}\eqno(2b.14)$$
Thus, $0 \leq vv^\dagger \leq \1$ as an operator on $\H$
with purely discrete spectrum, thanks to $\Tr[vv^\dagger] < \infty$.
Furthermore, if
$h$ is a normalized eigenvector of $vv^\dagger$ with eigenvalue $0 <
\zeta < 1$ we find, using (2b.14), that $k = (\zeta - \zeta^2)^{-1/2}
u \v^\dagger \h$ satisfies
$$\eqalignno{vv^\dagger k &= (\zeta - \zeta^2)^{-1/2} vv^\dagger u
\v^\dagger \h = - (\zeta - \zeta^2)^{-1/2} v
\,\u^\dagger \v\phantom{|} \v^\dagger \h \cr
&= - \zeta (\zeta - \zeta^2) v \u^\dagger \h =
\zeta k \qquad&(2b.15) \cr }$$
and thus $k$ is also an eigenvector of $vv^\dagger$.  Moreover,
$$\eqalignno{\langle h \ \vert \ k \rangle &= (\zeta - \zeta^2)^{-1/2}
\langle h \ \vert\ u\v^\dagger \h \rangle = - (\zeta -
\zeta^2) \langle h \ \vert \ v \u^\dagger \h \rangle
\cr
&= - (\zeta - \zeta^2) \langle \u v^\dagger h \ \vert \
\h \rangle = - \langle h \ \vert \ k \rangle \qquad&(2b.16) \cr}$$
and hence $\langle h \ \vert \ k \rangle = 0$.  Likewise, we see that $\langle
k \ \vert \ k\rangle = 1$.  Iterating the map from $h$ to $k$ will not
produce more eigenvectors since $(\zeta - \zeta^2)^{-1/2} u
\v^\dagger \k = (\zeta - \zeta^2)^{-1} u\overline
v^\dagger \u v^\dagger h = (\zeta - \zeta^2)^{-1} uu^\dagger
vv^\dagger h = - h$.

We can thus find an orthonormal basis for $\H$ of the form $g_1, \dots ,
g_r, h_1, k_1, h_2, k_2, \dots ,\break l_1, l_2, \dots$, where
$g_1, g_2, \ldots$ and $l_1, l_2, \dots$
are the eigenvectors of $vv^\dagger$ of eigenvalue 1 and 0,
respectively.  Another orthonormal
basis is given by $\v^\dagger \g_1, \dots , \overline
v^\dagger \g_r, \zeta^{-1/2}_1 \v^\dagger \overline
h_1, \zeta^{-1/2}_1 \v^\dagger \k_1, \dots , \break u^\dagger
\l_1, u^\dagger \l_2, \dots$.  To prove this we first note, as is easily
seen, that this is an orthonormal family.  We then note that if $f \in \H$
is orthogonal to all members of this family
 then $\v^\dagger \v f = 0$
(because $\v f$ is orthogonal to $\g_1, \dots , \g_r, \h_1, \k_1, \dots$
but not necessarily to $\l_1, \dots$ ; however $\v^\dagger$ annihilates
$\l_1, \dots$).  Hence $v^\dagger u f
= - \u^\dagger \v f =0$.  Thus $uf$ is an eigenvector for
$vv^\dagger$ of eigenvalue 0, i.e.,
$uf \in {\rm span} \{ l_1, l_2, \dots
\}$, but then $f = (1 - \v^\dagger \v) f = u^\dagger uf
\in \ {\rm  span} \{ u^\dagger l_1, u^\dagger l_2, \dots \}$, and thus $f =
0$.

Using this latter basis we define, in agreement with (2b.12),
$$d (\v^\dagger \g_i) := c (u \v^\dagger
\g_i) + c^\dagger (vv^\dagger g_i) = c^\dagger (g_i) \quad i = 1,
\dots , r, \eqno(2b.17)$$
where we have used $u \v^\dagger \g_i = - v \overline
u^\dagger \g_i = 0$, which follows from
$uu^\dagger g_i = (\1 - vv^\dagger) g_i = 0$.  We also make the
definitions, for $i = 1,2, \dots$,
$$\eqalignno{d (\zeta^{-1/2}_i \v^\dagger \h_i) &:= \zeta^{-1/2}_i c(u
\v^\dagger \h_i) + \zeta^{-1/2}_i c^\dagger (vv^\dagger h_i) = (1 -
\zeta_i)^{1/2} c(k_i) + \zeta^{1/2}_i c^\dagger (h_i), \qquad&(2b.18)\cr
d (\zeta^{-1/2}_i \v^\dagger \k_i) &:= \zeta^{-1/2}_i c (u \v^\dagger
\k_i) + \zeta^{-1/2}_i c^\dagger (vv^\dagger k_i) = - (1 -
\zeta_i)^{1/2} c (h_i) + \zeta^{1/2}_i c^\dagger (k_i),
\qquad&(2b.19)\cr
d (u^\dagger l_i) &:= c(uu^\dagger l_i) + c^\dagger (v\u^\dagger \l_i) =
c(\l_i). \qquad&(2b.20)\cr}$$

We shall now show that (2b.13) defines a normalized vector $\vert \psi
\rangle$ in $\F$ annihilated by all the operators in (2b.18--20).  To show
that the somewhat formal expression on the right side
in (2b.13) defines a vector in $\F$ we
expand the (possibly infinite) product, thereby arriving at a (possibly
infinite) sum of (possibly infinite) products.  Each term in this sum that
contains a product of infinitely many $c^\dagger$'s is zero since it will
also contain a product of infinitely many $\zeta^{1/2}_i$, and $\zeta_i
\rightarrow 0$ as $i \rightarrow \infty$.  Hence, the sum is at most a
countable sum of orthogonal simple vectors in $\F$.  An appropriate
truncation of this sum gives
$$\vert \psi_N \rangle := \prod \limits_{i > N} (1 - \zeta_i)^{1/2} \prod
\limits_{i \leq N} [(1 - \zeta_i)^{1/2} + \zeta^{1/2}_i c^\dagger (h_i)
c^\dagger (k_i)] \prod \limits^r_{i=1} c^\dagger (g_i) \vert 0 \rangle.
\eqno(2b.22)$$
We note that $\langle \psi_N \ \vert \ \psi_N \rangle = \prod \nolimits_{i
> N} (1 - \zeta_i)
\geq 1 - \sum \nolimits_{i > N} \zeta_i$,
which is non-zero and converges to 1 as $N
\rightarrow \infty$ since $\sum \nolimits_i \zeta_i < \infty$.  If $M >
N$ then $\vert \psi_N \rangle$ is orthogonal to $\vert \psi_M - \psi_N
\rangle$ and
$$\langle \psi_M - \psi_N \ \vert \ \psi_M - \psi_N \rangle = \langle
\psi_M \ \vert \ \psi_M \rangle - \langle \psi_N \ \vert \ \psi_N \rangle =
\prod \limits_{i > M} (1 - \zeta_i) - \prod \limits_{i > N} (1 -
\zeta_i) \rightarrow 0 \eqno(2b.23)$$
as $M \rightarrow \infty$.  Hence there is a normalized $\vert \psi \rangle
\in \F$ such that $\vert \psi_N \rangle \rightarrow \vert \psi \rangle$ as
$N \rightarrow \infty$.  It is easy to check that any operator in the list
(2b.18--20) will annihilate $\vert \psi_N \rangle$ if $N$ is large enough.
Thus $\vert \psi \rangle$ is annihilated by all the operators in
(2b.18--20),
and hence by any operator $d (f) := c^\dagger (uf) + c(v \overline f)$.

Denoting $d^\dagger_k := c^\dagger (uf_k) + c(vf_k)$ (recall $\overline f_k
= f_k$) we define $\W: \F \rightarrow \F$ by
$$\W c^\dagger_1 \dots c^\dagger_n \vert 0 \rangle = d^\dagger_1 \dots
d^\dagger_n \vert \psi \rangle \eqno(2b.24)$$
for all $n \geq 0$.  The unitarity of $W$ implies that $d^\dagger_k$
satisfy the CAR and hence that $\W$ is an isometry.  It is also clear that
$\W$ satisfies (2b.12).

To show that $\W$ is unitary we first observe that $\vert 0 \rangle$ can be
written as a linear combination of vectors of the form (2b.24).  This is
easily seen by interchanging the roles of $\vert 0 \rangle$ and $\vert \psi
\rangle$ and of $W$ and $W^\dagger$ in the argument which led to the
construction, (2b.13), of $\vert \psi \rangle$.  From this it follows
that all simple vectors $c^\dagger_1 \dots c^\dagger_n \vert 0 \rangle$ can
be written as combinations of the vectors of the form (2b.24).  Indeed, from
(2b.14), $c^\dagger_k = c^\dagger ((uu^\dagger + vv^\dagger) f_k) + c((u
\v^\dagger + v\u^\dagger) \overline f_k) = d^\dagger (u^\dagger f_k) + d
(\v^\dagger \overline f_k)$.
Thus $\W$ is invertible and hence unitary. \lanbox

{\bf 2.3  THEOREM:}  {\it Let $0 \leq \Gamma \leq \1$ be an operator on $\H
\oplus \H$ of the form
(2b.5), subject to (2b.8), and assume furthermore
that $\Tr [\gamma] < \infty$.  Then there exists a unique quasi-free state
$\uprho$ with finite particle number such that $\Gamma$ is the 1-$pdm$
of $\uprho$.}

{\it Remark:}  Suppose we are given a $\gamma$ satisfying $0 \leq \gamma
\leq 1$.  We can set $\Gamma = \pmatrix{\gamma &0 \cr 0&1 - \gamma \cr}$,
i.e. set $\alpha = 0$ and, according to Theorem~2.3, we can extend $\Gamma$
to a quasi-free state $\uprho$.  In other words, given $\gamma$ we can find
a {\bf particle conserving} quasi-free state having this $\gamma$ as its
1-$pdm$ (whether or not $\Tr \gamma$ is an integer).  By a {\bf particle
conserving state} we mean a state (such as a Gibbs state for a particle
conserving Hamiltonian) that is a convex combination of states, each having
a definite particle number, i.e. there are no $cc$ or $c^\dagger c^\dagger$
matrix elements in this state, which is the same thing as saying $\alpha =
0$.  However, unless $\gamma$ is a projection (i.e., a usual HF state)
$\Gamma$ cannot have a {\it definite\/} particle number, i.e. $\uprho
(\N^2) > \uprho (\N)^2$.  In fact, defining $\uprho (\N^2) = \sum
\nolimits_N N^2 \uprho (\pi^{(N)}) = \sum \nolimits_{k,l} \uprho
(c^\dagger_k c_k c^\dagger_l c_l)$, we can use (2a.12) to compute
$$
        \uprho(\N^2)=\uprho(\N)^2 +\sum_{k,l}
        \left(\gamma^{\phantom{\dagger}}_{lk}
        \gamma^{\phantom{\dagger}}_{kl}
        +\alpha^\dagger_{lk}\alpha^{\phantom{\dagger}}_{lk}\right)
+\sum_k\gamma^{\phantom{\dagger}}_{kk}
        =\uprho(\N)^2+\Tr[\gamma-\gamma^2]+\Tr[\alpha^\dagger\alpha],
        \eqno(2b.25)
$$
from which we see that $\uprho (\N^2) = \uprho (\N)^2$ requires {\it both}
$\alpha =0$ {\it and} $\gamma^2 = \gamma$.

{\it Proof:}  Since $\Tr [\gamma]$ is finite it is clear from the form in
(2b.5) that although $\Tr [\Gamma]$ may be infinite, which is the case when
$\dim \H = \infty$, we have $\Tr [\Gamma (\1 - \Gamma)] < \infty$.  Thus,
there is an orthonormal basis of eigenvectors for $\Gamma (\1 - \Gamma)$.
If $\varphi$ is an eigenvector for $\Gamma (\1 - \Gamma)$ of eigenvalue
$\mu$ then so is $\Gamma \varphi$ and, since $\Gamma^2 \varphi
= \Gamma \varphi - \mu \varphi$, it follows that $\Gamma$ leaves
invariant the subspace space $\{ \varphi, \Gamma \varphi \}$, which is at
most 2-dimensional.  We conclude that there is an orthonormal basis of
eigenvectors for $\Gamma$.

If $\varphi = f \oplus g$ is a normalized eigenvector for $\Gamma$ of
eigenvalue $\lambda$ then, by (2b.5) and
(2b.8), we find that $\widetilde
\varphi = \g \oplus \overline f$ is a normalized eigenvector for $\Gamma$ of
eigenvalue $(1 - \lambda)$.  Thus, we can find a unitary $W$ on $\H \oplus
\H$ of the form (2a.9) such that (using the basis $f_1, f_2, \dots$ for both
copies of $\H$) the four blocks of the transformed $\Gamma$ have the
form
$$W^\dagger \Gamma W = \pmatrix{\lambda_1 &&&&&\cr &\lambda_2 &&&0&\cr
&&\ddots &&&\cr &&&1 - \lambda_1 &&\cr &0&&&1 - \lambda_2 \cr &&&&&\ddots
\cr}, \eqno(2b.26)$$
with $0 \leq \lambda_i \leq 1/2$ for $i = 1,2, \dots$.
Using this and $\Tr [\Gamma
(\1 - \Gamma)] < \infty$, we obtain $\sum \nolimits_i \lambda_i <
\infty$.

We shall now prove that $W$ satisfies the condition of Theorem~2.2, i.e., that
$\Tr [vv^\dagger] < \infty$ and hence that $W$ corresponds to a Bogoliubov
transformation $\W$ on $\F$.  Indeed, from (2b.26) we know that the upper
left block of the matrix $W^\dagger \Gamma W$ has finite trace, i.e.,
$$\Tr [u^\dagger \gamma u + \v^\dagger \alpha^\dagger u + u^\dagger \alpha
\v + \v^\dagger (1 - \overline \gamma) \v] < \infty. \eqno(2b.27)$$
Since $0 \leq \Gamma \leq \1$ we have that $\Gamma^2 \leq \Gamma$.  The
upper left block of this inequality reads $\gamma^2 + \alpha \alpha^\dagger
\leq \gamma$.  Thus $\Tr [\alpha \alpha^\dagger] \leq \Tr [\gamma] <
\infty$.  By Cauchy-Schwarz we estimate
$$\eqalignno{\Tr [u^\dagger \gamma u + \v^\dagger \alpha^\dagger u +&
u^\dagger \alpha \v + \v^\dagger (1- \overline \gamma) \v] - \Tr
[\v^\dagger \v] \cr
&= \Tr [u^\dagger \gamma u + \v^\dagger \alpha^\dagger u + u^\dagger \alpha
\v - \v^\dagger \overline \gamma \v] \cr
&\geq - \Tr [uu^\dagger \gamma] - 2 \Tr [\v^\dagger \v]^{1/2} \Tr
[uu^\dagger \alpha \alpha^\dagger]^{1/2} - \Tr [\v \,\v^\dagger \overline
\gamma] \cr
&\geq - 2 \Tr [\gamma] - 2 \Tr [\v^\dagger \v]^{1/2} \Tr [\gamma]^{1/2}.
\qquad&(2b.28) \cr}$$
We conclude from (2.31) that $\Tr [\v^\dagger \v] - 2 \Tr [\v^\dagger
\v]^{1/2} \Tr [\gamma]^{1/2} < \infty$ and hence that $\Tr [vv^\dagger] =
\Tr [\v^\dagger \v] < \infty$.

If we prove that the diagonal matrix $W^\dagger \Gamma W$ is the 1-$pdm$ of
a quasi-free state $\widetilde{\uprho}$ then we know that $\Gamma$ is the
1-$pdm$ of the state $\uprho$ with $\uprho (A) = \widetilde{\uprho}
(\W^\dagger A \W)$.  We may therefore assume that $\Gamma$ is itself
diagonal of the form (2.26).

Let $\Pi_0$ be the projection onto the subspace of $\F$ on
which $\sum \nolimits_{i: \lambda_i =0} c^\dagger_i c_i =0$.

For each $i$ such that
$\lambda_i > 0$ choose $e_i$ to satisfy
$$\left(1 + \exp (e_i)\right)^{-1} = \lambda_i, \eqno(2b.29)$$
(note that $0\leq e_i<\infty$, since $0<\lambda_i\leq1/2$) and consider the
following (possibly unbounded) operator $H$ on $\F$.
$$H = \sum \limits_{i: \lambda_i \not= 0} e_i c^\dagger_i c_i.
\eqno(2b.30)$$
We shall now prove that the operator
$$G :=  \Pi_0 \exp ( -H ) \eqno(2b.31)$$
has finite trace on $\F$ and that the state
$$\uprho (A) = \Tr [G]^{-1} \Tr [AG] \eqno(2b.32)$$
is quasi-free and has $\Gamma$ as its 1-$pdm$.  It is easy to see that the
trace of $G$ is
$$\Tr [G] = \prod \limits_{i: \lambda_i \not= 0} [1 + \exp (-e_i)] = \prod
\limits_i (1 - \lambda_i)^{-1} < \infty, \eqno(2b.33)$$
since $\sum \limits_i \lambda_i < \infty$.

The operator $G$ looks peculiar but, by introducing the operator $H^\prime =
\sum \limits_{i: \lambda_i = 0} c^\dagger_i c_i$ which commutes with $H$, we
can write the state $\uprho$ as a limit of Gibbs states:
$$\uprho (A) = \lim \limits_{\beta \rightarrow \infty} Z^{-1}_\beta \Tr [A
\exp (-\beta H^\prime - H)]. \eqno(2b.34)$$
It is well-known (see [GM] for a simple proof)
that the Gibbs state for
an operator of the form $\sum \limits_i e_i c^\dagger_i c_i$ is quasi-free.
Hence we conclude (by taking the limit $\beta \rightarrow \infty$)
that $\uprho$ is quasi-free.

The fact that $\Gamma$ is the 1-$pdm$ of $\uprho$ follows from the
computation
$$\eqalignno{\uprho ((c^\dagger_k + c^{\phantom{\dagger}}_l) (
c^{\phantom{\dagger}}_m + c^\dagger_n)) &= \Tr
[G]^{-1} \Tr [(c^\dagger_k + c^{\phantom{\dagger}}_l)
(c^{\phantom{\dagger}}_m + c^\dagger_n) G] \cr
&= \delta_{km} \Tr [G]^{-1} \Tr [c^\dagger_k c^{\phantom{\dagger}}_k G]
+ \delta_{ln} \Tr
[G]^{-1} \Tr [c^{\phantom{\dagger}}_l c_l^\dagger G] \cr
&= \prod \limits_i (1 - \lambda_i) \left( \delta_{km} \lambda_k (1 -
\lambda_k)^{-1} \prod \limits_{i: i \not= k} (1 - \lambda_i)^{-1} +
\delta_{ln} \prod \limits_{i: i\not= l} (1 - \lambda_i)^{-1}\right) \cr
&=\delta_{km} \lambda_k + \delta_{ln} (1 - \lambda_l),
\qquad&(2b.35) \cr}$$
where we have used the fact that $\exp [-e_k] = \lambda_k (1 -
\lambda_k)^{-1}$ if $\lambda_k \not= 0$. Finally $\uprho$ has finite
particle number since $\uprho (\N) = \Tr [\gamma]$.

The uniqueness of $\uprho$ follows as in the remark after (2b.8):  The
1-$pdm$ of a quasi-free state determines the state.  \lanbox

We call an operator $\Gamma$ {\bf admissible} if it
satisfies the properties in Theorem~2.3, i.e., is of the
form (2b.5) subject to (2b.8) with $\Tr[\gamma]<\infty$ and
$0\leq\Gamma\leq\1$. We then have that:
\smallskip
\centerline{\it $\Gamma$ is admissible if and only if it is the
1-$pdm$ of a generalized Hartree-Fock state.}
\smallskip

In the above proof we not only proved the existence of a quasi-free state
having $\Gamma$ as its 1-$pdm$, we also gave the explicit form of $\uprho$.
To make this more explicit we introduce the following notion:

{\bf Quadratic Hamiltonian:}  {\it
A self-adjoint operator $H$ (bounded or unbounded) on $\F$ is
said to be a quadratic Hamiltonian if the unitary
operators
$\W (t) := \exp(iHt)$ are Bogoliubov transformations for all $t$.}

If $H$ is a quadratic Hamiltonian there correspond operators
$W(t)$ on $\H\oplus\H$ of the form (2a.9)
corresponding to the Bogoliubov transformations $\W(t)$. Since the
anticommutator satisfies
$$
        \left\{c^\dagger(\h_2)+c(h_1), \exp(iHt)\left(c^\dagger(g_1)+c(\g_2)
        \right)
        \exp(-iHt)\right\}
        =\left\langle {h_1 \choose h_2} \ \biggl\vert \ W(t)
        {g_1 \choose g_2} \right\rangle \1,\eqno(2b.36)
$$
it follows that $W(t)$ is a strongly continuous one-parameter group
of unitaries on $\H \oplus \H$. Hence there is a self-adjoint
operator $A$ (bounded or unbounded) such that $W(t)=\exp(iAt)$.

The operator $A$ has the block structure
$$A = \pmatrix{a&b\cr\vrule height15ptwidth0pt b^\dagger &-\overline a \cr},
\eqno(2b.37)$$
where $a$ and $b$ are operators on $\H$ with $a^\dagger = a$ and $b^T =
-b$.

We call\footnote{$^\dagger$}{\eightpoint Our terminology is far from being
conventional, but it is descriptive.} the operator $A$ the
{\bf first quantization of} $H$ and we call $H$ {\it a }
{\bf second quantization of} $A$. Notice that from (2b.36)
$A$ is determined uniquely by $H$. If $H$ is a bounded operator
on $\F$ then $A$ is bounded on $\H \oplus \H$ and by differentiating
(2b.36) we can, in this case,  write
$$
\{ c^\dagger (\h_2) + c(h_1), [H, c^\dagger (g_1) + c(\g_2)]\} =
\left\langle {h_1 \choose h_2} \ \biggl\vert \ A {g_1 \choose g_2}
\right\rangle \1. \eqno(2b.38)$$
Here we have introduced the commutator $[K_1, K_2] := K_1 K_2 - K_2
K_1$.

The operator $H$ , however, is only determined by $A$ up to addition of
a multiple of the identity. If $A$ is bounded we may write the
{\it unique} second quantization, $H$, of $A$ satisfying
$\langle0\vert H\vert0\rangle=0$ in terms of
the matrix elements of $a$ and $b$, as
$$
        H=\sum_{i,j}a_{ij}c_i^\dagger c^{\phantom{\dagger}}_j
        +\mfr1/2\sum_{i,j}
        \left (b_{ij}c_i^\dagger c_j^\dagger +
        b^\dagger_{ij}c^{\phantom{\dagger}}_ic^{\phantom{\dagger}}_j
        \right).\eqno(2b.39)
$$

If $H$ is a quadratic Hamiltonian and if $\W$ is a Bogoliubov
transformation then $\W H \W^\dagger$ is also a quadratic Hamiltonian.  If
$H$ is a second quantization of $A$ then $\W H \W^\dagger$ is a second
quantization of $WAW^\dagger$, where $W$ is the unitary given in (2a.9).

The proof of Theorem~2.3 implies the following result about the structure of
quasi-free states.

{\bf 2.4  LEMMA:}  {\it Let $\uprho$ be a quasi-free state with finite particle
number, i.e., $\uprho (\N) < \infty$ (in terms of its 1-$pdm$ this means
$\Tr [\gamma] < \infty$).  Then there exist two commuting quadratic
Hamiltonians $H$ and $H^\prime$ (possibly $H=0$ or $H^\prime=0$, but not both)
such that
$$\uprho (B) = \lim \limits_{\beta \rightarrow \infty} \Tr [\exp (-\beta
H^\prime - H)]^{-1} \Tr [B \exp (-\beta H^\prime - H)]. \eqno(2b.40)$$}

This means that the state $\uprho$ is a product of the ground state (zero
temperature state) for $H^\prime$ and the Gibbs state for $H$.  Thus if
$H^\prime = 0$, $\uprho$ is a Gibbs state and if $H = 0$, $\uprho$ is a pure
state.  In the next two lemmas we discuss quasi-free Gibbs states and
quasi-free pure states in more detail.

{\bf 2.5  LEMMA:}  {\it If $A$ is an operator on $\H \oplus \H$ of the form
(2b.37) then
$$\Gamma := (\1 + \exp(A))^{-1} \eqno(2b.41)$$
is of the form (2b.5) subject to (2b.8).  Furthermore, if this $\Gamma$
satisfies $\Tr [\gamma ] < \infty$, so that it defines a quasi-free state
$\uprho$ by Theorem~2.3, then
this $\uprho$ is given by
$$\uprho (B) = \Tr [\exp(-H)]^{-1} \Tr [B\exp(-H)], \eqno(2b.42)$$
where $H$ is any second quantization of $A$.

Conversely, if $\uprho$ in (2b.42) is a quasi-free state of finite particle
number with 1-$pdm$ $\Gamma$, then $H$ is quadratic with first quantization
satisfying (2b.41).}

{\it Proof:}  Let $U$ be the unitary
$$U = \pmatrix{0 &\1\cr \1&0\cr} \eqno(2b.43)$$
on $\H \oplus \H$.  Then $UAU^\dagger = - \overline A$, since $A$ has the
form (2b.37).  Hence, $U \Gamma U^\dagger = \1 - \overline \Gamma$ which
proves that $\Gamma$ is of the form (2b.5) subject to (2b.8).

If $\Gamma$ satisfies $\Tr [\gamma] < \infty$ it follows from the proof of
Theorem~2.3 that there is a Bogoliubov transformation $\W$ with corresponding
$W$ such that the $W^\dagger \Gamma W$, which is the 1-$pdm$ of the
transformed state $\uprho_\W$, is diagonal in our chosen basis.  We
denote by $\lambda_1, \lambda_2,\ldots$, those eigenvalues of
$W^\dagger \Gamma W$ in the interval $[0,1/2)$ together with
half the eigenvalues equal to $1/2$ (if any). The other half of the
eigenvalues of $W^\dagger \Gamma W$ are then given by
$1-\lambda_1,1-\lambda_2,\ldots$.
The operator $W^\dagger AW$ is
also diagonal with the first half of the eigenvalues given by
$e_1,e_2,\ldots$ and the second half by $-e_1,-e_2,\ldots$, according
to the definition (2b.41).

Since
$$\exp \Bigl(i \sum \limits_k e_k c^\dagger_k c_k\Bigr) c^\dagger_l \exp
\Bigl( -i
\sum \limits_k e_kc^\dagger_k c_k \Bigr) = \exp(ie_l) c^\dagger_l
\eqno(2b.44)$$
we see that the second quantizations of $W^\dagger AW$ are of the form
$\widetilde H_\tau = \sum\limits_k e_k c^\dagger_k c_k + \tau \1$, where
$\tau$ is any real number.  Since the Gibbs states are independent of
$\tau$ we see that all the operators $\widetilde H_\tau$ define the same
Gibbs state as the operator in (2.33), i.e., the state $\uprho_\W$.  (As in
(2b.42) we are referring to Gibbs states with the inverse temperature $\beta
= 1$.)

Since $\uprho_\W$ is the Gibbs state for $\widetilde H_\tau$, the state
$\uprho$ is given by
$$\uprho (B) = \uprho_\W (\W^\dagger B \W) =
\Tr [\exp({-\widetilde H_\tau})
]^{-1} \Tr [\W^\dagger B \W \exp({-\widetilde H_\tau})], \eqno(2b.45) $$
which agrees with (2b.42) if $H = \W \widetilde H_\tau \W^\dagger$.  Such an
$H$ is however, a second quantization of $WW^\dagger AWW^\dagger = A$.

The converse statement is also a simple consequence of the proof of
Theorem~2.3.
One only has to realize that the state $\uprho$ in (2b.42) uniquely
determines the operator $H$. \lanbox

{\bf 2.6 THEOREM:}  {\it A quasi free state $\uprho$ with finite particle
number is a pure state $\uprho (B) = \langle \psi \vert B \vert \psi
\rangle$ if and only if the 1-$pdm$ $\Gamma$ with $\Tr [\gamma] < \infty$
is a projection on $\H \oplus \H$, i.e., $\Gamma^2 = \Gamma$.

In terms of $\Gamma$ the vector $\vert \psi \rangle$ is of the form (2b.13)
but this time with $g_1, \dots , g_r, h_1, k_1, \dots$ being orthonormal
eigenvectors of $\gamma$; the vectors $g_1, \dots g_r$ with eigenvalue 1
and the pair $h_1, k_i := - (\zeta_i - \zeta_i^2)^{-1/2} \alpha
\overline{h_i}$ with eigenvalue $\zeta_i$, where $0 < \zeta_i < 1$.}

{\it Proof:}
Let $\uprho$ be a pure state. Since $\uprho$ is uniquely determined
by its 1-$pdm$ $\Gamma$, we may assume it to be of the form (2b.32).
The purity of $\uprho$, then, is equivalent to $G$ being rank one and
hence $\lambda_i =0$ for all $i$,
which is equivalent to $\Gamma$ to be a projection.

Again from the proof of Theorem~2.3 the diagonal form of $\Gamma$ is
$$W^\dagger \Gamma W = \pmatrix{0 &0\cr 0&\1 \cr}. \eqno(2b.46)$$
Hence
$$\Gamma = W \pmatrix{0&0\cr 0&\1\cr} W^\dagger = \pmatrix{vv^\dagger &v
\u^\dagger \cr \u v^\dagger &\u\,\u^\dagger \cr} \eqno(2b.47)$$
and we have $\gamma = vv^\dagger$ and $\alpha = v\u^\dagger = -
u\v^\dagger$.  Since $W^\dagger \Gamma W$ is the 1-$pdm$ of the pure state
corresponding to the vacuum, $\Gamma$ is the 1-$pdm$ of the pure state
corresponding to $\vert \psi \rangle = \W \vert 0 \rangle$.  Here $\W$ is
the Bogoliubov transform defined by $W$.  The last statement of the lemma
now follows by comparison with Theorem~2.2.  \lanbox

Using this lemma we can find a basis for $\H\oplus\H$
where the blocks  $\gamma$ and $\alpha$ of  $\Gamma$ take
a particularly simple form when $\Gamma$ is a projection.
In fact, if we choose the basis
consisting of
$g_i \oplus 0, l_i \oplus 0, h_i \oplus k_i$ and
$0 \oplus \g_i, 0 \oplus  \l_i, \k_i \oplus \h_i$
for all $i = 1,2, \ldots$,
we find (with $\alpha_i
=(\zeta_i-\zeta_i^2)^{-1/2}$)
$$
        \gamma=\pmatrix{\ddots&&&\cr&\zeta_i&&\cr&&\zeta_i&\cr
                 &&&\ddots}
\quad\hbox{and}\quad
        \alpha=\pmatrix{\ddots&&&\cr
        &0&\alpha_i&\cr&-\alpha_i&0&\cr&&&\ddots}. \eqno(2b.48)
$$

{\bf 2.7 LEMMA:} {\it Let $\uprho$ be a pure and quasi-free
state of finite particle number with \hfill\break
1-$pdm\ \Gamma$.
Then $\uprho (\N^2) < \infty$ and it is given by (2b.49) with $\alpha$ given
in (2b.5).
$$
        \uprho(\N^2)-\uprho(\N)^2=2\Tr[\alpha^{\dagger}\alpha].
        \eqno(2b.49)
$$
This equation shows that $\uprho$ is not necessarily a fixed
particle number state.}

{\it Proof:} Since $\uprho$ is pure, $\Gamma$ is a projection and
hence $\gamma=\gamma^2+\alpha\alpha^\dagger$ which
together with (2b.25) implies (2b.49).\lanbox

{F}rom  Lemma~2.7 we see that a generalized HF state
with 1-$pdm\ \Gamma$ has conserved particle
number if and only if the component $\alpha$ of $\Gamma$
vanishes. We call a generalized
HF state for which this holds a {\bf normal state}.

\bigskip
\noindent
{\bf II.c THE GENERALIZED HF FUNCTIONAL}

In this section we shall introduce the generalized Hartree-Fock approximation
for a self-adjoint operator $H$ on Fock space of the form
$$
        H=\widehat h+\widehat V,\eqno(2c.1)
$$
where $\widehat h$ is a quadratic (particle number preserving) operator
$$
        \widehat h=\sum_{i,j}h_{ij}c^\dagger_ic^{\phantom{\dagger}}_j,
        \eqno(2c.2)
$$
and $\widehat V$ is a {\bf quartic operator} (again particle number preserving)
$$
        \widehat V=\mfr1/2\sum_{k,l,m,n}V_{kl;mn}c^\dagger_k
        c^\dagger_lc^{\phantom{\dagger}}_nc^{\phantom{\dagger}}_m.
        \eqno(2c.3)
$$

By (2c.2) the operator $\widehat h$ is defined in terms of matrix elements
$h_{ij}$ of a self-adjoint operator $h$ on $\H$.
The operator $h$ is the restriction of $\widehat h$
to the one-body space $\H^{(1)}=\H$, i.e.,
$$
        h_{ij}=\langle f_i\vert h\vert f_j\rangle=\langle 0\vert
        c^{\phantom{\dagger}}_i \widehat hc_j^\dagger\vert 0\rangle.
        \eqno(2c.4)
$$

By (2c.3) the operator $\widehat V$ is defined in terms of matrix elements
$V_{kl;mn}$ of a self-adjoint operator $V$ on $\H\otimes\H$.
Note that we are not restricting $V$ to
be an operator on the antisymmetric two-body space $\H^{(2)}=\H\wedge\H$.
The restriction of $V$ to the antisymmetric subspace is equal to the
restriction of $\widehat V$ to $\H^{(2)}$.
$$
\eqalignno{\langle0\vert c_lc_k \widehat V c^\dagger_mc^\dagger_n\vert 0\rangle
&=\mfr1/2(V_{kl;mn}+V_{lk;nm}-V_{kl;nm}-V_{lk;mn})\cr
&=\mfr1/2\langle f_k\wedge f_l\vert V\vert f_m\wedge f_n\rangle.
\qquad&(2c.5)\cr}
$$

The operators $V$ and $h$ defining $\widehat V$ and $\widehat h$ may be
bounded or unbounded.
We shall, however, assume that the operator $H$ is bounded below
This is the case if, for example,
$\widehat h$ and $\widehat V$ are bounded below.
One way to ensure this is to assume that $h$ and $V$ are
bounded below by (negative) operators of finite trace (it is not
enough to assume that $h$ and $V$ are bounded below
in order to have $\widehat h$ and $\widehat V$ bounded below.).

The expressions (2c.2--3) are somewhat formal.
A more precise definition can be given as follows. On each $\H^{(N)}$
we can, in the obvious way, define the sum
$h^{(N)}=\sum_{i=1}^N h_i$ of $N$ commuting copies of $h$ and
the corresponding sum
$V^{(N)}=\sum_{1\leq i<j\leq N}V_{ij}$ of
$N$ commuting copies of $V$. Then $H=\sum_N(h^{(N)}
+V^{(N)})\Pi^{(N)}$.

In discussing Hamiltonians of the form (2c.1) we have two particular
examples in mind.  The first is the
{\it Hubbard Hamiltonian} defined in Sect.~III. The second is the {\it
atomic Hamiltonian}
with  $\H$ being the square integrable (spinor valued) functions on $\R^3$
and $h=-(\hbar^2/2m)\Delta-Z/|x|-\mu$, where $\mu\leq0$
is a chemical potential
and $V=e^2|x-y|^{-1}$. Both $h$ and $V$ are independent of (diagonal in) spin.
The Hubbard Hamiltonian is a bounded operator (in fact a finite
dimensional matrix)
while the atomic Hamiltonian is unbounded but bounded below.
[If $\mu<0$ then $h$ and $V$ are bounded below by operators
of finite trace,
but if $\mu=0$, $h$ will not be bounded below by such an
operator (e.g., the negative eigenvalues of hydrogen are
not summable). This, however, is
not a real problem because the operator $H$ is still bounded below.]

For both the Hubbard Hamiltonian and the atomic Hamiltonian we shall be
interested in the ground
state and its energy. The ground state is simply the (maybe not unique)
state $\uprho_0$ (with finite particle number) for which
$\uprho(H)$ takes
on the smallest possible value -- the ground state energy --
provided this smallest value is attained for some state at all.
Otherwise the ground state does not exist.

If $H$ is unbounded
the expectation $\uprho(H)$ is not necessarily well-defined. If,
however, $H$ is bounded below we can define $\uprho(H)$.
This is easy to see for states that can be written as
$\uprho(B)=\Tr[GB]$ for some positive operator $G$
of finite trace on Fock space (this is not true for all states, but we are only
interested
in states for which it holds).
If $H$ is bounded from below, we can without loss of generality assume that
$H$ is positive. Then $\Tr[GH]$ is, when expanded in the
eigenvector basis for $G$, an infinite
sum of positive terms. This sum then defines $\uprho(H)$ (possibly to be $+
\infty$).

The expected number of particles in the ground state is $\uprho(\N)$.
Since both
Hamiltonians are particle number preserving there is a ground state $\uprho$
with a fixed number of
particles, i.e., $\uprho(\N^2)=\uprho(\N)^2$. The number of particles
$N:=\uprho(\N)$ is thus
an integer and $\uprho$ is, in fact, a state on $\H^{(N)}$
(i.e., $\uprho(\Pi^{(N)})=1$).

Alternatively to specifying a chemical potential $\mu$,
we could also have specified the
number $N$ and then considered the problem on $\H^{(N)}$. The equivalence of
the two descriptions by Legendre transform
(equivalence of the grand  canonical and canonical ensembles)
requires that the ground state energy is a convex function
of the particle number. While this is
believed to be the case there is, to the best of our knowledge, no rigorous
proof of this fact in the two models discussed. In the grand canonical
picture the ground state energy {\it is} a concave function of the chemical
potential, but as long as we do not know the convexity of
the canonical energy as a function of $N$ we cannot assert that the
two energy functions are Legendre transforms of each other.
Here we shall mostly work in the grand canonical framework,  i.e.,
specify a chemical potential, except at the end of the section
where we shall discuss the canonical picture when $V$ is assumed to be
positive.

In addition to the ground state energy we shall also
be interested in the grand canonical Gibbs states
$\uprho(B)=Z^{-1}\Tr[B\exp(-\beta H)]$.
The Gibbs state, however, is not well defined for the atomic
Hamiltonian since the
operator $\exp(-\beta H)$ will not have finite trace in
this case.

The object of study in this section is not the real ground states
and Gibbs states
but rather their (generalized) Hartree-Fock approximations which we
shall now define.

The Hartree-Fock approximation to the ground state is simply the generalized
Hartree-Fock state with least possible energy.
By Theorem~2.3 there is a one-to-one correspondence between a generalized HF
state $\uprho$
and its 1-$pdm$ $\Gamma$. We may therefore define the {\bf generalized
Hartree-Fock energy functional},
$$\E(\Gamma)=\uprho(H), \eqno(2c.6)$$
on the set of all admissible density matrices.

The energy of a generalized Hartree-Fock state
can be computed in terms of the \hfill\break
1-$pdm\ \Gamma$ as follows. The expectation of the
quadratic part is $ \uprho(\widehat h)=\Tr[h \gamma]$. In computing the
expectation of the quartic part we apply (2.12) and obtain
$$
        \uprho(\widehat V)=\mfr1/2\sum_{k,l,m,n}
        V^{\phantom{\dagger}}_{kl;mn}
        \left(\gamma_{mk}^{\phantom{\dagger}}
        \gamma^{\phantom{\dagger}}_{nl}
        -\gamma^{\phantom{\dagger}}_{ml}
        \gamma^{\phantom{\dagger}}_{nk}+\alpha^\dagger_{lk}
        \alpha^{\phantom{\dagger}}_{mn}\right).\eqno(2c.7)
$$
The operator $G^{(2)}$ on $\H\otimes\H$ with matrix elements
$G^{(2)}_{mn;kl}=\gamma_{mk}^{\phantom{\dagger}}
        \gamma^{\phantom{\dagger}}_{nl}
        -\gamma^{\phantom{\dagger}}_{ml}
        \gamma^{\phantom{\dagger}}_{nk}+\alpha^\dagger_{lk}
        \alpha^{\phantom{\dagger}}_{mn}$
has finite trace. In fact, if we choose
$V=I$ in (2c.7) we obtain $\widehat V=\N(\N-1)$ and
$\Tr[G^{(2)}]=\uprho(\N^2)-\uprho(\N)$ which is finite by Lemma~2.7.

Equation (2c.7) states that $\uprho(\widehat V)=\mfr1/2\Tr[VG^{(2)}]$.
Both $\Tr[VG^{(2)}]$ and $\Tr[h\gamma]$ are well-defined since
$V$ and $h$ are bounded from below and $G^{(2)}$ and $\gamma$
are positive operators of finite trace.

As mentioned
after (2a.12) the three terms in (2b.9) are called respectivley the
{\bf direct energy}, the {\bf exchange energy} and the {\bf pairing energy}.

We can thus write the HF energy functional as
$$
\E(\Gamma)=\Tr[h \gamma]+\mfr1/2\sum_{k,l,m,n}V_{kl;mn}
\left(\gamma_{mk}\gamma_{nl}
-\gamma_{ml}\gamma_{nk}+\alpha^\dagger_{lk}\alpha_{mn}\right).\eqno(2c.8)
$$
The {\bf Hartree-Fock energy} is given by
$$
E^{\HF}:=\inf\left\{\E(\Gamma)\ \vert\ \Gamma\ \hbox{is an admissible density
matrix}\right\}. \eqno(2c.9)
$$

As we have proved the one-to-one correspondence between
quasi-free states and admissible density matrices, (2c.9)
is evidently equivalent to
$$
E^{\HF}:=\inf\left\{\uprho(H) \ \vert\ \uprho\ \hbox{is a quasi-free
state}\right\}. \eqno(2c.10)
$$

We shall not discuss, in general, whether the infimum
in (2c.9) is attained.
For the Hubbard Hamiltonian, however, it is clearly the case that
the infimum in (2c.9) is attained since the set
of admissible denisity matrices is a compact subset of a finite dimensional
space.  In case of the
{\it atomic Hamiltonian} it is also true that the infimum is
attained. This result was proved in [LS],
where it was assumed
that $\alpha=0$, but this follows from Theorem~2.11 below).
A 1-$pdm$ for which
the infimum (2c.9) is attained defines a {\bf HF ground state}.

To define the finite temperature {\bf HF Gibbs state} we must introduce the
{\bf entropy of a quasi-free state}:
$$
        S(\Gamma):=-\mfr1/2\Tr[\Gamma\ln\Gamma]-
        \mfr1/2\Tr[(\1-\Gamma)\ln(\1-\Gamma)]
        =-\Tr[\Gamma\ln\Gamma].\eqno(2c.10)
$$
The last equality in (2c.10)
holds because $\Gamma$ and $\1-\overline\Gamma$ are
unitarily equivalent (cf. proof of Lemma~2.5) and $\Gamma$ has real
eigenvalues. Notice that
by Theorem~2.6, $S(\Gamma)=0$ if and only if $\Gamma$ is the 1-$pdm$ of a pure
state.

We define the {\bf Hartree-Fock pressure functional} $\P_\beta$ at inverse
temperature $\beta$ as

$$
        -\P_\beta(\Gamma)=\E(\Gamma)-\beta^{-1} S(\Gamma).\eqno(2c.11)
$$
The {\bf Hartree-Fock pressure}
is defined by
$$
        \P^{\HF}(\beta)=\sup\left\{\P_\beta(\Gamma)\ \vert\ \Gamma\
        \hbox{is an admissible density matrix}\right\}.  \eqno(2c.12)
$$

As explained above we only consider positive temperature in case of the Hubbard
Hamiltonian and as for the energy it is then clear that the supremum is
attained.  A HF Gibbs state is defined by a 1-$pdm$ maximizing (2c.11).

In the next lemma we show that the HF
energy gives an upper bound to the true energy and the HF pressure
gives a lower bound to the true pressure.

{\bf 2.8 THEOREM:} {\it We have the inequality
$$
        E^{\HF}\geq E^{\Q}:=\inf_{\uprho}\uprho(H),\eqno(2c.13)
$$
where the infimum is over all states $\uprho$
(not just HF states).
If $\widehat h$ and $\widehat V$ are
bounded
$$
        \P^{\HF}(\beta)\leq \P^{\Q}:=\beta^{-1}\ln\Tr[\exp(-\beta H)].
	\eqno(2c.14)
$$
}

{\it Proof:} The inequality (2c.13) is obvious since $E^{\HF}$ is defined
according to (2c.10)
as an infimum over a restricted class of states, namely the
generalized HF states.

Inequality (2c.14) is more complicated. The aim is to show that
for any generalized HF state $\uprho$ with 1-$pdm\ \Gamma$ we
have
$$
        \exp(-\beta\uprho(H)+S(\Gamma))\leq \Tr[\exp(-\beta H)].
        \eqno(2c.15) 
$$

According to Lemma~2.4 any generalized HF state is
a limit (in the sense of convergence of expectation values
of bounded operators) of quasi-free Gibbs states.
Moreover, it is clear that the  entropies of the approximating
Gibbs states are greater than the entropy of the limiting
state. We can therefore assume that the
quasi-free state $\uprho$ in (2c.15)
is a Gibbs state.

Since $\uprho$ is a Gibbs state we can define an operator
$A$ on $\H\oplus\H$ as in (2b.41). This operator is
then the first quantization of a quadratic operator $h_A$
and
$$
        \uprho(B)
=\Tr[\exp(-\beta h_A)]^{-1}\Tr[B\exp(-\beta h_A)]. \eqno(2c.16)
$$
To specify $h_A$ uniquely we assume that
$\langle 0\vert h_A\vert 0\rangle=0$.

By the Peierls-Bogoliubov inequality
[TW] we infer
$$
        \eqalignno{\Tr[\exp(-\beta H)]=&\Tr[\exp(-\beta(h_A+H-h_A
)]\cr
        \geq &\Tr[\exp(-\beta h_A)]
        \exp(-\beta\uprho(H-h_A)). \qquad&(2c.17) \cr}
$$

The inequality (2c.14) follows if we can prove
$$
        S(\Gamma)=\beta \uprho(h_A)+\ln\Big(
        \Tr[\exp(-\beta h_A)]\Big).\eqno(2c.18) 
$$
By diagonalizing the operator $A$ as in the proof
of Lemma~2.5 we find
$$
        \Tr[\exp(-\beta h_A)]=\prod_k (1+\exp(-\beta e_k)),
\eqno(2c.19) 
$$
where $e_1, e_2,\ldots$ again denote the positive
eigenvalues of $A$ and half the zero eigenvalues (if any).
We also find
$$
        \uprho(h_A)=\Tr[\exp(-\beta h_A)]^{-1}
                 \Tr[h_A\exp(-\beta h_A)]
        = \sum_k e_k
        [1+\exp(\beta e_k)]^{-1}.\eqno(2c.20) 
$$
Using (2b.41), (2c.19) and (2c.20) we finally obtain
that the right side of (2c.18) is
$$
        \sum_k\ln\Big(1+\exp(\beta e_k)\Big)
        +\sum_k\beta e_k\left(\Big(1+\exp(\beta e_k)\Big)^{-1}
        -1\right)=\mfr1/2\Tr\big[-\ln(\Gamma)
        -(\1-\Gamma)\ln(\Gamma^{-1}-\1)\Big], \eqno(2c.21)
$$
which is exactly $S(\Gamma$).\lanbox

It is always possible to choose the quantum ground state
to be a pure state. The same is true in generalized HF
theory. This follows from the next lemma when compared
with Theorem~2.6.

{\bf 2.9 THEOREM:} {\it The infimum of $\E$ over all
admissible density matrices agrees
with the infimum over all admissible projections, i.e.,
$$
        E^{\HF}=\inf\{\E(\Gamma)\ \vert\ \Gamma
        \hbox{ is admissible and }\Gamma^2=\Gamma\}.\eqno(2c.22)
$$
}

{\it Proof:}  We shall show that for any admissible $\Gamma$ there is a
projection $\Gamma_0$ such that $\E(\Gamma_0) \leq \E (\Gamma)$.  For
this purpose it suffices to approximate $h$ and $V$ by bounded
operators --- for which there is obviously no difficulty with
convergence of the following sums.

For any admissible $\Gamma$,
as in the proof of Theorem~2.3,
we can find a Bogoliubov transformation $\W$ with
corresponding $W$ such that $W^\dagger\Gamma W$
is diagonal. If $\Gamma$ is the 1-$pdm$ of the state
$\uprho$ then $W^\dagger\Gamma W$ is the 1-$pdm$ of the
transformed state $\uprho_{\W}$. We have the relation
$\E(\Gamma)=\uprho(H)=\uprho_{\W}(\W^\dagger H\W)$.
The transformed Hamiltonian $\W^\dagger H \W$
is also a sum of a quadratic operator
$\W^\dagger \widehat h\W$ and a quartic operator
$\W^\dagger \widehat V\W$, but these are not
necessarily number preserving.
If we (anti)commute all $c^\dagger$ to the left
(normal ordering) we obtain
$$
        \W^\dagger H\W=\sum_{i,j}
        \widetilde h^{\phantom{\dagger}}_{ij}c^\dagger_i
        c^{\phantom{\dagger}}_j+\mfr1/2\sum_{kl;mn}
        \widetilde V^{\phantom{\dagger}}_{kl;mn}c^\dagger_kc^\dagger_l
        c^{\phantom{\dagger}}_nc^{\phantom{\dagger}}_m
        +\kappa\1 +R, \eqno(2c.23)
$$
where $\widetilde h^{\phantom{\dagger}}_{ij}$ and
$\widetilde V^{\phantom{\dagger}}_{kl;mn}$ are new
matrix elements and $\kappa$ is some constant. The operator
$R$ contains all particle non-conserving terms of the form
$$
        cc,\quad c^\dagger c^\dagger,\quad cccc, \quad c^\dagger ccc,
        \quad  c^\dagger c^\dagger c^\dagger c,\quad
        c^\dagger c^\dagger c^\dagger c^\dagger. \eqno(2c.24)
$$

Since $\uprho_{\W}(R)=0$, we obtain
$$
        \E(\Gamma)=\uprho_{\W}(\W^\dagger H\W)
        =\kappa+\sum_{i}\widetilde h_{ii}\lambda_i
        +\mfr1/2\sum_{kl} (\widetilde V_{kl;kl}-\widetilde V_{kl;lk})
        \lambda_k\lambda_l,\eqno(2c.25) 
$$
where $\lambda_1,\lambda_2,\ldots,$ are as usual
the eigenvalues of $\Gamma$ smaller than $1/2$.

The important fact to observe about the expression in (2c.25)
is that, although it is a
quadratic form in
$\lambda_1,\lambda_2,\ldots$, it is linear in each variable.
Hence $\partial \E(\Gamma)/\partial \lambda_j$ is independent
of $\lambda_j$.
Therefore, we do not increase the energy expectation by replacing
$\lambda_j \vert \varphi_j \rangle \langle \varphi_j \vert
+ (1-\lambda_j) \vert U \overline\varphi_j \rangle
\langle U \overline\varphi_j \vert$ in $\Gamma$ by
$\vert \varphi_j \rangle \langle \varphi_j \vert$ in case
$\partial \E(\Gamma) / \partial \lambda_j <0$ and by
$\vert U \overline\varphi_j \rangle
\langle U \overline\varphi_j \vert$ otherwise.
Proceeding this way, we arrive at a 1-$pdm$
 $\Gamma_0$ with energy no greater
than before and with all eigenvalues either $0$ or $1$.
This means that $\Gamma_0$ is a projection.\lanbox

If $\uprho$ is a generalized HF ground state
for $H$ we define a corresponding
{\bf HF mean field Hamiltonian}.
It is the following quadratic Hamiltonian
written in terms of the blocks $\gamma$ and $\alpha$
of the 1-$pdm\ \Gamma$ of $\uprho$.
$$
        \eqalignno{H_\rho:=\sum_{i,j}h^{\phagger}_{ij}
        c^\dagger_i c^\phagger_j
        +\mfr1/2\sum_{kl;mn} V^\phagger_{kl;mn}\biggl[&
        \gamma^\phagger_{km}c^\dagger_lc^\phagger_n
        +\gamma^\phagger_{nl}c^\dagger_kc^\phagger_m
        -\gamma^\phagger_{ml}c^\dagger_kc^\phagger_n
        -\gamma^\phagger_{nk}c^\dagger_lc^\phagger_m\cr
        &+\alpha^\dagger_{lk}c^\phagger_nc^\phagger_m
        +\alpha^\phagger_{mn}c^\dagger_kc^\dagger_l\biggr].
\qquad&(2c.26) \cr}         
$$

A HF ground state of $H$ is {\bf self-consistent} in the sense
given in the next lemma.

{\bf 2.10 LEMMA:} {\it If $\uprho$ is a HF ground state
for the Hamiltonian $H$, i.e., a HF minimizer,
then $\uprho$ is a true
(not just HF) ground state for the Hamiltonian $H_\rho$.}

{\it Proof:} We must show that $\uprho(H_\rho)\leq
\uprho'(H_\rho)$ for any state $\uprho'$
with finite particle number.

{F}rom the 1-$pdm$ $\Gamma'$ of $\uprho'$ and the
1-$pdm$ $\Gamma$ of $\uprho$ we can for $0\leq t\leq1$
construct a new 1-$pdm$ $\Gamma_t=(1-t)\Gamma+t\Gamma'$.
Then $\Gamma_t$ is admissible and since $\Gamma$ is a
minimizer for $\E$ we have

$$
        0\leq{d\E(\uprho_t)\over dt}_{\Big\vert t=0}
        =\uprho'(H_\rho)-\uprho(H_\rho), \eqno(2c.27)
$$
which proves the claim. It is important here that
since $H_\rho$ is quadratic,
$\uprho'(H_\rho)$ depends only on the 1-$pdm$
of $\uprho'$.  \lanbox

Although, we can always find a state
with fixed particle number among the quantum
ground states for the Hamiltonian $H$,
this may
not be the case for the generalized HF ground states, as discussed in
the introduction (and proved for the attractive
Hubbard Hamiltonian in Sect.~III).  Here there need not be a normal
ground state.

It is often stated in the literature that if
the HF ground state is not a normal state, (i.e.,
it is a BCS state) the potential
$V$ must have a negative component.
It is now very easy to state this precisely.

{\bf 2.11 THEOREM:} {\it If the operator $V$ is positive (semi) definite
on $\H\otimes\H$
$$
        E^{\HF}=\inf\{\E(\Gamma)\ \vert\ \Gamma
        \hbox{ is admissible and normal, i.e.,}\
        \alpha=0\}.\eqno(2c.28) 
$$
Moreover, if $V$ is strictly positive (i.e., positive definite)
then any ground state (if it exists) must be a normal state.
Likewise, we have for the pressure
$$
        \P^\HF(\beta)=\sup\{\P_\beta(\Gamma)\ |\
	\Gamma\hbox{ is admissible and normal }\}.
        \eqno(2c.29)
$$
}
{\it Proof:} That $V$ is positive means that for all
$g\in\H\otimes\H$ we have
$$
        \sum_{kl;mn}V_{kl;mn}\g_{kl}g_{mn}\geq0. \eqno(2c.30) 
$$
Strict positivity means that if $g\ne0$ then (2.67) is
a strict inequality.
The pairing energy is exactly of the form (2.67) hence
$$
        \sum_{kl;mn}V^{\phantom\dagger}_{kl;mn}\alpha^{\dagger}_{lk}
        \alpha^{\phantom{\dagger}}_{mn}=
        \sum_{kl;mn}V^{\phantom\dagger}_{kl;mn}
        \overline{\alpha}_{kl}\alpha_{mn}\geq0.\eqno(2c.31) 
$$

Assume $\Gamma$ is admissible with non-zero $\alpha$ and form a new
operator $\widetilde\Gamma$ by replacing $\alpha$ by zero.
It is clear that $\widetilde\Gamma$ is still admissible and
by (2c.31) that $\E(\widetilde\Gamma)$ is (strictly) smaller than
$\E(\Gamma)$ if $V$ is (strictly) positive.
To prove the result on the pressure we notice that the 1-$pdm$
$\Gamma'$ obtained by changing $\alpha$ to $-\alpha$ is unitarily
equivalent to $\Gamma$. Hence, since the entropy is a concave function we
find
$$
        S(\widetilde\Gamma)=S(\mfr1/2\Gamma+\mfr1/2\Gamma')\geq
        \mfr1/2S(\Gamma)+\mfr1/2S(\Gamma')=S(\Gamma). \eqno(2c.32)
$$
\lanbox

In the remainder of this section we shall specialize
to the case of positive $V$ and discuss the canonical
picture. Thus, instead of a chemical potential we now
fix the value $\Tr[\gamma]=N$. Since we have $\alpha=0$
it is enough to consider the component $\gamma$ of $\Gamma$.

The discussion is of special interest in the atomic
case where the potential $V$ {\it is} strictly positive.

There is a version of Theorem~2.9
for the canonical case of fixed particle number $N$.
We state it below without proof.
It is, in fact, more difficult to prove than Theorem~2.9
because
in deforming $\gamma$ to a projection we have to keep
$\Tr[\gamma]=N$.  The proof was first given in
[LE1] (see [BV] for a simple proof).

{\bf 2.12 THEOREM (Variational principle):}  {\it If $V$ is positive then
$$
        \eqalignno{E^{\HF}(N):&=\inf\{\E(\gamma)\ \vert\ \gamma
        \hbox{ is admissible with } \Tr[\gamma]=N\}\cr
        &=\inf\{\E(\gamma)\ \vert\ \gamma
        \hbox{ is an admissible projection with } \Tr[\gamma]=N]\}
        .&(2c.33) } 
$$
Moreover, if $V$ is strictly positive then any HF minimizer must
be a projection.}

Combining Theorem~2.11 and Theorem~2.12 we see that a HF ground state
$\uprho$ for a system with strictly positive $V$ is not only particle
conserving $(\alpha = 0)$ but, in fact, has fixed particle number, i.e.,
$\uprho (\N^2) = \uprho (\N)^2$.

Theorem 2.12 is useful in many cases for obtaining upper bounds to the
Hartree-Fock energy, $E^{\HF}$, and hence an upper bound to the true quantum
energy, $E^Q$.  It allows one to deal conveniently with what is sometimes
called the ``orthogonality problem''.  Take any matrix $\gamma$ satisfying
$0 \leq \gamma \leq \1$ (as an operator) and $\Tr \gamma = N$, and compute
the one-body energy
$$E_1 (\gamma) = \Tr \gamma h = \sum \limits_{i,j} \gamma_{ij} h_{ji}
\eqno(2c.34) 
$$
and the two body energy
$$E_2 (\gamma) = \mfr1/2 \sum \limits_{k,l,m,n} V_{kl; mn} (\gamma_{mk}
\gamma_{nl} - \gamma_{ml} \gamma_{nk}) \eqno(2c.35) 
$$
Then
$$E^Q \leq E^{\HF} \leq E_1 (\gamma) + E_2 (\gamma). \eqno(2c.36)
$$

The important point here is that the matrix $\gamma_{mk} \gamma_{nl} -
\gamma_{ml} \gamma_{nk}$ is not, in general, the two-body reduced density
matrix of {\it any} $N$-body density matrix {\it unless} $\gamma$ comes from a
Slater determinant, i.e., $\gamma_{ij} = \sum \nolimits^N_{\alpha =1}
\varphi^\alpha_i \overline \varphi^\alpha_j$, with $\varphi^1, \dots ,
\varphi^N$ being $N$ orthonormal functions.  Nevertheless, (2c.36) continues
to be true.

Contrary to the proof of Theorem~2.9
the proof of Lemma~2.10 is unchanged in the
canonical case  (because the 1-$pdm$ $\gamma_t=(1-t)\gamma
+t\gamma'$,
automatically satisfies $\Tr[\gamma_t]=N$
if $\Tr[\gamma]=N$ and $\Tr[\gamma']=N$).

{\bf 2.13 LEMMA:} {\it If $V$ is positive
and if $\uprho$ is a HF ground state
for the Hamiltonian $H$ under the constraint $\uprho(\N)=N$
then $\uprho$ is a true
ground state for the Hamiltonian $H_\rho$ under
the same constraint.}

The mean field Hamiltonian has a simpler form
when restricted to fixed particle number.
$$
        {H_\rho:=\sum_{i,j}h^{\phagger}_{ij}
        c^\dagger_i c^\phagger_j
        +\mfr1/2\sum_{kl;mn} V^\phagger_{kl;mn}\biggl[
        \gamma^\phagger_{km}c^\dagger_lc^\phagger_n
        +\gamma^\phagger_{nl}c^\dagger_kc^\phagger_m
        -\gamma^\phagger_{ml}c^\dagger_kc^\phagger_n
        -\gamma^\phagger_{nk}c^\dagger_lc^\phagger_m
        \biggr]},\eqno(2c.37) 
$$
which is simply a (number preserving) independent
particle Hamiltonian.

When $V$ is strictly positive we can prove a much stronger
result than Lemma~2.13.
Namely, not only is
a HF ground state, $\uprho$, for $H$ a true ground state for the
mean field operator $H_\rho$, but it is the
{\it unique} ground state for $H_\rho$ satisfying
the constraint $\uprho(\N)=N$.

This uniqueness result is equivalent to
the striking statement made in the introduction,
that {\it no degenerate energy levels of $H_\rho$ (if they exist)
can be only partially filled in the HF state.}
We empasize again how contradictory this is
to what is taught in elementary chemistry courses.
We learn in the theory of chemical binding how two unpaired
electrons in separate atoms can pair up and create a strongly
bound molecule. The concept of unpaired electrons
relies of course essentially on the independent orbital
picture of HF theory. What we prove is that there are {\it no
unpaired} electrons in HF theory.

The obvious question is of course:
{\it But what then has happened to the spin degeneracy?}
The answer is, as already discussed, that there may not
be spin degeneracy in HF theory, because even the spin symmetry
can be broken.
Likewise for the angular momentum degeneracy; there may not
be spherical symmetry.

The following is the theorem that there are no unfilled shells.  It was
proved by M. Loss and the authors [BLL].  Since the proof is very
short we repeat it here.

{\bf 2.14 THEOREM: (Shells are always closed.)}
{\it Assume that the two-body
potential $V$ in the Hamiltonian $H$ is strictly positive (as in Coulomb
case).
If $\uprho$ is a HF ground state (a Slater determinant state)
for $H$, i.e., a HF minimizer subject to the constraint
$\uprho(\N)=N$, then $\uprho$ is
the unique ground state of the mean field operator
$H_\rho$ satisfying the constraint $\uprho(\N)=N$ .
}

{\it Proof:}  {F}rom Theorem~2.12 we know that
the 1-$pdm$ $\gamma$ of $\uprho$ is an $N$-dimensional
projection, which we can write
$$
        \gamma=\sum_{j=1}^N\vert g_j\rangle\langle g_j\vert,
        \eqno(2c.38) 
$$
where $g_1,g_2,\dots$ are $N$ orthonormal eigenvectors of
$\gamma$. Since $\uprho$ is a ground state for
$H_\rho$ we can assume that
$g_1,g_2,\ldots$ are eigenvectors of $H_\rho$.

If there is another  ground state with the same number
of electrons there must be a degenerate level containing
a vector from $\gamma$, say $g_N$, and at the same time a
normalized eigenvector $g'$ which is not in $\gamma$, i.e.,
with $\gamma g'=0$. We can then define a new $N$-dimensional
projection by
$$
        \gamma'= \vert g'\rangle
        \langle g'\vert+
\sum_{j=1}^{N-1}\vert g_j\rangle
        \langle g_j\vert .\eqno(2c.39) 
$$
The HF state $\uprho'$ with 1-$pdm$ $\gamma'$ is
then also a ground state for $H_\rho$, i.e.,
$\uprho(H_\rho)=\uprho'(H_\rho)$.

Since $\gamma$ is a minimizer for the HF functional we have
\goodbreak
$$
        \eqalignno{0\leq&\E(\gamma')-\E(\gamma)
        -\left(\uprho'(H_\rho)-\uprho(H_\rho)\right)\cr
         =&\mfr1/2\sum_{kl;mn}V_{kl;mn}\biggl[
        (\gamma-\gamma')_{mk}(\gamma-\gamma')_{nl}
        -(\gamma-\gamma')_{nk}(\gamma-\gamma')_{ml}\biggr].
        &(2c.40) } 
$$
The operator $\gamma-\gamma'=\vert g_N\rangle \langle g_N\vert
-\vert g'\rangle \langle g'\vert$. Hence, since $V$ is
strictly positive,
$$
        \eqalignno{\mfr1/2\sum_{kl;mn}V_{kl;mn}\biggl[&
        (\gamma-\gamma')_{mk}(\gamma-\gamma')_{nl}
        -(\gamma-\gamma')_{nk}(\gamma-\gamma')_{ml}\biggr]\cr
        =&\langle g_N\otimes g_N\vert V\vert g_N\otimes g_N\rangle
        +\langle g'\otimes g'\vert V\vert g'\otimes g'\rangle
        -\langle g_N\otimes g'\vert V\vert g_N\otimes g'\rangle\cr
        &-\langle g'\otimes g_N\vert V\vert g'\otimes g_N\rangle
        -\langle g_N\otimes g_N\vert V\vert g_N\otimes g_N\rangle
        -\langle g'\otimes g'\vert V\vert g'\otimes g'\rangle\cr
        &+\langle g_N\otimes g'\vert V\vert g'\otimes g_N\rangle
        +\langle g'\otimes g_N\vert V\vert g_N\otimes g'\rangle\cr
        =&\langle g_N\wedge g'\vert V\vert g'\wedge g_N\rangle
        <0, \qquad&(2c.41) \cr}
$$
which contradicts (2c.40).\lanbox

To illustrate this result let us consider the
simple, but not altogether trivial example of $N=1$
for the atomic Hamiltonian.
In this case the HF ground state energy is in fact
the true ground state energy.
It is simply the ground state energy
of hydrogen (with nuclear charge
$Z$), i.e, $-\mfr1/4Z^2$ in our units. Owing to the spin
degeneracy the ground state is doubly degenerate.
We consider the spin up ground state and write
it $\psi=|\varphi(x)\uparrow\rangle$. Here we have used the
hydrogen ground state wave function
$\varphi(x)=CZ^{3/2}\exp(-Z|x|)$.
It is important to realize that
the mean field operator corresponding to  $\psi$
is not just the hydrogen operator, but rather
the operator $H_{\psi}$ that acts on a general
$\psi'=|\varphi'(x)\sigma\rangle$, with $\sigma=\uparrow$ or
$\sigma=\downarrow$ according to
$$
        \eqalignno{H_{\psi}|\varphi'(x)\sigma\rangle
        =&\left(-\Delta-{Z\over|x|}+
        \int\varphi(y)^2|x-y|^{-1}dy\right)
        |\varphi'(x)\sigma\rangle\cr
        &-\delta_{\sigma\uparrow}\int \varphi(y)
        \varphi'(y)|x-y|^{-1}dy
        |\varphi'(x)\sigma\rangle. \qquad&(2c.42) \cr}
$$
Notice that $H_{\psi}$ is spin dependent --- it does
not commute with spin rotations.

The mean field operator and the hydrogen operator
agree on their common ground state
$\psi$, i.e.,
$$
        H_{\psi}\psi=\left(-\Delta-{Z\over|x|}\right)\psi
        =-\mfr1/4 Z^2\psi. \eqno(2c.43)
$$
On the other ground state for hydrogen, namely,
$|\varphi\downarrow\rangle$ we find, however,
$$
        \eqalignno{\langle\varphi\downarrow|
        H_{\psi}|\varphi\downarrow\rangle=&
        \langle\varphi\downarrow|
        \left(-\Delta-{Z\over|x|}\right)
        |\varphi\downarrow\rangle+\int\!\!\int
        \varphi(x)^2\varphi(y)^2|x-y|^{-1}dxdy\cr
        =&-\mfr1/4 Z^2+cZ.&(2c.44) } 
$$

Thus the mean field operator induces a gap
in the energy between the two degenerate
ground states of hydrogen.

Theorem~2.14 states that this is not only true for
the state $|\varphi\downarrow\rangle$, but in
fact, the ground state
for $H_{\psi}$ is {\it unique}.
More importantly, by Theorem~2.14 this is
not particular to the case $N=1$.

\bigskip\bigskip
\noindent
{\bf III.  THE GENERALIZED HF THEORY FOR THE
HUBBARD MODEL WITH ATTRACTIVE INTERACTION}
\bigskip\noindent
{\bf III.a  DEFINITIONS}

In this section the generalized HF theory will be applied to the Hubbard
model with attractive interaction.  Our {\it main result} (Theorem 3.12) in
Sect.~III.e) will be that the HF ground state and positive temperature
Gibbs state are unique, modulo global gauge transformations.
First, we recall the definition of the
Hubbard model.  Let $\Lambda$ be a finite lattice, i.e., a finite
collection of points, and let $\vert \Lambda \vert$ be the number of these
points.  The one-particle Hilbert space $\H$ is the $2 \vert \Lambda
\vert$-dimensional space of spinor-valued functions on $\Lambda$, i.e., the
set of
complex-valued functions on $\Lambda \times \{ -1, 1 \}$.  The value of
such a function at $(y, \tau)$ (with $y \in \Lambda, \tau \in \{ -1, 1
\}$) is $f(y, \tau)$ and the inner product is $\langle f \vert g \rangle =
\sum \nolimits_{y, \tau} \overline f(y, \tau) g(y, \tau)$.  A
canonical orthonormal basis in $\H$ is given by the delta functions, which
we denote by $|x, \sigma\rangle$.  Thus, $|x,\sigma\rangle \in \H$ is the
function $f_{x,\sigma} (y, \tau) = \delta_{y,x} \delta_{\tau, \sigma}$.
We define our complex conjugation in this basis.
In the Fock space $\F$
corresponding to $\H$ we refer to $c^\dagger (\vert x, \sigma \rangle)$ as
$c^\dagger_{x, \sigma}$.  We often use the abbreviation $\uparrow := +1$
and $\downarrow :=
-1$.  An orthonormal basis in $\F$ is given by $\{
c^\dagger_{x_1, \sigma_1} \dots c^\dagger_{x_N, \sigma_N} \vert 0 \rangle
\ \big\vert \ (x_i, \sigma_i) \not= (x_j, \sigma_j)$ if $i \not= j\}$, which
implies that $\F$ is $4^{\vert \Lambda \vert}$-dimensional.

The Hubbard Hamiltonian for negative coupling is
$$H_- = \sum \limits_{\scriptstyle x,y \in \lambda \atop \scriptstyle
\sigma} t_{xy} c^\dagger_{x,\sigma} c_{y,\sigma} - \sum \limits_{x \in i}
U_x \big(c^\dagger_{x,\uparrow} c^{{\phantom{\dagger}}}_{x,\uparrow} -
\mfr1/2 \big) \big( c^\dagger_{x,\downarrow}
c^{{\phantom{\dagger}}}_{x,\downarrow} - \mfr1/2 \big).\eqno(3a.1)$$

The second term in $H_-$ is an attractive interaction among the electrons
with position dependent coupling $-U_x < 0$.  (Our notation here and
elsewhere is that $U_x \geq 0$.)  Other authors frequently replace the last
two factors in (3a.1) by $(c^\dagger_{x\uparrow} c^{{\phantom{\dagger}}}_{x
\uparrow}) (c^\dagger_{x \downarrow} c^{{\phantom{\dagger}}}_{x
\downarrow}$), but we prefer our form (also used in [LLM]) because it
preserves hole-particle symmetry; if $U_x$ is independent of $x$ the
distinction is unimportant.  In the first term, the $\vert \Lambda
\vert \times \vert \Lambda \vert$ self-adjoint matrix $t = \{ t_{xy}
\}_{x,y \in \Lambda}$ is called the {\bf hopping matrix}.  Since
we shall later include a chemical potential we can and will assume without
loss of generality that $\Tr [t] = 0$.  We do not assume
$t_{xx} = 0$, but when $t$ is bipartite (defined below) the condition
$t_{xx} = 0$ is automatic.  We emphasize
that up to this point $\Lambda$ was an arbitrary set and need not have any
topological structure.  It is the structure of $t$, linking different
points in $\Lambda$, that makes the embedding of $\Lambda$ into $\R^d$
sometimes useful.  Indeed, in the original model introduced by Hubbard,
Kanamori and Gutzwiller [HJ], [KJ], [GMC],
$\Lambda$ is a finite cube of lattice points in $\Z^d$ and $t_{xy}
:= \tau$ when $x$ and $y$ are nearest neighbors, and $t_{xy} = 0$
otherwise.  The
imposition of periodic boundary conditions makes $\Lambda$ into a
$d$-dimensional discrete torus on which $t$ is translation invariant.  Let
us remark that,
except in Sect.~III.g,
we shall neither assume that $t$
is translation invariant nor that $U_x$ is constant.  However,
connectedness, reality and bipartiteness of $t$ will play an important role
in our analysis.  These notions can be conveniently characterized by means
of path.  A {\bf path} is an ordered sequence $\{ x_1, x_2, \dots , x_n \}$
of points in $\Lambda$ such that $t_{x_1x_2}, t_{x_2 x_3}, \dots ,
t_{x_{n-1} x_n}$ are all non-vanishing.  We will always assume $t$ to be
{\bf connected}, i.e., any two points $x,y \in \Lambda$ are linked by a
path $\{ x, x_1, \dots , x_n, y\}$.  We say $t$ is {\bf real} if $t_{xy}
\in \R$ for all $x,y \in \Lambda$.  For real $t$ self-adjointness implies
that $t_{xy} = \overline t_{yx} = t_{yx}$.  For every closed path $\{ x_1,
x_2, \dots , x_n, x_1 \}$ the product $t_{x_1 x_2} t_{x_2 x_3} \dots t_{x_n
x_1}$ obviously yields a real number provided $t$ is real.  Conversely, if
$t_{x_1 x_2} \dots t_{x_n x_1}$ is real for every closed path then $t$ is
unitarily equivalent to some real hopping matrix $t^\prime = W t W^\dagger$
where $W$ is a gauge transformation (see [LL] Lemma~2.1).  Hence,
unless there is at least one closed path for which $t_{x_1 x_2} \dots t_{x_n
x_1}$ is not real, we may as well assume that $t$ is real.

The matrix $t$ is said to be {\bf bipartite} if there are two disjoint
subsets $A,B \subseteq \Lambda$ with $A \cup B = \Lambda$, such that
$t_{xy} = 0$ whenever both $x,y \in A$ or $x,y \in B$.
Evidently, $t$ is bipartite if and
only if all paths contain an even number of points.

One important property of a bipartite $t$ is that it is unitarily
equivalent to $-t$.  Indeed, defining the unitary $\vert \Lambda \vert
\times \vert \Lambda \vert$-matrix $(-1)^x = [(-1)^x]^{-1} =
[(-1)^x]^\dagger$ by
$$(-1)^x := \sum \limits_{x\in A} \vert
x \rangle \langle x \vert - \sum \limits_{x \in B}
\vert x \rangle \langle x \vert, \eqno(3a.2)$$
one easily checks that
$$(-1)^x t (-1)^x =- t . \eqno(3a.3)$$
If it is unambigously clear from the context what is meant,
we will use the same symbol $(-1)^x$ to denote the function
which takes the value $1$ on the A sublattice and $-1$ on the
B sublattice.
The Dirac notation is used in (3a.2), whereby $\vert \varphi \rangle
\langle \varphi \vert$ denotes the projection onto a normalized vector
$\varphi$.  The vector $x$ is here the delta function
$\delta_{x, \cdot}$ at $x \in \Lambda$ in the space of
complex-valued functions $\H_\Lambda$ on $\Lambda$ (not $\Lambda \times
\{ \uparrow, \downarrow \}$).
Let us now make a few remarks about the different Hilbert spaces we will
encounter in what follows. We start with $\H_\Lambda$.
Mathematically we may view the
one-particle Hilbert space $\H$ as
$$ \H = \H_\Lambda \oplus \H_\Lambda = {\bf C}^2 \otimes \H_\Lambda
\eqno(3a.4)$$
and, consequently, the space on which the 1-$pdm$ are defined as
$$ \H \oplus \H = {\bf C}^4 \otimes \H_\Lambda.
\eqno(3a.5)$$
(We wrote equalities in (3a.4) and (3a.5) despite of our
awareness that
an isomorphism would have been mathematically more appropriate.)
It is thus clear that operators on $\H = \H_\Lambda \oplus \H_\Lambda$
can be written as $2\times 2$-matrices with operators on $\H_\Lambda$
as entries and, likewise, operators on $\H \oplus \H$ as
$4 \times 4$-matrices with operators on $\H_\Lambda$ as entries.
The usual conventions for matrix algebras are understood and,
in particular,
$$ q \pmatrix{ m_{11} & m_{12} \cr m_{21} & m_{22} \cr} :=
     \pmatrix{ qm_{11} & qm_{12} \cr qm_{21} & qm_{22} \cr} , \eqno(3a.6)$$
where $q$ and $m_{ij}$ are complex numbers or operators on $\H_\Lambda$,
and likewise for $4 \times 4$-matrices.
It will often be convenient for us to change between
the Hilbert spaces $\H_\Lambda$, $\H$, and $\H \oplus \H$,
and to take traces over these various spaces. To simplify
notations we shall use a common symbol, Tr, for these traces;
the Hilbert space in question will be evident from the operator
whose trace is being computed. Likewise, we shall denote the identity
operator on these different spaces be the common symbol $\1$.
It will always be clear from the context which identity we are refering to.

We are now in a position to write down the pressure functional for the
Hubbard model.  We choose to define the 1-$pdm$ in terms of the orthonormal
basis of delta functions $\vert x, \sigma \rangle$.  Hence, the energy
expectation for
a 1-$pdm$ $0 \leq \Gamma = \pmatrix{\gamma &\alpha \cr \alpha^\dagger &\1 -
\overline \gamma \cr} \leq \1$ is given by
$$\eqalign{\E(\Gamma) = \Tr[t \gamma] &- \sum \limits_{x \in \Lambda}
U_x \big\{ [\langle x\uparrow \vert \gamma \vert x \uparrow \rangle -
\mfr1/2 ] [\langle x \downarrow \vert \gamma \vert x \downarrow \rangle
- \mfr1/2 ] \cr
&- \vert \langle x, \uparrow \vert \gamma \vert x,\downarrow \rangle \vert^2
+ \vert \langle x \uparrow \vert \alpha \vert x\downarrow \rangle \vert^2
\big\}. \cr}\eqno(3a.7
)$$
Here, we identified $t$ with the operator $\pmatrix{t&0\cr 0&t\cr}$ on
$\H$.

The thermodynamic pressure of $\Gamma$ (actually, the pressure multiplied
by the ``volume'') is given by (with $\beta = 1/k_{{\rm Boltzmann}} T$)
$$\eqalignno{-\P_{\beta, \mu} (\Gamma) &= \E (\Gamma) - \beta^{-1}
S(\Gamma) - \mu N (\Gamma) \cr
&= \E (\Gamma) + \mfr1/2 \beta^{-1} \Tr [\Gamma \ln
\Gamma + (\1 - \Gamma) \ln(\1 - \Gamma)] - \mu \Tr [\gamma].
\qquad&(3a.8)\cr}$$
We have here introduced the chemical potential $\mu$. We could equivalently
have replaced the Hamiltonian $H_-$ by $H_--\mu\N$, where $\N$ is the
particle number operator. Also we have introduced the notation $N(\Gamma):=
\Tr[\gamma]$ for the particle number expectation
in the state described by $\Gamma$.

Our aim is to characterize the set of maximizing 1-$pdm$'s for the
pressure (which we typically denote by $\Gamma_0$) as explicitly as possible
and to determine the pressure $\P (\beta, \mu)$ of the system, i.e.,
$$\P (\beta, \mu) := \max \{ \P_{\beta, \mu} (\Gamma) \ \big\vert \ 0 \leq
\Gamma = \pmatrix{\gamma &\alpha \cr \alpha^\dagger &\1 - \overline \gamma
\cr} \leq \1 \} = \P_{\beta,\mu} (\Gamma_0). \eqno(3a.9)$$
A priori, the {\it max} in (3a.9) should be replaced by {\it supremum} but,
since the underlying Hilbert space is finite dimensional, the existence of at
least one maximizing 1-$pdm$ is assured.

\bigskip\noindent
{\bf III.b  LINEARIZATION OF THE PRESSURE FUNCTIONAL}

As a first illustration of the notation introduced above,
a 1-$pdm \ \Gamma$ as an operator
on $\C^4\otimes\H_\Lambda$, is written as
$$\Gamma = \pmatrix{\gamma_\uparrow &\gamma_* &\alpha_\uparrow &\alpha_* \cr
&&&\cr
\gamma^\dagger_* &\gamma_\downarrow &-\alpha_*^T &\alpha_\downarrow \cr
&&&\cr
\alpha^\dagger_\uparrow &-\overline{\alpha_*} &\1 - \overline{\gamma_\uparrow}
&\overline{\gamma_*} \cr
&&&\cr
\alpha^\dagger_* &\alpha^\dagger_\downarrow &\overline{\gamma}^\dagger_* & \1 -
\overline{\gamma_\downarrow} \cr}, \eqno(3b.1)$$
where $\langle x \vert \gamma_\sigma \vert y \rangle := \langle x, \sigma \vert
\gamma \vert y,\sigma \rangle$, $\langle x \vert \gamma_* \vert y \rangle :=
\langle x,\uparrow \vert \gamma \vert y,\downarrow \rangle, \ \langle x \vert
\alpha_\sigma
\vert y \rangle := \langle x,\sigma \vert \alpha \vert y,\sigma \rangle$,
and \hfill\break
$\langle x \vert \alpha_* \vert y \rangle := \langle x,\uparrow \vert \alpha
\vert y,\downarrow \rangle$.  Now, observe that $\E(\Gamma)$ depends neither on
$\alpha_\uparrow$ nor on $\alpha_\downarrow$, and that $\E(\Gamma)$ would be
lowered, if we
were allowed to replace $\gamma_*$ by 0.  Indeed, as the following lemma
shows, a restriction of our attention to the 1-$pdm$ with $\alpha_\uparrow,
\alpha_\downarrow$ and $\gamma_*$ all equal to zero and, moreover,
$\gamma_\uparrow = \gamma_\downarrow$
and $\alpha_*=\alpha_*^T$ is justified.  Such
matrices are of the form (with empty spaces denoting zeros)
$$\Gamma = \pmatrix{\gamma^\prime & & &\alpha' \cr
&&&\cr
&\gamma^\prime&-\alpha' & \cr
&&&\cr
&-\alpha'^\dagger &\1 - \overline{\gamma^\prime} & \cr
&&&\cr
\alpha'^\dagger & & &\1 - \overline{\gamma^\prime} \cr}, \eqno(3b.2)$$
where $\gamma'=\gamma'^\dagger$ and
$\alpha'^T=\alpha'$.

{\bf 3.1  LEMMA:} For all $\beta>0$ (including the $\beta=\infty$) and all
$\mu$
we have
$$- \P (\beta, \mu) = \min \left\{ - \P_{\beta, \mu} (\Gamma)
\ \vert \ \Gamma \
\hbox{of the form (3b.2) and satisfying } 0\leq\Gamma\leq\1 \ \right\}.$$

{\it Proof:}  Define two orthogonal projections on $\H \oplus \H$ by
$$P := \pmatrix{0&&&\cr &\1&&\cr &&\1&\cr &&&0\cr}, \qquad \widetilde
P := \pmatrix{\1&&&\cr &0&&\cr &&0&\cr &&&\1\cr}. \eqno(3b.3)$$
Clearly, $P \widetilde P = 0$ and $P + \widetilde P = \1$.  Observe
that for any 1-$pdm \ \Gamma$ written as in (3b.1), we have
$$\widetilde \Gamma := P\Gamma P + \widetilde P \Gamma \widetilde P =
\pmatrix{\gamma_\uparrow & & &\alpha_* \cr
&\gamma_\downarrow &-\alpha_* & \cr
&-\alpha^\dagger_* &\1 - \overline \gamma_\uparrow & \cr
\alpha^\dagger_* &
& &\1-\overline \gamma_\downarrow \cr}. \eqno(3b.4)$$
This operator $\widetilde \Gamma$ is also of the desired form (3b.1).
Moreover, $0\leq\widetilde\Gamma=P\Gamma P+\widetilde P \Gamma\widetilde P
\leq P+\widetilde P=\1$ and hence $\widetilde \Gamma$ is a
1-$pdm$.  As remarked above, $\E( \widetilde \Gamma) \leq \E (\Gamma)$ and
$N (\widetilde \Gamma) = N(\Gamma)$.  It remains to show that $-S
(\widetilde \Gamma) \leq - S (\Gamma)$.  We recall that $A \mapsto \Tr \{
f(A) \}$ is a concave function of the self-adjoint operator $A$ if $f: \R
\rightarrow \R$ is concave, i.e., $f(\lambda x + (1 - \lambda) y) \geq
\lambda f(x) + (1 - \lambda) f(y)$ for all $x,y$ and all $0 \leq \lambda
\leq 1$.  Applying this to $f(x) := x \ln x + (1 - x) \ln
(1-x)$, we observe that $S (\Gamma)$ is concave in $\Gamma$;
$$S (\mfr1/2 \Gamma_1 + \mfr1/2 \Gamma_2) \geq \mfr1/2 S (\Gamma_1) +
\mfr1/2 S (\Gamma_2), \eqno(3b.5)$$
for all 1-$pdm \ \Gamma_1$ and $\Gamma_2$.  We evaluate (3b.5) on $\Gamma_1 :=
\Gamma$ and $\Gamma_2 := (P - \widetilde P) \Gamma (P - \widetilde P)$, the
latter being a 1-$pdm$ since $P - \widetilde P$ is unitary and
preserves the form (3b.1).  But,
using the unitarity of $P - \widetilde P$ again, (3b.5) yields
$$S(\widetilde \Gamma) = S (\mfr1/2 \Gamma_1 + \mfr1/2
\Gamma_2) \geq \mfr1/2 S(\Gamma) + \mfr1/2 S ((P - \widetilde P) \Gamma (P -
\widetilde P)) = S (\Gamma). \eqno(3b.6)$$
Hence, we may restrict the variation to 1-$pdm \ \widetilde \Gamma$ of the form
(3b.4).

Moreover, defining the unitary
$$W = \pmatrix{0&\1 &&\cr -\1&0&&\cr &&0&\1\cr &&-\1&0\cr} , \eqno(3b.7)$$
we observe that $W \widetilde \Gamma W^\dagger$ coincides with $\widetilde
\Gamma$ except that
$\gamma_\uparrow$ and $\gamma_\downarrow$ are interchanged as
are $\alpha_*$ and $\alpha_*^T$.  Hence,
$$\widehat{\Gamma} := \mfr1/2 \widetilde \Gamma + \mfr1/2 W \widetilde \Gamma
W^\dagger = \pmatrix{\gamma^\prime
&&&\alpha'\cr &\gamma^\prime &-\alpha' &\cr &- \alpha'^\dagger &\1 -
\overline{\gamma^\prime} &\cr \alpha'^\dagger &&&\1 - \overline{\gamma^\prime}
\cr}, \eqno(3b.8)$$
where $\gamma^\prime := \mfr1/2 (\gamma_\uparrow + \gamma_\downarrow)$
and $\alpha':=\mfr1/2 (\alpha_*+\alpha_*^T)$.

As before, concavity of $S$  implies
$$S (\widehat{\Gamma}) = S (\mfr1/2 \widetilde \Gamma +
\mfr1/2 W \widetilde \Gamma W^\dagger) \geq \mfr1/2 S
( \widetilde \Gamma) + \mfr1/2 S (W \widetilde \Gamma W^\dagger)
= S(\Gamma), \eqno(3b.9)$$
with equality if and only if $\widehat{\Gamma}= \widetilde \Gamma$.
We have $N(\Gamma) = N(\widehat{\Gamma})$ and
$$\eqalign{\E (\Gamma) &- \E (\widehat\Gamma) = \sum \limits_x U_x \bigl\{
\mfr1/4 [\langle x \vert \gamma_\uparrow \vert x \rangle + \langle x \vert
\gamma_\downarrow \vert x \rangle - 1]^2 \cr
&- [\langle x \vert \gamma_\uparrow \vert x \rangle - \mfr1/2] [\langle x
\vert \gamma_\downarrow \vert x \rangle - \mfr1/2]\bigr\} \geq 0
\cr}\eqno(3b.10)$$
with equality if and only if $\langle x \vert \gamma_\uparrow \vert x
\rangle = \langle x \vert \gamma_\downarrow \vert x \rangle$ for all $x \in
\Lambda$.  \lanbox

Thanks to Lemma~3.1 we may now restrict ourselves to 1-$pdm$ of the form
(3b.2) for which the $4 \times 4$-matrix formalism is clearly redundant.
Indeed, introducing the unitary operator
$$ Y := \pmatrix{ \1 & 0 & 0 & 0 \cr
                  0 & 0 & 0 & \1 \cr
                  0 & \1 & 0 & 0 \cr
                  0 & 0 & -\1& 0 \cr} \eqno(3b.11)$$
on $\H \oplus \H$, one easily checks that $\Gamma$ in (3b.2) becomes
$$ Y \Gamma Y^\dagger =
\pmatrix{
\gamma^\prime & \alpha' &&\cr
\alpha'^\dagger & \1 - \overline{\gamma}^\prime &&\cr
&&      \gamma^\prime & \alpha' \cr
&&      \alpha'^\dagger & \1 - \overline{\gamma}^\prime \cr }
=:
\pmatrix{ \Gamma^\prime &\cr & \Gamma^\prime \cr} . \eqno(3b.12)$$
Although it seems that
the conditions that $\Gamma$ be a 1-$pdm$ on $\H \oplus \H$ are
equivalent to the conditions for $\Gamma^\prime$ to be a 1-$pdm$ on
$\H_\Lambda \oplus \H_\Lambda$, there is one important difference.
We must require $\Gamma^\prime$ to satisfy $0\leq \Gamma'\leq\1$
and to be of the form
$$ \Gamma^\prime = \pmatrix{ \gamma^\prime & \alpha' \cr
\vrule height15ptwidth0pt
\alpha'^\dagger & {\bf 1}- \overline{\gamma}^\prime \cr}, \eqno(3b.13)$$
with
$$ ( \gamma^\prime )^\dagger = \gamma^\prime, \ \ \ \
             \alpha'^T = \alpha' . \eqno(3b.14)$$
Because of this difference
we will distinguish between
$\H_\Lambda \oplus \H_\Lambda$  and $\H$,
even though they are isomorphic.
Physically,
$\H$ is the space of spin-up and spin-down
particles whereas
the action of $Y$ shows that $\H_\Lambda \oplus \H_\Lambda$,
on which $\Gamma^\prime$ is defined, is rather the space of spin-up
particles and spin-down holes.

We shall denote the projection operator in $\H_\Lambda \oplus \H_\Lambda$
which projects onto (functions non-vanishing only at) the site $x \in
\Lambda$ by $\1_x$.
The diagonal part of a 1-$pdm \ \Gamma^\prime$ is denoted by
$$\Gamma^\prime_x := \1_x \Gamma^\prime \1_x
= \pmatrix{\gamma^\prime (x) &\alpha'(x) \cr\vrule height15ptwidth0pt
\overline \alpha'(x) &\1 - \gamma^\prime (x) \cr} \1_x, \eqno(3b.15)$$
where $\gamma^\prime (x) := \langle x \vert \gamma \vert x
\rangle$ and $\alpha'(x) := \langle x  \vert \alpha' \vert
x \rangle$.
Note that the product of any two operators on
$\H_\Lambda \oplus \H_\Lambda$
of the form $A(x)\1_x$ and $B(x)\1_x $, where $A$ and $B$ are $2 \times 2$
matrices as in (3b.15), is given by the operator $\left( A(x) B(x)\right)
\1_x$.
According to this rule one easily checks that
the trace on $\H_\Lambda \oplus \H_\Lambda$
of $\Gamma^2_x$ satisfies
$$\eqalign{\mfr1/2 \Tr [\Gamma^{\prime \ 2}_x] &= \gamma^\prime (x)^2 + \vert
\alpha'(x) \vert^2 + \mfr1/2 - \gamma^\prime (x) \cr
&= [ \gamma^\prime (x) -
\mfr1/2]^2 + \vert \alpha'(x) \vert^2 + \mfr1/4 . \cr}\eqno(3b.16)$$

By means of (3b.16) we may rewrite the pressure functional (3a.8) as
$$\eqalignno{-\P_{\beta, \mu} (\Gamma^\prime) := -\P_{\beta, \mu} (\Gamma)
= & 2 \Tr [t \gamma] - \mfr1/2 \sum
\limits_x U_x \left(\Tr [\Gamma^{\prime \ 2}_x] + \mfr1/2\right)\cr
&- 2 \mu  \Tr[ \gamma^\prime ] +
\beta^{-1} \Tr [\Gamma^\prime \ln \Gamma^\prime
+ (\1 - \Gamma^\prime) \ln (\1 -\Gamma^\prime)].
\qquad&(3b.17)\cr}$$

Finally, we embed the hopping matrix into the $2 \times 2$-matrix
formalism.  We define for any real number $\lambda$
$$T_\lambda := \pmatrix{t - \lambda &\cr
&-(\overline t - \lambda) \cr}. \eqno(3b.18)$$
With this definition we obtain
$$\mfr1/2 \Tr [T_\lambda \Gamma^\prime] = \Tr [t \gamma^\prime]
- \lambda \Tr [\gamma^\prime] + \mfr1/2\lambda \vert \Lambda \vert,
\eqno(3b.19)$$
and, hence,
$$\eqalignno{-\P_{\beta, \mu} (\Gamma^\prime) = &\Tr [T_\mu \Gamma^\prime] -
\mfr1/2 \sum \limits_x U_x \Tr [\Gamma_x^{\prime \ 2}] \cr
&+  \beta^{-1} \Tr [\Gamma^\prime  \ln \Gamma^\prime + (\1 - \Gamma^\prime)
 \ln (\1 - \Gamma^\prime)]
+ \mfr1/4 \sum_xU_x - \mu \vert \Lambda \vert . \qquad&(3b.20)\cr}$$

We rewrite $\Gamma^\prime_x$ as
$$\Gamma_x = (\gamma'(x)-\mfr1/2) \pmatrix{1&\cr &-1 \cr} \1_x +
\pmatrix{& \alpha'(x)  \cr
\overline \alpha'(x) &\cr} \1_x
+ \mfr1/2 \1_x
.\eqno(3b.21)$$
The cross terms in $\Gamma^{\prime \ 2}_x$ are traceless, which
implies that
$$\eqalignno{\Tr [\Gamma^{\prime \ 2}_x] &=
\Tr [(\Gamma^\prime_x - \mfr1/2 \1_x)^2] + \mfr1/4 \Tr[\1_x] \cr
&= \Tr [(\Gamma^\prime_x - \mfr1/2 \1_x)^2] + \mfr1/2.
\qquad&(3b.22) \cr}$$
Using (3b.22), (3b.20) becomes
$$\eqalignno{-\P_{\beta, \mu} (\Gamma^\prime)
&= \Tr [T_\mu \Gamma^\prime] -
\mfr1/2 \sum \limits_x U_x \Tr [(\Gamma^\prime_x - \mfr1/2 \1_x)^2] \cr
&- \mu  \vert \Lambda \vert +
\beta^{-1} \Tr [\Gamma^\prime \ln \Gamma^\prime
 + (\1 - \Gamma^\prime) \ln (\1 - \Gamma^\prime)].
\qquad&(3b.23)\cr}$$

We associate a multiplication operator $UD := \sum \limits_x
U_x D_x\1_x$ with any real function $d(x)$ and complex function $\delta (x)$ by
$$D_x = \pmatrix{d(x) & \delta (x) \cr\vrule height15ptwidth0pt
\overline{\delta (x)} & -d(x) \cr}.
\eqno(3b.24)$$
Notice that $D_x$ has the same form as $\Gamma'_x-\mfr1/2 \1_x$.
The operator $UD$ is really a matrix-valued
potential which will enable us to linearize the quadratic trace in (3b.23)
by means of the identity
$$\eqalignno{-\Tr [(\Gamma^\prime_x - \mfr1/2\1_x)^2] &= \min
\limits_{d,\delta}
\big\{ - 2 \Tr [D_x (\Gamma^\prime_x - \mfr1/2\1_x)] + \Tr [D^2_x] \big\} \cr
&= \min \limits_{d,\delta} \big\{ - 2 \Tr [D_x \Gamma^\prime_x] + \Tr [D^2_x]
\big\}. \qquad&(3b.25)\cr}$$
Indeed it is possible to convert the variation over 1-$pdm$'s in (3a.9)
into a variation over all matrix-valued potentials $D$,
which we will show by means of the following lemma.

{\bf 3.2 LEMMA:} {\it
Let $D$ be any matrix-valued potential as in (3b.24),
$Q$ be a self-adjoint operator on $\H_\Lambda \oplus \H_\Lambda$
and $V$ the unitary given by
$$ V := \pmatrix{ 0 & \1 \cr -\1 & 0 \cr}. \eqno(3b.26) $$
Suppose that $F$ is an odd real-valued function,
i.e., $ F[x] = - F[-x] $.
Then $\widetilde\Gamma$ defined by
$$\widetilde\Gamma - \mfr1/2 := F\left[ T_\mu - U D \right]
  + Q - V \overline{Q} V^\dagger, \eqno(3b.27)$$
fulfills (3b.13), (3b.14). }

{\it Proof:}
It is easily checked that $V T_\mu V^\dagger = - \overline{T}_\mu$
and $V UD V^\dagger = - \overline{UD}$.
Then by the spectral theorem,
$$\eqalignno{
V ( \widetilde\Gamma - \mfr1/2 ) V^\dagger
=&
F\left[ V ( T_\mu - UD ) V^\dagger \right] + VQV^\dagger - \overline{Q} \cr
=&
F\left[ -( \overline{T_\mu - UD }) \right]
-( \overline{ Q - V \overline{Q} V^\dagger } ) \cr
=&
-\overline{ \left( F\left[ T_\mu - UD  \right]
 + Q - V \overline{Q} V^\dagger \right)} \cr
=&
 - ( \overline{ \widetilde\Gamma - \mfr1/2 } ) ,
\qquad&(3b.28) \cr } $$
which is equivalent to (3b.13) and (3b.14). \lanbox

{\bf 3.3 THEOREM (Positive temperature pressure):} {\it For all
$0<\beta<\infty$
and all $\mu$ we can write the pressure $\P (\beta, \mu)$ as the following
variation over the functions $d$ and $\delta$.

$$
        -\P (\beta, \mu)=\min_{d,\delta}
         \sr_{\beta,\mu}(d,\delta)- \left( 2 \beta^{-1} \ln 2 + \mu
\right) \vert \Lambda \vert,\eqno(3b.29)
$$
where
$$\sr_{\beta, \mu}(d,\delta): = \sr_{\beta, \mu}(D):=
 -  \beta^{-1} \Tr [\ln \cosh  \mfr{\beta}/2 (T_\mu - UD)] +
\mfr1/2 \sum \limits_x U_x \Tr [D^2_x].\eqno(3b.30)$$
If a potential $D$ minimizes $\sr_{\beta, \mu}$ then the operator
$$\Gamma^\prime = \left( \1 + \exp \biggl[ \beta  (T_\mu -
UD) \biggr] \right)^{-1} ,\eqno(3b.31)$$
is a minimizer for (3b.23), (i.e.,
defines a HF Gibbs state) and satisfies the consistency equation
$$\Gamma^\prime_{x} = \left(D_{x} + \mfr1/2\right )\1_x.\eqno(3b.32)$$
Conversely, if $\Gamma'$ is a minimizer for (3b.23) then the potential $D$
defined by (3b.32) minimizes $\sr_{\beta,\mu}$ and satisfies (3b.31).}

{\it Proof:} By means of (3b.25) we write
$$\eqalignno{-\P (\beta, \mu) &= \min \limits_{\Gamma^\prime} \min
\limits_{d, \delta} \Biggl\{  \Tr [(T_\mu - UD) \Gamma^\prime] + \mfr1/2
\sum \limits_x U_x \Tr [D^2_x] \cr
&\phantom{= \min \limits_{\Gamma^\prime} \min
\limits_{d, \delta} \Biggl\{ }
- \mu \vert \Lambda \vert + \beta^{-1} \Tr [\Gamma^\prime \ln \Gamma^\prime
+ (\1 - \Gamma^\prime) \ln (\1 - \Gamma^\prime) ] \Biggr\} \cr
& = \min \limits_{d, \delta} \Biggl\{\min_{\Gamma^\prime} \Bigl\{
\Tr [(T_\mu - UD)\Gamma']
+ \beta^{-1} \Tr [\Gamma' \ln\Gamma' + (\1 - \Gamma')  \ln (\1 - \Gamma')]
\Bigr\}\cr
&\phantom{= \min
\limits_{d, \delta} \Biggl\{ }
+ \mfr1/2 \sum \limits_x U_x \Tr [D^2_x] \Biggr\}
- \mu  \vert \Lambda \vert.
\qquad&(3b.33) \cr }$$
The minimum is over all $0\leq\Gamma^\prime\leq \1$ satisfying (3b.13) and
(3b.14).
Note that all we did in (3b.33) was to interchange the two minimizations this
is
of course allowed since we are simply looking for the minimum in the set
of all $\Gamma'$, $d$ and $\delta$.

First relaxing the minimization
to be over all $0\leq\Gamma'\leq\1$,
we can explicitly compute the minimum over
$\Gamma'$ in the second line in (3b.33).
We see that the minimum is uniquely achieved
for the $\Gamma'$ defined in (3b.31).
We observe, however, that $\Gamma' - \mfr1/2 =
- \mfr1/2 \tanh\beta ( T_\mu - UD)$ is of the form suitable for Lemma~3.2
(choosing $F[x] = - \mfr1/2 \tanh \left[ \beta x \right]$ and $Q=0$).
The operator $\Gamma'$ defined by (3b.31) therefore automatically satisfies
(3b.13) and
(3b.14) and we have, indeed, found the minimizer in the second line in
(3b.33).
Moreover, a simple computation then shows that the right side of (3b.33)
is identical to the left side of (3b.29).

If $D$ minimizes $\sr_{\beta,\mu}$ and we define $\Gamma'$ by (3b.31) we have
from (3b.25) and (3b.29) that
$$
        \eqalignno{-\P (\beta, \mu)=\sr_{\beta,\mu}(D)-(2\beta^{-1}\ln 2
	+\mu)|\Lambda|=&
        \Tr [(T_\mu - UD) \Gamma^\prime] + \mfr1/2
        \sum \limits_x U_x \Tr [D^2_x] \cr
        &- \mu \vert \Lambda \vert + \beta^{-1}
        \Tr [\Gamma^\prime \ln \Gamma^\prime + (\1 -
        \Gamma^\prime) \ln (\1 - \Gamma^\prime) ]\cr \geq& -\P_{\beta, \mu}
        (\Gamma')
	. &(3b.34)}
$$
Since $-\P (\beta, \mu)\leq -\P_{\beta, \mu}(\Gamma')$ we must have equality
in (3b.34)
but it follows from (3b.25) that this can only happen if (3b.32) holds.

Conversely, if $\Gamma'$ minimizes $-\P_{\beta, \mu}$
and we define $D$ by (3b.32) we see that
$$
	\eqalignno{-\P (\beta, \mu)=-\P_{\beta, \mu}(\Gamma')=&
	\Tr [(T_\mu - UD) \Gamma^\prime] + \mfr1/2
        \sum \limits_x U_x \Tr [D^2_x] \cr
        &- \mu \vert \Lambda \vert + \beta^{-1}
        \Tr [\Gamma^\prime \ln \Gamma^\prime + (\1 -
        \Gamma^\prime) \ln (\1 - \Gamma^\prime) ]\cr\geq&\sr_{\beta,\mu}(D)
	-(2\beta^{-1}\ln 2 +\mu)|\Lambda|& (3b.35)}
$$
and we conclude that $D$ is a minimizer for $\sr_{\beta,\mu}$ and that
(3b.31) must hold.
\lanbox

We remark that the minimum of $\sr_{\beta,\mu}(d,\delta)$
will be attained for $d$ and $\delta$ satisfying
$$
        d(x)^2 + \vert \delta (x) \vert^2 \leq \mfr1/4
        \quad\hbox{for all } x\in\Lambda. \eqno(3b.36)
$$
In fact, since $0 \leq \Gamma^\prime \leq \1$ we see from (3b.32)
that $D^2_x \leq \mfr1/4$ which implies (3b.36).

\bigskip\noindent
{\bf III.c  GAP AND ZERO TEMPERATURE LIMIT}

Our discussion will turn in this section to the zero temperature limit
$\beta \to \infty$. Recall that the generalized HF energy
is given by
$$ E^{\HF}(\mu) = \inf\{ \E(\Gamma^\prime) - \mu N(\Gamma^\prime)
\ \vert  \ \
\Gamma^\prime \hbox{ is \ a \ 1-}pdm \}. \eqno(3c.1)$$
A minimizing 1-$pdm$ $\Gamma^\prime_0$, that is one that satisfies,
$E^{\HF}(\mu) =
\E(\Gamma^\prime_0) - \mu N(\Gamma^\prime_0)$, is called a HF ground state.
To make contact with our previous notation let us denote
$$ - \P_{\infty, \mu}(\Gamma^\prime) := \E( \Gamma^\prime) - \mu
N(\Gamma^\prime)
= \lim \limits_{\beta \to \infty} - \P_{\beta, \mu} (\Gamma^\prime),
\eqno(3c.2)$$
and $- \P(\infty, \mu) := E^{\HF}(\mu) =
\lim \nolimits_{\beta \to \infty} - \P(\beta, \mu)$.

We can derive an analog of Theorem~3.3 for the zero temperature pressure
by simply dropping the term $- \beta^{-1} S(\Gamma)$, and essentially
repeating the whole positive temperature discussion.
There is, however, one subtle difference. Given the potential $D$,
the minimizing
$\Gamma'$ in (3b.33) was uniquely defined by (3b.31). In the zero temperature
case the $\Gamma'$ we are looking for is simply the minimizer of
$\Tr[(T_\mu-UD)\Gamma']$. If the operator $(T_\mu-UD)$ has zero eigenvalues
this minimizer would not be uniquely defined. In our case, however, we
prove in Lemma~3.5 below that $(T_\mu-UD)$ has no zero eigenvalues. In fact,
$$ \vert e_j \vert \geq \mfr1/4
\min \limits_{x} \{ U_x \} \vert \Lambda \vert^{-1} >0
$$
for any eigenvalue $e_j$ of $T_\mu -UD$.
We can therefore write that the minimizer of $\Tr[(T_\mu-UD)\Gamma']$
is $\Gamma'= \chi(T_\mu - UD)= \lim_{\beta\to\infty}
\left( \1 + \exp[ \beta  (T_\mu - UD)] \right)^{-1}$, where
$\chi(a) := 1$ if $a<0$ and zero otherwise. The point is that the value
$\chi(0)$ is unimportant.
Let us note that there is another way of characterizing the
zero temperature states, thereby avoiding the repetition of the whole
positive temperature discussion. Namely, the main theorem~3.12
of this chapter allows us to obtain all zero temperature states
as limits of positive temperature states.

{\bf 3.4 THEOREM (Ground state pressure):}
{\it We have
$$
        -\P (\infty, \mu)=E^{\HF}(\mu)=\min_{d,\delta}
        \sr_{\infty,\mu}(d,\delta)- \mu \vert \Lambda \vert,
$$
where
$$\sr_{\infty,\mu}(d,\delta):=\sr_{\infty,\mu}(D):=
 -\mfr1/2\Tr |T_\mu - UD| +
\mfr1/2 \sum \limits_x U_x \Tr [D^2_x]. \eqno(3c.3)$$
If a potential $D$ minimizes $\sr_{\infty,\mu}$ then the operator
(with $\chi$ defined as above)
$$\Gamma^\prime = \chi(T_\mu - UD) ,\eqno(3c.4)$$
minimizes (3c.2), (i.e., defines a HF ground state) and satisfies the
consistency equation
$$\Gamma^\prime_{x} = (D_{x} + \mfr1/2) \1_x . \eqno(3c.5)$$
Conversely, if $\Gamma^\prime$ minimizes (3c.2) then the potential $D$
defined by (3c.5) minimizes $\sr_{\infty,\mu}$ and satisfies (3c.4).}

The reader may wonder why the pressure depends on the absolute value
of $T_\mu - UD$ and not only on the negative eigenvalues. In this context
it should be kept in mind that because of the special form of the operator
$T_\mu - UD$ the trace $-\mfr1/2\Tr|T_\mu - UD|$ is, in fact, equal to
the sum of the negative eigenvalues.
In comparing (3c.3) to (3b.30) we notice that
$\lim_{\beta\to\infty}\beta^{-1}\ln\cosh(\beta x/2)=\mfr1/2|x|$
it is therefore natural to write the absolute value in (3c.3).
It remains to prove the absence of zero eigenvalues.

{\bf 3.5 LEMMA (Gap estimate):}
{\it Let $D$ be a minimizing matrix-valued potential for $\sr_{\beta,\mu}$
with $0<\beta\leq\infty$ and denote
the eigenvalues of $T_\mu - UD$ by $e_1, e_2, \ldots, e_{ 2
\vert \Lambda \vert
}$. Then, for any $j=1,2, \ldots, 2 \vert \Lambda \vert$, we have
$$ \vert e_j \vert \geq \mfr1/4 U_{\min} \vert \Lambda \vert^{-1} -
   \beta^{-1}  2 \ln 2 , \eqno(3c.6)$$
where $U_{\min} := \min \limits_{x \in \Lambda} \{ U_x \}$. }

{\it Proof:}
If $D$ minimizes $\sr_{\beta,\mu}$ for finite $\beta$
we know that the operator $\Gamma'$
defined in (3b.31) minimizes the functional $-\P_{\beta,\mu}$ in (3b.23).
Likewise, if $D$ minimizes $\sr_{\infty,\mu}$ we know that the operator
$ \Gamma'=\lim_{\beta\to\infty}(\1+\exp[\beta(T_\mu-UD)])^{-1}$ minimizes the
energy functional (3c.2). Notice that these definitions of $\Gamma'$
imply (3b.13) and (3b.14). What we do not know a-priori is that if
$\beta=\infty$ (i.e., zero temperature), then $\Gamma'$ agrees with
$\chi(T_\mu-UD)$.

Let $Q := \vert \varphi \rangle \langle \varphi \vert$ denote the
projection onto some eigenvector $\varphi$ of $T_\mu - UD$ with
corresponding eigenvalue $e$.
Note that $V\overline{Q} V^\dagger$ is the projection onto the eigenvector
$V\overline{\varphi}$ of ${ T_\mu - UD}$ with eigenvalue $-e$,
provided $V$ is the unitary given by (3b.26).

In terms of the operators $\Gamma'$ discussed above we define
$$ \widetilde\Gamma^\prime := \Gamma^\prime + \delta Q
- \delta V\overline{Q} V^\dagger.  \eqno(3c.7)$$
We choose $ -1/2 \leq \delta \leq 1/2$ such that $e \delta \geq 0$.
With this choice $0\leq \widetilde\Gamma^\prime\leq\1$.
Note that $\varphi$ is an eigenvector of $\Gamma'$. We denote its
eigenvalue by $\lambda$. Then $\varphi$ is also an eigenvector of
$\widetilde\Gamma^\prime$ with eigenvalue $\lambda+\delta$. With our
choice of $\delta$ we have $0\leq \lambda,\lambda+\delta\leq1$.

Lemma~3.2 now implies that $\widetilde\Gamma^\prime$ is
admissible on $\H_\Lambda \oplus \H_\Lambda$, as it corresponds to
the 1-$pdm$
$$ Y^\dagger \pmatrix{ \widetilde\Gamma^\prime & \cr &
\widetilde\Gamma^\prime \cr} Y . \eqno(3c.8) $$

Since $\Gamma^\prime$ minimizes $- \P_{\beta, \mu}$ it follows that
$$\eqalignno{ 0
\geq &
\P_{\beta, \mu}(\widetilde\Gamma^\prime) - \P_{\beta, \mu}(\Gamma^\prime) \cr
=&
- \delta \Tr[ T_\mu (Q- V\overline{Q} V^\dagger )]
 - \mfr1/2 \sum \limits_{x} U_x
\Tr[ ( \Gamma^\prime_{x} - \mfr1/2 \1_x)^2 -
     ( \Gamma^\prime_{x} - \mfr1/2 \1_x + \delta Q_x
     - \delta V\overline{Q}_x V^\dagger )^2 ] \cr
&
+ 2  \beta^{-1} [ \lambda \ln \lambda + (1- \lambda ) \ln (1- \lambda )
- (\lambda + \delta) \ln (\lambda + \delta)
- (1- \lambda - \delta ) \ln (1- \lambda - \delta ) ] \cr
=&
- 2 \delta e + \mfr1/2 \delta^2 \sum \limits_{x} U_x
\Tr[ (Q_x - V\overline{Q}_x V^\dagger )^2 ]
+ 2 \beta^{-1} [ f(\lambda) - f(\lambda + \delta)], \qquad&(3c.9) \cr}$$
where $f(\lambda) := \lambda \ln \lambda + (1- \lambda ) \ln (1- \lambda )$.
Clearly, $-\ln2\leq f(\lambda)\leq0$ for $0\leq\lambda\leq1$. Hence
$|f(\lambda) - f(\lambda + \delta)|\leq \ln2$. We insert this estimate into
(3c.9) and obtain, using $\delta e=|\delta||e|$
$$ 2 \vert e \vert
\geq \mfr1/2
\vert \delta \vert \sum \limits_x U_x
\Tr[ (Q_x - V\overline{Q}_x V^\dagger )^2 ]
- 2|\delta|^{-1}\beta^{-1} \ln 2. \eqno(3c.10) $$

Furthermore,
$$ \Tr[ (Q_x - V\overline{Q}_x V^\dagger )^2 ]
= 2 \left( \vert \varphi(x, \uparrow) \vert^2 +
\vert \varphi(x, \downarrow) \vert^2 \right)^2, \eqno(3c.11)$$
leads us by the Schwarz inequality to
$$ \eqalignno{ \sum \limits_x U_x
\Tr[ (Q_x - V\overline{Q}_x V^\dagger )^2 ]
\geq &
2 U_{\min} \sum \limits_x
 \left( \vert \varphi(x, \uparrow) \vert^2 +
\vert \varphi(x, \downarrow) \vert^2 \right)^2 \cr
\geq &
2 U_{\min} \vert \Lambda \vert^{-1} \left(\sum \limits_x
 \vert \varphi(x, \uparrow) \vert^2 +
\vert \varphi(x, \downarrow) \vert^2\right)^2 \cr
=& 2 U_{\min} \vert \Lambda \vert^{-1}. \qquad&(3c.12) \cr } $$
Inserting (3c.12) into (3c.10) and choosing $|\delta|=\mfr1/2$
concludes the proof. \lanbox

We interpret any minimizing $\Gamma$ in Theorems~3.3 and 3.4 as the physical
state of the system at $\beta, \mu$. Since $\Gamma$, subject to (3b.31),
is the Fermi distribution of the corresponding operator
$T_\mu - UD$,
we might as well interpret its eigenstates as the only orbitals the
electrons can possibly occupy. This is nothing but viewing
$T_\mu - UD$
as the relevant quasiparticle Hamiltonian for the considered system.
{F}rom the BCS theory of superconductivity the question arises whether or not
the quasiparticle spectrum, i. e., the spectrum of
$T_\mu - UD$
has a gap around $0$ for low enough temperatures.
The answer to this question is always positive as we have just pointed out
in Lemma~3.5.
We remark, however, that our estimate for the size of the gap
becomes trivial in the thermodynamic limit. In Theorem~3.15 below
we give a formula for the gap in the translation invariant case.

\bigskip\noindent
{\bf III.d  BROKEN GAUGE SYMMETRIES}

We pause to remark that so far no assumption was made about the hopping
matrix $t$, except that it be self-adjoint, connected and traceless.
In Theorems~3.3 and 3.4 in the previous section we established a
unique correspondence between the HF ground and Gibbs states (described
by $\Gamma'$) and the potentials $D$ that minmize $\sr_{\beta,\mu}$. We
can therefore now entirely concentrate on the determination of the
functions $d$ and $\delta$ that yield a minimizer $D$ for
$\sr_{\beta,\mu}$, where $0<\beta\leq\infty$.

A key ingredient in the following analysis are the representations
$$\eqalign{\ln \cosh x =& \sum \limits^\infty_{k=0} \ln \left[ 1 + \biggl(
{x \over \pi (k + 1/2)} \biggr)^2 \right] , \cr
        |x|=&{1 \over\pi} \int_0^\infty\ln\left(1+{x^2\over c^2}\right)dc.
        }\eqno(3d.1)
$$
The significance of (3d.1), as examples of {\it integrated
Pick functions}, was emphasized in Lieb and Loss [LL] and Kennedy and
Lieb [KL].  The
virtue of (3d.1) is, roughly speaking, that it allows us to convert traces
into determinants. In fact, using these representations we may write
$$
        \eqalignno{\sr_{\beta,\mu}(D)=&
        - \beta^{-1}\sum \limits^\infty_{k=0}\Bigl[ \ln \Det
        \{ 4 \pi^2 (k + 1/2)^2 \beta^{-2} + (T_\mu - UD)^2 \}\cr
        &\phantom{- \beta^{-1}\sum \limits^\infty_{k=0}\Bigl[ }
        -2|\Lambda|\ln \{ 4 \pi^2 (k + 1/2)^2 \beta^{-2} \}\Bigr]
        +\mfr1/2 \sum \limits_x U_x \Tr [D^2_x],\qquad&(3d.2)\cr
        \sr_{\infty,\mu}(D)=&-{1 \over\pi}\int_0^\infty
        \Bigl[\mfr1/2 \ln\Det\{ c^2+(T_\mu - UD)^2\}-2|\Lambda|\ln c\Bigr]dc\cr
        & +\mfr1/2 \sum \limits_x U_x \Tr [D^2_x].\qquad&(3d.3)}$$
Since we are dealing with finite dimensional matrices,
convergence of the above expressions are evident for any choice of $D$.

{\bf 3.6 LEMMA (Phase alignment):}  {\it Let $t$ be real.  Then,
$$\sr_{\beta, \mu} (d, \delta) \geq \sr_{\beta, \mu} (d, e^{i \theta}
\vert \delta \vert). \eqno(3d.4)$$
If $\delta (x) \not= 0$ for
all $x \in \Lambda$, equality holds in (3d.4) only if
$\delta (x) = \vert \delta (x) \vert e^{i\theta}$ for all points $x \in
\Lambda$ and for some fixed $0 \leq \theta < 2 \pi$.  }

Remark: Lemma 3.8 below will show that, for the ground state, either $\delta
(x) \not= 0$ for all $x$ or else $\delta (x) = 0$ for all $x$.

{\it Proof:}
In order to prove (3d.4) we may assume that $\delta(x)\ne0$ for all
$x\in\Lambda$. In fact, if we have proved (3d.4) for
non-vanishing $\delta$ we conclude for general $\delta$ by the continuity of
$\sr_{\beta,\mu}$ that
$$
\sr_{\beta,\mu}(d,\delta)=\lim_{\varepsilon\to0}
        \sr_{\beta,\mu}(d,\delta+\varepsilon)\geq \lim_{\varepsilon\to0}
        \sr_{\beta,\mu}(d,|\delta+\varepsilon|)
        =\sr_{\beta,\mu}(d,|\delta|).\eqno(3d.5)
$$
For the cases of equality, however, we have to assume
$\delta$ non-vanishing.

Just as $D$ corresponds to $d$ and $\delta$ via (3b.24), we denote the
operator corresponding to $\widetilde d = d$ and $\widetilde \delta :=
e^{i\theta} \vert \delta
\vert$ by $\widetilde D$.  In view of (3d.2--3), it suffices to prove, for
every real $c$, that
$$\ln \Det [c^2 + (T_\mu - UD)^2] \leq \ln \Det [c^2 + (T_\mu - U
\widetilde D)^2] \eqno(3d.6)$$
and to show that equality in (3d.6) implies $\delta (x) =
e^{i \theta} \vert \delta (x) \vert$ for some $0 \leq \theta < 2 \pi$.

To this end, we define an operator
$$B := \sum \limits_x B_x,
\eqno(3d.7)$$
with $B_x > 0$ (to be chosen below).
Furthermore, we rewrite $UD$ in the
following way.  Let $F := \sum \nolimits_x F_x\1_x$ and $G := \sum\nolimits_x
G_x\1_x$, where
$$\eqalign{F_x &:=  U_x \ d(x)
\pmatrix{1& \cr
        &-1\cr} ,\cr
G_x &:= U_x \pmatrix{&\delta (x) \cr
\overline{\delta (x)} & \cr}  .
}\eqno(3d.8)$$
Hence,
$$T_\mu - UD = T_\mu - F - G. \eqno(3d.9)$$
Notice that both $F$ and $G$ commute with $B$, and that
$F$ and $G$ anticommute.
  Thus, abbreviating $B^{1/2} Q B^{1/2} = : \widehat
Q$ for any operator $Q$
we have the following identity.
$$\eqalignno{\ln \Det &[c^2 + (T_\mu - UD)^2] + 2 \ln \Det [B] \cr
&= \ln \Det [B^{1/2} (ic + T_\mu - F - G) B (-ic + T_\mu - F -
G) B^{1/2}] \qquad&(3d.10) \cr
&= \ln \Det [ A(c,D) - \{ \widehat T_\mu, \widehat G \}], \cr}$$
where $\{A,B\}:=AB+BA$ is the anticommutator and where
$$A(c,D) := c^2 B^2 + (\widehat T_\mu - \widehat F)^2 + {\widehat G}^2
+ ic [B, \widehat T_\mu].  \eqno(3d.11)$$

We shall now prove that
$$\ln \Det [c^2 + (T_{\mu}-UD)^2 ] \leq
\ln \Det [A(c,D)] - 2\ln\Det [B], \eqno(3d.12)$$
and that equality holds in (3d.12) if and only if $\{ \widehat T_\mu
,\widehat G\}$ vanishes.

We will prove this claim by using a concavity argument.
Introducing a unitary
$$V := \pmatrix{0 &-\1 \cr \1 & 0 \cr} , \eqno(3d.13)$$
one easily checks that $VBV^\dagger = B$,   $V \widehat T_\mu V^\dagger = -
\widehat T_\mu = - \overline{\widehat T_\mu}$, using the reality of $t$,
and $VFV^\dagger = - F$,
$VGV^\dagger = \overline
G$.  Hence, we obtain
$$VA(c,D) V^\dagger = A(-c,D) = \overline{A(-c,D)}, \qquad \qquad V
\{\widehat T_\mu,
\widehat G\} V^\dagger = - \overline{\{\widehat T_\mu ,\widehat G\}}
\eqno(3d.14)$$
Now, from (3d.10) we see that the determinant in question is real and
depends only on $c^2$.  Therefore,
$$\eqalignno{ \ln \Det &[A(c,D) - \{ \widehat T_\mu, \widehat G \}]
= \ln \Det [A(-c,D) - \{ \widehat T_\mu, \widehat G \}] \cr
&= \ln \Det [\overline{A(-c,D)} -
\overline{ \{ \widehat T_\mu, \widehat G \} }]
= \ln \Det [V (A(c,D) + \{ \widehat T_\mu, \widehat G \}) V^\dagger] \cr
&= \ln \Det [A(c,D) + \{ \widehat T_\mu, \widehat G \}].
\qquad&(3d.15)\cr }$$
By the strict concavity of $\ln \Det [\cdot]$, i.e., $\mfr1/2 \ln \Det [a]
+ \mfr1/2 \ln \Det [b] \leq \ln \Det [\mfr1/2 (a+b)]$, with strict
inequality unless $a = b$, we arrive at the assertion in (3d.12).

We now choose $B_x = U_x^{-1}|\delta(x)|^{-1}$, keeping in mind our
assumptions that $\delta(x)\not= 0$ and $U_x >0$. The key observation
is that
$$\widehat G_x = (B G)_x = {1 \over \vert \delta (x) \vert}
\pmatrix{&\delta (x) \cr
\overline{\delta (x)} &\cr}  \eqno(3d.16)$$
is unitary for all $x \in \Lambda$, i.e., $B^2 G^2 = \1$.
Thus, for any two points $x,y \in
\Lambda$ the $2 \times 2$ matrices belonging to $(\widehat G)_x$ and
$(\widehat G)_y$ are identical modulo a phase factor $\delta (x)
\overline{\delta (y)} \vert \delta (x) \delta (y) \vert^{-1}$.
The last term $\{ \widehat T_\mu , \widehat G \}$ vanishes if
and only if
$$t_{xy} \delta (x) \vert \delta (x) \vert^{-1} = t_{xy}
\delta (y) \vert \delta (y) \vert^{-1} \eqno(3d.17)$$
for all $x,y \in \Lambda$.
It is essential for (3d.17) that $t$ be real.  {F}rom (3d.17) we learn by
the connectedness of $t$ that $\widehat T_\mu \widehat G + \widehat G
\widehat T_\mu$ vanishes if and only if $\delta$ is of the desired
form, $\delta (x) = e^{i\theta} \vert \delta (x) \vert$.

Notice now that $A(c,D) = A(c,\tilde D)$ and that for $\tilde D$ we
have equality in (3d.12). Inequality (3d.6) is therefore a consequence of
(3d.12). \lanbox

We remark that by (3d.12) we also get a lower bound on $\sr_{\beta, \mu}
(D)$ that depends only on $d$ and $\vert \delta \vert$, even in cases
in which $t$ is not real.  However, this lower bound cannot then be
expressed as $\sr_{\beta, \mu} (D^\prime)$ for some matrix-valued
potential $D^\prime$.  The reason is that (3d.17) now reads
$t_{xy}\delta(x)|\delta(x)|^{-1}=
\overline{t_{xy}}\delta(y)|\delta(y)|^{-1}$. This equation cannot be
satisfied unless $\delta\equiv0$ as shown in (3d.26) below.

The strategy for proving Lemma~3.6 can actually be applied to other types
of hopping matrices, as the following lemma shows.
Using the notion of pseudo-spin discussed below (in (3d.50--53))
the conclusion of the next lemma is that minimizers will have aligned
pseudo-spin.

{\bf 3.7  LEMMA (Pseudo-spin alignment):}  {\it Let $t$ be bipartite
(but not necessarily real) and $\mu = 0$. Then
$$\sr_{\beta,0} (d , \delta) \geq \sr_{\beta,0} \left( (-1)^x
\sqrt{d^2 + \vert \delta \vert^2}, 0 \right). \eqno(3d.18)$$
In case $t$ is not real and $n(x) := \sqrt{d^2 (x) + \vert \delta (x)
\vert^2}\ne0$ for all $x \in\Lambda$ equality holds in (3d.18) if and only
if $\delta (x) = 0$ and $d (x) =  (-1)^x n(x)$ or
$d(x) =  (-1)^x n(x)$ for all $x \in
\Lambda$.  In case $t$ is real and $n(x)\ne0$ for all $x\in\Lambda$
equality holds in (3d.18) if and only if for all $x \in \lambda$
$$
        \pmatrix{d(x) & \delta (x) \cr \vrule height15ptwidth0pt
	\overline \delta (x) &-d(x)\cr} =
        w_x\pmatrix{ (-1)^x n(x) & \cr\vrule height15ptwidth0pt
        &- (-1)^x n(x) \cr} w_x^\dagger,\eqno(3d.19)
$$
where
$$
        w_x=\pmatrix{(-1)^x & \cr & 1 \cr}w\pmatrix{(-1)^x & \cr & 1 \cr}
$$
for some unitary $2 \times 2$-matrix $w$ independent of $x \in \Lambda$.}

{\it Proof:}  Since the main idea of the proof is the same as that in Lemma
3.6, we shall use the notation therein.
In proving (3d.18) we can assume $n(x)$ non-vanishing for all $x$,
otherwise we use a continuity argument as in Lemma~3.6.
For simplicity we write $T_{\mu=0}=T$.
Again, our assertion follows by
showing that for all real numbers $c$ the inequality
$$\ln \Det [c^2 + (T  - UD)^2] \geq \ln \Det [c^2 + (T  - U
\widetilde D)^2] \eqno(3d.20)$$
holds, and there is equality if and only if $d$ and $\delta$ fulfill (3d.19)
and $\widetilde D$ is assumed to be some matrix-valued potential for which
the corresponding functions $\widetilde d$ and $\widetilde \delta$ do obey
(3d.19).

Again, we define  $\widehat Q := B^{1/2} QB^{1/2}$.
In analogy with (3d.10) we obtain
$$\eqalignno{\ln \Det &[c^2 + (T  - UD)^2] + 2 \ln \Det [B]
\cr
&= \ln \Det [B^{1/2} (ic + T  - UD) B (-ic + T  - UD) B^{1/2}]\qquad&(3d.21)
\cr
&= \ln\Det [A_*(c,D) - \{ \widehat T, U\widehat{D} \} ],
\cr}$$
where now $$A_*(c,D) := (U\widehat{D})^2 + c^2 B^2 + \widehat T^2
+ ic [B ,\widehat T]. \eqno(3d.22)$$

Thanks to our assumption of bipartiteness of $t$ and the special choice of
$\mu = 0$ for all $x \in \Lambda$, we can
now repeat the concavity argument above, using the unitary $(-1)^x$.
Indeed, $(-1)^x \widehat T  (-1)^x = - \widehat T , \ (-1)^x B
(-1)^x = B,\ (-1)^x UD(-1)^x = UD , \ (-1)^x A_*(-c,D)
(-1)^x = A_*(c,D)$, and thus
$$\ln \Det[A_*(c,D) - \{\widehat T, U\widehat D\}]= \ln \Det [A_*(-c,D) -
\{ \widehat T, U\widehat{D} \}] =
\ln \Det [A_*(c,D) + \{ \widehat T, U\widehat{D} \}], \eqno(3d.23)$$
Hence, concavity of $\ln \Det [\cdot ]$ again implies that
$$\ln \Det [c^2 + (T  - UD)^2] \leq \ln \Det [A_*(c,D)] -2 \ln \Det[B]
\eqno(3d.24)$$
with equality if and only if $\{ \widehat T, U\widehat{D} \}= 0$.
We now make the choice $B_x = {U_x}^{-1}n(x)^{-1}$. The condition
$\{\widehat T, U\widehat D\}=0$ becomes
$$\eqalignno{d(x) /n(x) = - d(y) / n(y)\phantom{\overline{t_{xy}}}
\qquad&\hbox{for} \ x \in A, y \in B, \qquad&(3d.25a) \cr
t_{xy} \delta (x) / n(x) = \overline{t_{xy}} \delta (y) / n(y)\phantom{-}
\qquad&\hbox{for} \ x \in A, y \in B. \qquad&(3d.25b)\cr}$$
But (3d.25) is equivalent to (3d.19) provided $t$ is real.  Conversely, in
case $t$ is not real, as we pointed out in the definition of the hopping
matrix, there exists at least one closed path $\{ x, x_1, \dots x_n, x \}$
such that $t_{xx_1} \dots t_{x_n x}$ is not real.  Thus, iterating (3d.25b)
along this closed path we obtain
$${t_{xx_1} \dots t_{x_n x_1} \over \overline t_{xx_1} \dots \overline
t_{x_nx_1}} {\delta (x) \over n(x)} = {\delta (x) \over n(x)} ,
\eqno(3d.26)$$
whose only solution is $\delta (x) = 0$, implying $\delta \equiv 0$ on the
entire lattice $\Lambda$ because $t$ is connected.
Finally, we remark that we may achieve
the bound in (3d.20) by choosing $\widetilde d (x) = (-1)^x n(x)$ and
$\widetilde \delta (x) = 0$. \lanbox

In order to use Lemma 3.6 to conclude that a potential $D$ minimizing
$\sr_{\beta, \mu}$ is phase aligned we must show that for such a potential
$\delta (x) \not= 0$ for all $x$ (or otherwise vanishes everywhere).
Likewise, in
order to use Lemma 3.7 we must show that a minimizing potential has $n(x)
\not= 0$ for all $x$.  We prove these two results
in parts (a) and (b) of the next lemma, again using the
method of choosing an appropriate normalizer, $B$.

{\bf 3.8 LEMMA (Nonvanishing of minimizers):}  {\it (a) If $t$ is real and
if $D = (d, \delta)$ minimizes $\sr_{\beta, \mu}$ then either $\delta (x)
\not= 0$ for all $x$ or $\delta (x) = 0$ for all $x$.

(b) If $t - \mu$ is
bipartite (but not necessarily real) and if $D = (d, \delta)$ minimizes
$\sr_{\beta, \mu}$ then either $n(x) \not= 0$ for all $x$ or $n(x) = 0$ for
all $x$.}

{\it Proof:}  (a)  Let $D = (d, \delta)$ be a potential such that both sets
$$\Lambda_0 =\{ x \in \Lambda \ \vert\ \delta (x) = 0\} \quad \hbox{and} \quad
\Lambda \setminus\Lambda_0 = \{ x \in \Lambda\  \vert\ \delta (x) \not= 0 \}$$
are non-empty.  We shall prove that $D$ cannot be minimizing
for $\sr_{\beta, \mu}$.  For $0 <\tau \leq 1$ we define a new potential
$D_\tau$ by
$$d_\tau (x) = d(x) \quad \hbox{and} \quad \delta_\tau (x) = \cases{\sqrt{1
- \tau^2} \vert \delta (x) \vert,& $x \not\in \Lambda_0$ \cr
\tau \delta_{\rm av},& $x \in \Lambda_0$ \cr}, \eqno(3d.27)$$
where $\delta_{\rm av}^2 =
\left( \sum\nolimits_{x \in \Lambda_0} U_x \right)^{-1}
\sum \limits_{x \not\in \Lambda_0} U_x \vert \delta (x) \vert^2 > 0$.  We
shall show that for $\tau$ small enough $\sr_{\beta, \mu} (D) >
\sr_{\beta, \mu} (D_\tau)$.

According to (3d.2--3) we can write
$$\sr_{\beta, \mu} (D) = - \int^\infty_0 \Bigl[\ln \Det [c^2 + (T_\mu - UD)^2]
-
4 \vert \Lambda \vert \ln c\Bigr] d \mu_\beta (c)
+\sum \nolimits_x U_x \Tr [D^2_x], \eqno(3d.28)$$
where the measure $d\mu_\beta $ for $\beta < \infty$ is a sum of Dirac
delta functions \hbox{$d \mu_\beta (c)= \sum \limits^\infty_{k = 0}
\beta^{-1}\delta[(2k+1)$} $\pi \beta^{-1} - c] dc$,
while for $\beta = \infty$ it is $d \mu_\infty
= (2 \pi)^{-1} dc$.

We chose $\delta_{\rm av}$ to make the last term
in $\sr_{\beta, \mu}$ independent of $\tau$, i.e., $\sum \nolimits_x U_x
\Tr [D^2_x] = \sum \nolimits_x U_x \Tr [D^2_{\tau, x}]$. To conclude
that $\sr_{\beta, \mu} (D) >
\sr_{\beta, \mu} (D_\tau)$ we must therefore show that for
$\tau$ small enough
$$\int \limits^\infty_0 \Bigl\{ \ln \Det [c^2 + (T_\mu - UD)^2] - \ln \Det
[c^2 + (T_\mu - UD_\tau)^2]\Bigr\} d \mu_\beta (c) <0\eqno(3d.29)$$
To prove (3d.29) we again appeal to (3d.12).  We choose $B_x = U^{-1}_x
[\delta_\tau (x) \vert^{-1}$ and obtain that
$$\ln \Det [c^2 + (T_\mu - UD)^2] \leq \ln \Det [A (c,D)] - 2 \ln \Det [B]
\eqno(3d.30)$$
and
$$\ln \Det [c^2 + (T_\mu - UD_\tau)^2] = \ln \Det [A (c, D_\tau)] - 2 \ln
\Det [B] . \eqno(3d.31)$$
Since $\{\widehat T_{\mu}, \widehat G_{\tau}\}=0$ (here $G_{\tau}$ is
defined as in (3d.8) but with $\delta$ replaced by $\delta_\tau$)
we see from (3d.10) that
$$A(c, D_\tau) = (icB + \widehat T_\mu - U \widehat D_\tau) (-i c +
\widehat T_\mu - U \widehat D_\tau)$$
and $A(c, D) = A (c, D_\tau) - \1 + \widehat G^2$.  Therefore (3d.30) and
(3d.31) imply
$$\eqalign{\ln \Det &[c^2 + (T_\mu - UD)^2] - \ln \Det [c^2 + (T_\mu -
UD_\tau)^2] \cr
&\leq \ln \Det [\1 - (icB + \widehat T_\mu - U\widehat D_\tau)^{-1} (\1 -
\widehat G^2) (-icB + \widehat T_\mu - U\widehat D_\tau)^{-1}] \cr
&= \Tr \ln [\1 - (icB + \widehat T_\mu - U\widehat D_\tau)^{-1} (\1 -
\widehat G^2) (-ic B + \widehat T_\mu - U \widehat D_\tau)^{-1} ].\cr}
\eqno(3d.32)$$
Using the inequality $\ln (\1 - A) \leq - A$ we obtain
$$\eqalignno{
&\ln \det [c^2 + (T_\mu - UD)^2] - \ln \Det [c^2 + (T_\mu - UD_\tau)^2]\cr
&\leq - \Tr [(icB + \widehat T_\mu - U\widehat D_\tau)^{-1} (\1 - \widehat
G^2) (-icB + \widehat T_\mu - U \widehat D_\tau)^{-1}] \cr&
\leq - \Tr [K_\tau (c)],&(3d.33)}$$
where we have denoted
$$\eqalignno{K_\tau (c) :=& (icB + \widehat T_\mu
- U \widehat D_\tau)^{-1} (\1 -
\widehat G^2) (-icB + \widehat T_\mu - U \widehat D_\tau)^{-1}\cr
 =& B^{-1/2} (ic + T_\mu - U D_\tau)^{-1} (B^{-1} - BG^2)
(-ic + T_\mu - UD_\tau)^{-1} B^{-1/2}.&(3d.34)}$$
The estimate (3d.29) follows if we show that for $\tau$ small enough
$$\int \Tr [K_\tau (c)] d\mu_\beta (c) > 0.\eqno(3d.35)$$

We denote by $P_0$ the projection onto the sites in
$\Lambda_0$, i.e., $P_0 =
\sum \limits_{x \in \Lambda_0} \1_x$ and $\widetilde P_0 = \1 - P_0 = \sum
\limits_{x \not\in \Lambda_0} \1_x$.  If we assume $\tau \leq \mfr1/2$ we
then have
$$B^{-1} - BG^2 = \sum \limits_{x \in \Lambda_0} \tau U_x \delta_{\rm av}
\1_x
- \sum \limits_{x \not\in \Lambda_0} {\tau^2 \over \sqrt{1 - \tau^2}} U_x
\vert \delta (x) \vert \1_x \geq a \tau P_0 - b \tau^2 \widetilde P_0,
\eqno(3d.36)$$
where $a$ and $b$ are strictly positive constants depending on the values
of $\vert \delta (x) \vert$ and $U_x$ for all $x$.

We therefore find
$$\eqalign{\Tr [K^{(c)}_\tau] \geq&
\tau a \Tr [B^{-1/2} (ic + T_\mu - U D_\tau)^{-1}
P_0 (-i c + T_\mu - UD_\tau)^{-1} B^{-1/2}]\cr
& - \tau^2 b \Tr [B^{-1/2} (ic +
T_\mu - UD_\tau)^{-1} \widetilde P_0 (-ic + T_\mu - UD_\tau)^{-1}
B^{-1/2}].}$$
Since $a^\prime \widetilde P_0 \leq B^{-1} \leq b^\prime \1$ for constants
$a^\prime$ and $b^\prime$ (again depending only on $\vert \delta (x) \vert$
and $U_x$ for all $x$) we have the estimate, with $|A|^2=AA^{\dagger}$,
$$\Tr [K^{(c)}_\tau] \geq \tau aa^\prime \Tr [ \vert \widetilde P_0 (ic +
T_\mu - UD_\tau)^{-1} P_0 \vert^2] - \tau^2 bb^\prime \Tr [(c^2 + (T_\mu -
UD_\tau)^2)^{-1} ]. \eqno(3d.37)$$

Since the problem is finite dimensional it is clear that the eigenvalues of
$T_\mu - UD_\tau$ converge to the eigenvalues of $T_\mu - UD$.  Hence,
the gap estimate (3c.6) implies that for $\tau$ small (depending on
$U_x$, $D_x$ and $t_{xy}$)
there is a constant $g_\beta$, satisfying $g_\beta>0$ for $\beta$ large
enough (in particular for $\beta=\infty$)
such that $(T_\mu - UD_\tau)^2\geq g_\beta^2$.
Therefore
$$[c^2 + (T_\mu - UD_\tau)^2]^{-1} \leq (c^2 + g_\beta^2)^{-1} \1.
\eqno(3d.38)$$

Recall now that the measure $d\mu_\beta$ appearing in the Pick
representation (11.2) is supported away from zero when $\beta < \infty$.
It therefore follows from (3d.38) that for small \hbox{enough $\tau$}
$$\int \Tr \Bigl[\bigl(c^2 + (T_\mu + UD_\tau)^2\bigr)^{-1}\Bigr]
d \mu_\beta (c)$$
is bounded independently of $\tau$ for all $\beta$.

In order to conclude (3d.35) (and hence the lemma) from (3d.37)
it only remains to prove that
$$\liminf \limits_{\tau \rightarrow 0} \int \Tr [\vert \widetilde P_0 (ic +
T_\mu - UD_\tau)^{-1} P_0 \vert^2 ] d \mu_\beta(c) \not= 0.\eqno(3d.39)$$
Since $t$ is connected we know that
$$\widetilde P_0 (ic + T_\mu - UD) P_0 = \widetilde P_0 T_0 P_0 \not=
0 \quad \hbox{and} \quad P_0 (ic + T_\mu - UD) \widetilde P_0 = P_0
T_0 \widetilde P_0 \not= 0.$$
Therefore we must have $\widetilde P_0 (ic + T_\mu - UD)^{-1} P_0
\not= 0$
(because if we have a matrix $\pmatrix{a&b\cr c&d}$ with operator valued
entries and $c\ne0$ and $b\ne0$ then its inverse $\pmatrix{\alpha&\beta\cr
\gamma&\delta}$ must have $\beta\ne0$ and $\gamma\ne0$).
Since $\lim_{\tau\to0}\Tr [\vert \widetilde P_0 (ic +
        T_\mu - UD_\tau)^{-1} P_0 \vert^2 ]= \Tr [\vert \widetilde P_0 (ic +
        T_\mu - UD)^{-1} P_0 \vert^2 ]$ we obtain by Fatou's Lemma that
$$
	\liminf \limits_{\tau \rightarrow 0} \int
	\Tr [\vert \widetilde P_0 (ic +
	T_\mu - UD_\tau)^{-1} P_0 \vert^2 ] d \mu_\beta(c)
	\geq\int \Tr [\vert \widetilde P_0 (ic +
        T_\mu - UD)^{-1} P_0 \vert^2 ] d \mu_\beta(c)>0
$$
and part (a) follows.

(b) Part (b) is proved in the same way as part (a). We this time
choose $D_\tau$  with $\delta_\tau(x)=0$ and $d_\tau(x)=(-1)^xn_\tau(x)$,
where
$$
	n_\tau(x)=\cases {\sqrt{1-\tau^2}n(x),&$x\not\in\Lambda_0$\cr
	\tau n_{\rm av},&$x\in\Lambda_0$}.
$$
Here $n_{\rm av}$ is again defined such that $\sum_xU_xn(x)^2=
\sum_xU_xn_{\tau}(x)^2$.
The rest of the proof is identical to the proof of part (a)
except that we use (3d.24) instead of (3d.12) and we choose
$B_x=U_xn_\tau(x)^{-1}$.\lanbox

We pause to discuss the symmetry aspects of the Hubbard model and how they
are reflected by Lemmas~3.6 and 3.7.  Independent of the hopping matrix $t$
and the chemical potential $\mu$ the Hamiltonian $H$ will always be
invariant under a global (i.e., all spins are equally transformed)
{\bf spin transformation} $\W_\s=\W_\s(w)$
$$\W_\s c^\dagger_{x,\sigma} \W^\dagger_\s := \sum \limits_{\sigma^\prime}
w_{\sigma', \sigma} c^\dagger_{x, \sigma^\prime} \eqno(3d.40)$$
for all $x \in \Lambda$, where $w \in U(2)$ is a unitary $2 \times
2$-matrix.  Note that $\W_\s$ is a Bogoliubov transformation.

As always we shall be particularly interested in the spin {\it rotations}, i.e.
transformations $\W_\s(w)$ corresponding to $w$ in $SU(2)$. The full
$U(2)$ group is generated by the $SU(2)$ subgroup together with
the subgroup (isomorphic to $U(1)$) consisting of all
$w_\theta=\pmatrix{e^{i\theta}&\cr &e^{i\theta}}$, $0\leq\theta<2\pi$.
The Bogoliubov transformation $\W_{\theta}:=\W_\s(w_\theta)$ is
a global {\bf phase change}, i.e.,
$$\W^\phagger_\theta c^\dagger_{x, \sigma} \W^\dagger_\theta :=
e^{i \theta} c^\dagger_{x, \sigma} \eqno(3d.41)$$
for all $x \in \Lambda$.
Spin rotations and phase changes (i.e., the full $U(2)$ group) exhaust the
symmetries of the Hamiltonian $H$ unless we assume more about the hopping
matrix $t$.

The spin rotations are generated by the (quadratic) spin operators
$$
        \eqalign{\S_1=&\mfr1/2\sum_{x\in\Lambda}\left(c_{x,\uparrow}^\dagger
        c_{x,\downarrow}^{\phantom{\dagger}}+c_{x,\downarrow}^\dagger
        c_{x,\uparrow}^
        {\phantom{\dagger}}\right),
\quad
        \S_2=\mfr1/{2i}\sum_{x\in\Lambda}\left(c_{x,\uparrow}^\dagger
        c_{x,\downarrow}^{\phantom{\dagger}}-c_{x,\downarrow}^\dagger
        c_{x,\uparrow}^{\phantom{\dagger}}\right),
\quad\cr
        \S_3=&\mfr1/2\sum_{x\in\Lambda}\left(c_{x,\uparrow}^\dagger
        c_{x,\uparrow}^{\phantom{\dagger}}-c_{x,\downarrow}^\dagger
        c_{x,\downarrow}^
        {\phantom{\dagger}}\right).}
$$
The $U(1)$ phase change is generated by the number operator
$\N=\sum_{x,\sigma}c_{x,\sigma}^\dagger c_{x,\sigma}^
        {\phantom{\dagger}}$.

According to (2.a9) the Bogoliubov transformation $\W_\s(w)$ has its
counterpart acting on $\H \oplus \H$ which we will denote by
$W_\s(w)$ and in the special case of a phase change by $W_\theta$. Indeed,
$$W_\s(w) = \pmatrix{w &\cr &\overline w \cr} \quad \hbox{and} \quad W_\theta =
\pmatrix{e^{i \theta} &&&\cr &e^{i \theta} &&\cr &&e^{-i \theta} &\cr
&&&e^{-i\theta} \cr} .\eqno(3d.42)$$

{F}rom Lemma~3.1 it is easy to conclude that any $\Gamma$ minimizing
the energy functional (3.37b) corresponds to a HF ground state with total
spin zero, i.e., a state with $\uprho(\S_1^2+\S_2^2+\S_3^2)=0$.

{\bf 3.9 THEOREM (Zero total spin):} {\it  A 1-$pdm$ $\Gamma$ corresponds to an
$SU(2)$ invariant HF state if and only if it has the form (3b.2).
In particular, the HF ground states have total spin zero.
}

{\it Proof:} The condition that a state be $SU(2)$ invariant is
that $\Gamma$ commutes with $W_s(w)$ for all $w\in SU(2)$.
Recalling that the only $2\times2$ matrices commuting with all
elements in $SU(2)$ are multiples of the identity matrix it
follows easily that exactly the matrices of the form (3b.2) commute with all
$W_s(w)$. It then follows that the HF ground states have total
spin zero since they are pure states.\lanbox

The HF states are,  however, not necessarily invariant under the $U(1)$
phase symmetries. If we use the unitary operator $Y$ defined in (3b.11)
we can write
$$
        YW_\theta Y^\dagger=\pmatrix{W_\theta'&0\cr0&W_\theta'},
$$
where
$$
        W_\theta'=\pmatrix{e^{i\theta}&0\cr0&e^{-i\theta}}.\eqno(3d.43)
$$
A 1-$pdm$ $\Gamma$ will therefore commute with $W_\theta$ if and only
if the corresponding $\Gamma'$ commutes with $W_\theta'$.
In case of the minimizing $\Gamma_0'$ it follows from (3b.31) (for positive
temperature) or (3c.4) (for the zero temperature ground states)
that $\Gamma_0'$ commutes with $W_\theta'$ if and only if the corresponding
matrix-valued potential $D_0$ commutes with $W_\theta'$. From (3b.24) and
(3d.43) we find
$$
        W_\theta' D_xW_\theta'^\dagger=
        W_\theta' \pmatrix{d(x)&\delta(x)\cr\vrule height15ptwidth0pt
	\overline{\delta(x)}&-d(x)}
        W_\theta'^\dagger=\pmatrix{d(x)&e^{2i\theta}\delta(x)\cr\vrule
height15ptwidth0pt
        e^{-2i\theta}\overline{\delta(x)}&-d(x)}.\eqno(3d.44)
$$
Thus $D_0$ commutes with $W_\theta'$ if and only if $\delta(x)=0$, which not
surprisingly is exactly the condition that the state is normal.
Notice that (3d.44) agrees precisely with the characterization
of the possible minimizers given in Lemma~3.6.

Note also that (3d.44) yields the same 1-$pdm$ for $\theta + \pi (\mod \ 2
\pi)$
as for $\theta$, meaning that the representation of the group $U(1)$ by the
minimizing states is not faithful.  This reflects the fact, that
generalized HF-states $\uprho$ obey the particular restriction $\uprho
(e_1 \dots e_{2k+1}) = 0$, where $e_i$  is either a $c$ or a
$c^\dagger$.

Now, let us consider the case in which $t$ is bipartite and $\mu = 0$.
Whether $t$ is real or not the Hamiltonian $H$ is
invariant under the {\bf particle-hole transformation}
$$\eqalign{\W_\ph \alpha c^\dagger_{x, \sigma} \W_\ph &:= (-1)^x
\overline \alpha c_{x, \sigma}, \cr
\W_\ph \alpha c_{x, \sigma} \W_\ph &:= (-1)^x \overline \alpha
c^\dagger_{x, \sigma}, \cr} \eqno(3d.45)$$
where $\alpha$ is any complex number which
we insert to indicate that $\W_\ph$ is antiunitary.
In fact, for $t$ non-real the particle-hole symmetry of the Hamiltonian
cannot be
unitarily realized.  On $\H \oplus \H$, with the fixed basis $\{ (x,
\sigma) \vert x \in \Lambda, \sigma = \uparrow, \downarrow \}$ in $\H$,
$\W_\ph$ corresponds to (the $4\times4$ operator)
$W_\ph$, acting on operators as ($A$, $B$, $C$ and $D$ are here
$2\times2$ operators)
$$W_\ph \pmatrix{A&B\cr\vrule height15ptwidth0pt C&D \cr} W^{-1}_\ph
= (-1)^x \pmatrix{\overline D &  \overline C \cr \vrule height15ptwidth0pt
\overline B & \overline A \cr} (-1)^x. \eqno(3d.46)$$

As for the $U(1)$ symmetry we would like to write $W_\ph$ in terms of a
symmetry on $\H_\Lambda\oplus\H_\Lambda$. This is only possible if we
first compose it with a spin rotation which after all will leave our states
invariant by Theorem~3.9. In fact, if $w=\pmatrix{0&1\cr-1&0}$ we have
$$
        YW_s(w)W_\ph Y^\dagger
        =(-1)^x\pmatrix{W_\ph'&\cr&W_\ph'},
$$
where $W_\ph'$ is the antiunitary map on $\H_\Lambda\oplus\H_\Lambda$ with
matrix (in the standard basis) given by
$$
        W_\ph'=\pmatrix{0&1\cr-1&0}.
$$
In order to understand the transformation property of a minimizing
$\Gamma_0'$ it is again enough to consider the potential.
$$
        W_\ph'\pmatrix{d&\delta\cr\vrule height15ptwidth0pt
	\overline{\delta}&-d}W_\ph'^{-1}=
        \pmatrix{-d&-\delta\cr\vrule height15ptwidth0pt-\overline{\delta}&d}
	=-\pmatrix{d&\delta\cr\vrule height15ptwidth0pt
        \overline{\delta}&-d}\eqno(3d.47)
$$
since $(-1)^x D(-1)^x = D$.

If $t$ is not real
we compare this with our result from Lemma 3.7.
According to the condition on the minimizing $D$ therein, we must
have $\delta (x) = 0$ for all $x \in \Lambda$.
The minimizing $D$ is, hence, of the form $D = \sum \limits_x D_x\1_x$ or
$D = \sum \limits_x - D_x \1_x$ where
$$D_x = \pmatrix{d(x) &\cr &-d(x)}.
\eqno(3d.48)$$
Thus, assuming that $t$ is bipartite and non-real and $\mu = 0$, the
minimizing 1-$pdm \ \Gamma$ has no pairs and corresponds to a normal state.
It is unaffected by spin rotations and phase changes which
means it has spin zero and fixed particle number.  Only the particle-hole
symmetry is broken due to the global change $d(x) \rightarrow -d (x)$ for
all $x \in \Lambda$.

In case of a real, bipartite $t$ at $\mu = 0$ the symmetry group of the
Hamiltonian  is even larger.  Let us introduce the Bogoliubov
transformation $\W_{\bp}$ by
$$\eqalign{\W_{\bp} c^\dagger_{x \uparrow} \W^\dagger_{\bp} &= c^\dagger_{x
\uparrow}, \cr
\W_{\bp} c^\dagger_{x \downarrow} \W^\dagger_{\bp} &= (-1)^x
c^{\phantom{\dagger}}_{x
\downarrow}. \cr}\eqno(3d.49)$$
The global {\bf pseudo-spin rotation} $\W_\ps$ is given by
$$\W^{\phantom{\dagger}}_\ps(w) = \W^\dagger_{\bp} \W^{\phantom{\dagger}}_\s(w)
\W^{\phantom{\dagger}}_{\bp}, \eqno(3d.50)$$
for any spin rotation $\W_\s(w)$, with $w \in SU(2)$.  It
leaves the Hamiltonian invariant in the real bipartite case at $\mu=0$.

As in the spin case we could of course have considered the full group
of $U(2)$ pseudo-spin transformations. The transformation $\W_\ps(w_\theta)$
corresponding to $w_\theta=\pmatrix{e^{i\theta}&\cr&e^{i\theta}}$ is,
however, equal to the spin rotation corresponding
to $\widetilde{w}_\theta:=\pmatrix{e^{i\theta}&\cr&e^{-i\theta}}$, i.e.
$\W_\ps(w_\theta)=\W_\s(\widetilde{w}_\theta)$.
Conversely, the $U(1)$ group of phase change symmetries is really a subgroup
of the $SU(2)$ pseudo-spin rotations. In fact, the pseudo-spin rotations are
generated by the pseudo-spin operators
$$
\eqalign{\OS_1=&\mfr1/2\sum_{x\in\Lambda}(-1)^x\left(c_{x,\uparrow}^\dagger
        c_{x,\downarrow}^{\dagger}+c_{x,\downarrow}^{\phantom{\dagger}}
        c_{x,\uparrow}^{\phantom{\dagger}}\right),
\quad
        \OS_2=\mfr1/{2i}\sum_{x\in\Lambda}(-1)^x\left(c_{x,\uparrow}^\dagger
        c_{x,\downarrow}^{\dagger}-c_{x,\downarrow}^{\phantom{\dagger}}
        c_{x,\uparrow}^{\phantom{\dagger}}\right),
\quad\cr
        \OS_3=&\mfr1/2\sum_{x\in\Lambda}\left(c_{x,\uparrow}^\dagger
        c_{x,\uparrow}^{\phantom{\dagger}}-c_{x,\downarrow}^{\phantom{\dagger}}
        c_{x,\downarrow}^{\dagger}\right)=\mfr1/2\N-\mfr1/2|\Lambda| \1.}
$$
Notice that in terms of how they transform operators, i.e.,
$A\mapsto\W A\W^{\dagger}$ we cannot distinguish the unitaries
$$
        \exp(\mfr i/2\N\theta)\quad\hbox{and}\quad\exp(i\OS_3\theta).
$$

We also point out that for real $t$ the particle-hole
symmetry may be unitarily realized as $\widetilde{\W}^{\phantom{\dagger}}_\ph =
\W^{\phantom{\dagger}}_\s(w)\W_\bp^\dagger \W^{\phantom{\dagger}}_\s(w^\dagger)
\W_\bp^{\phantom{\dagger}}$
with $w = \pmatrix{0&1\cr-1&0\cr}$ and is, therefore,
contained in the group of spin- and pseudo-spin rotations.

The unitary operator $W_\ps(w)$ on $\H \oplus \H$ corresponding to
a $w\in SU(2)$ which we write as
$w=\pmatrix{w_{\uparrow \uparrow}&w_{\uparrow \downarrow}\cr\vrule
height12ptwidth0pt
-\overline{w}_{\uparrow \downarrow}&\overline{w}_{\uparrow \uparrow}}$ is
given by
$$W_\ps(w)= \pmatrix{w_{\uparrow \uparrow}
&&&(-1)^x w_{\uparrow\downarrow} \cr & w_{\uparrow \uparrow}
&-(-1)^x  w_{\uparrow \downarrow} &\cr &(-1)^x \overline w_{\uparrow
\downarrow} &\overline{w}_{\uparrow \uparrow} &\cr -(-1)^x
 \overline{w}_{\uparrow \downarrow} &&&\overline{w}_{\uparrow \uparrow} \cr} .
\eqno(3d.51)$$
Thus, $ Y W_\ps(w) Y^\dagger=\pmatrix{W_\ps'(w)&\cr&W_\ps'(w)}$, where
$$
W_\ps'(w)=\pmatrix{w_{\uparrow \uparrow}&(-1)^xw_{\uparrow \downarrow}\cr
\vrule height15ptwidth0pt
-(-1)^x\overline{w}_{\uparrow \downarrow}&\overline{w}_{\uparrow \uparrow}}.
\eqno(3d.52)
$$
In other words, $W_\ps'(w) D_x W_\ps'(w)^\dagger$ is the transformation
of the potential $D_x$.
Now observe that if $D_x$ fulfills (3d.19) then so does $W_\ps'(w) D_x
W_\ps'(w)^\dagger$,
indeed, starting from $D $ with $\delta = 0$, (3d.19)
simply states that all minimizers are of the form
$$W_\ps'(w) D  W_\ps'(w)^\dagger\eqno(3d.53)$$
with an arbitrary $SU(2)$-matrix $w$.  Hence, for real,
bipartite $t$ at $\mu = 0$ the situation is as follows.
Whenever $\delta$ is different from $0$,
the minimizing $\Gamma$ does not correspond to a normal state and the
particle number is broken.  Likewise, the pseudo-spin rotation is a
broken symmetry, as can be seen from (3d.53).

The complete tables to illustrate the broken symmetries can be found in
Chapter V.
\bigskip\noindent
{\bf III.e SPATIAL UNIQUENESS OF MINIMIZERS}

In the previous section we saw that if a minimizing potential
$D$ is nonvanishing there is a whole family of minimizers
related by global gauge transformations of $D$.
In fact, the phase of $\delta$ or the pseudo-spin must
be aligned over the lattice and we
can only allow global gauge transformations.
The aim in this section is to prove that all degeneracies of the minimizing
potentials $D$ are caused by these gauge transformations.
Since the HF ground and Gibbs states are uniquely determined by $D$ it shows
that the only degeneracies of these HF states are due to the symmetries
discussed in the previous section.

More precisely, we shall
show that if $D_1$ and $D_2$ are two minimizers for $\sr_{\beta,\mu}$ then
$D_1^2=D_2^2$, i.e., $n_1=n_2$, where as before $n_{1,2}=\sqrt{d_{1,2}^2
+|\delta_{1,2}|^2}$.  We show this in the case where $t$ is bipartite and
$\mu=0$.
If $t$ is not bipartite or $\mu\ne0$ we also get $d_1=d_2$ (see
Lemma~3.11).
The uniqueness statement in the bipartite case with $\mu=0$
is an immediate consequence of the strict convexity we now prove.

{\bf 3.10 LEMMA (Convexity of $\sr_{\beta,0}$):}
{\it If $t$ is bipartite
(not necessarily real) and $\mu=0$ then $\sr_{\beta,0}\Bigl((-1)^xn,0\Bigr)$,
regarded as a functional of the function $\eta=n^2$ is convex.  It is
strictly convex at $\eta$ if $\eta (x) \not= 0$ for all
$x$\footnote{$^\dagger$}{\ninepoint The convexity is, in fact, strict for all
$\eta$, but this is more complicated to prove.}.}

{\it Proof:} We shall prove that if $n_0=\sqrt{\lambda n_1^2+
(1-\lambda)n_2^2}$ for some $0<\lambda<1$ then
$$\sr_{\beta,0}\Bigl((-1)^xn_0,0\Bigr)\leq
\lambda\sr_{\beta,0}\Bigl((-1)^xn_1,0\Bigr)+
(1-\lambda)\sr_{\beta,0}\Bigl((-1)^xn_2,0\Bigr)\eqno(3e.1)
$$
with equality if and only if $n_1=n_2=n_0$. Since the term $\sum_x\Tr[D_x^2]$
is
linear in $n^2$ we may ignore it here and we only need to consider the
first term in (3d.2--3). Let $D_j$,  $j=0,1,2$
denote the potentials corresponding to
$(d_j,\delta_j)=((-1)^xn_j,0)$,  for $j=0,1,2$.

To prove the strict convexity we assume $n_0(x)\ne0$ (i.e.,
either $n_1(x)\ne0$ or $n_2(x)\ne0$) for all $x\in\Lambda$.  If $n_0$
vanishes somewhere we still get convexity (but not strict) by a continuity
argument.  For both $D_1$ and $D_2$
we use (3d.24) with $B = U_x^{-1} n_0(x)^{-1}$ and we find
$$
\ln \Det\Bigl[c^2 + (T  - UD_j)^2\Bigr] \leq \ln \Det [A_* (c, D_j)]
-2\ln \Det[B].\eqno(3e.2)
$$
Since $D_j^2=n_j^2\1$ we get from the definition (3d.22) that $A_* (c, D_1)
+ A_* (c, D_2) = A_* (c, D_0)$.  By strict
concavity of $\ln \Det[\ \cdot\ ]$ we obtain
$$
	\eqalignno{\lambda\ln \Det\Bigl[c^2+(T  - UD_1)^2\Bigr] &+
		(1-\lambda)\ln \Det\Bigl[c^2+(T  - UD_2)^2\Bigr]\cr
	&\leq  \ln \Det \Bigl[A_* (c, D_0)
        \Bigr] -2\ln \Det[B]\cr
	&=\ln \Det\Bigl[c^2+(T  - UD_0)^2\Bigr],&(3e.3)}
$$
with equality if and only if $n_1=n_2=n_0$. The last equality in (3e.3)
follows from $\{ \widehat T, U\widehat{D}_0 \} = 0$.
Inserting (3e.3) into (3d.2-d3) gives (3e.1).\lanbox

If we do not assume bipartiteness and $\mu=0$ the situation is somewhat
more compicated. We have the following convexity result.
Note that assuming $t-\mu$ not bipartite includes bipartite $t$
with $\mu\ne0$.

{\bf 3.11 LEMMA (Convexity of $\sr_{\beta,\mu}$):}
{\it If $t$ is real $\sr_{\beta,\mu}(d,\sqrt{n^2-d^2})$, regarded as a
functional of $d$ and $\eta=n^2$ is convex, but not always strictly convex.
If $t-\mu$ is not bipartite the functional
$\sr_{\beta,\mu}(d,\sqrt{n^2-d^2})$ is minimized by unique
functions $d$ and $n$.}

{\it Proof:} We first prove the convexity.
Note that $\sr_{\beta,\mu}$ as a functional of the functions
$\eta=n^2$ and $d$ is defined on the convex set $\{(\eta,d)\ |\ |d|^2
\leq \eta\}$.
Given $d_j$ and $n_j$ for $j=1,2$ satisfying $|d_j|\leq n_j$. Define
$d_0=\lambda d_1+(1-\lambda)d_2$ and $n_0=\sqrt{\lambda n_1^2+
(1-\lambda)n_2^2}$. Our aim is to prove that
$$\sr_{\beta,\mu}\Bigl(d_0,\sqrt{n_0^2-d_0^2}\Bigr)\leq
\lambda\sr_{\beta,\mu}\Bigl(d_1,\sqrt{n_1^2-d_1^2}\Bigr)+
(1-\lambda)\sr_{\beta,\mu}\Bigl(d_2,\sqrt{n_2^2-d_2^2}\Bigr).\eqno(3e.4)
$$
Let $D_j$  for $j=0,1,2$ correspond to $(d_j,\delta_j)
=(d_j, \sqrt{n_j^2-d_j^2})$.
Since we can otherwise use a continuity
argument we may assume that $\delta_0(x)=\sqrt{n_0(x)^2-d_0(x)^2}\ne0$
for all $x\in\Lambda$.

For both $D_1$ and $D_2$ we use (3d.12) with
$B = U_x^{-1}|\delta_0(x)|^{-1} = U_x^{-1}
\Bigl(n_0(x)^2-d_0(x)^2\Bigr)^{-1/2}.$  We obtain
$$
	\ln \Det\Bigl[c^2 + (T  - UD_j)^2\Bigr] \leq
	\ln \Det\Bigl[A(c, D_j) \Bigr] -2\ln \Det[B],
	\eqno(3e.5)
$$
with equality if and only if
$\{ \widehat T_\mu ,\widehat G_j\}$ vanishes.
Using $A(c, D_1) + A(c, D_2) = A(c, D_0)$
$$
	 \eqalignno{\lambda\ln& \Det\Bigl[c^2 + (T  - UD_1)^2\Bigr]
	+(1-\lambda)\ln \Det\Bigl[c^2 + (T  - UD_2)^2\Bigr]
	\cr&\leq \ln \Det\Bigl[A(c, D_0)
        \Bigr] -2\ln \Det[B]
	=\ln \Det\Bigl[c^2 + (T  - UD_0)^2\Bigr].&(3e.6)}
$$
The last equality in (3e.6) holds because $\{ \widehat T_\mu ,
\widehat G_0\}$ vanishes.
The convexity in (3e.4) is an immediate consequence of (3e.6).

We have equality in (3e.6) if and only if $A(c, D_1) = A(c, D_2) = A(c,
D_0)$, i.e.,
$$
	-\{\widehat T_\mu,\widehat F_1\}+\widehat G_1^2+\widehat F_1^2=
	-\{\widehat T_\mu,\widehat F_2\}+\widehat G_2^2+\widehat F_2^2=
	-\{\widehat T_\mu,\widehat F_0\}+\widehat G_0^2+\widehat F_0^2.
	\eqno(3e.7)
$$

To show that $\sr_{\beta,\mu}$ need not be strictly convex consider
the case of bipartite $t$ with $\mu\ne0$. Let $d_1(x)=-U_x^{-1}\mu+(-1)^xc$
and $d_2(x)=-U_x^{-1}\mu-(-1)^xc$ for some $c>0$
and let $\delta_1=\delta_2=0$.
If we take $\lambda=1/2$ we find
$d_0=\mfr1/2d_1+\mfr1/2d_2=-U_x^{-1}\mu$ and $n_0^2-d_0^2=
\mfr1/2 d_1^2+\mfr1/2 d_2^2-d_0^2=c^2$. On the other hand by pseudo-spin
invariance we see that strict convexity is violated:
$$
	\eqalign{\sr_{\beta,\mu}(d_1,0)=\sr_{\beta,\mu}(d_2,0)=&
	\sr_{\beta,0}((-1)^xc,0)+\sum_x U_x^{-1}\mu^2\cr=&\sr_{\beta,0}(0,c)
	+\sum_x U_x^{-1}\mu^2
	=\sr_{\beta,\mu}\left(d_0,\sqrt{n_0^2-d_0^2}\right).}
$$

Assume now that $D_1$ and $D_2$ are two minimizers for $\sr_{\beta,\mu}$
and define $D_0$ as above with $\lambda=1/2$. By convexity of
$\sr_{\beta,\mu}$ we conclude that $D_0$ is also a minimizer.
Moreover, we know from Lemma~3.5ab that either $\delta_0 (x) = 0$ for all $x$
or $\delta_0 (x) \not= 0$ for all $x$.  If $\delta_0 (x) = 0$ for all $x$
we have
$$0 = n^2_0 - d^2_0 = \mfr1/4 (d_1 - d_2)^2 + \mfr1/2 \delta^2_1 + \mfr1/2
\delta^2_2.$$
Hence $d_1 = d_2$ and $\delta_1 = \delta_2 = 0$ and thus $n_1 = n_2$.

If $\delta_0 (x) \not= 0$ for all $x$,
we know that (3e.7) is satisfied and that
$$
	\{ \widehat T_\mu , \widehat G_j\}=0\eqno(3e.8)
$$
for $j=0,1,2$.
{F}rom the off-diagonal part of (3e.7) we conclude that for all $x\ne y$
in $\Lambda$ with $t_{xy}\ne0$ we have
$$
	\widehat F_{1x}+\widehat F_{1y}=
	\widehat F_{0x}+\widehat F_{0y}.\eqno(3e.9)
$$
Hence, $ \widehat F_{1x}-\widehat F_{0x}=-(\widehat F_{1y}-\widehat F_{0y})$.
If the off-diagonal part $t'$   of $t$ (i.e.,
$t'_{xy}=t_{xy}-t_{xx}\delta_{xy}$) is not bipartite we
can find
a path $z_1, z_2, \ldots, z_{2k}=z_1$ with an odd number of points
such that $t'_{z_iz_{i+1}}\ne0$. Therefore $F_{1z_1}=F_{0z_1}$
and by connectedness of $t$ we then get $F_{1x}=F_{0x}$ for all $x$ in
$\Lambda$, i.e., $d_1=d_0=d_2$.
It then easily follows from (3e.7) that also $n_1=n_0=n_2$.

We are left with the case where the off-diagonal part $t'$
is bipartite. In this case we get from (3e.9):
$$
	|\delta_0(x)|^{-1}(d_1(x)-d_0(x))=c(-1)^x \eqno(3e.10)
$$
for some constant $c$. We shall show now that if $c\ne0$
then $D_0$ cannot be minimizing.
By connectedness of $t$ we see that (3e.8) implies that
$$
	\delta_1(x)=a_1\delta_0(x), \eqno(3e.11)
$$
where $a_1$ is a constant (recall that here
$\delta_0(x)$ is positive) .

We turn to the diagonal part of (3e.7):
$$
	(t_{xx}-\mu-U_xd_1(x))^2+U_x^2|\delta_1(x)|^2
	=(t_{xx}-\mu-U_xd_0)^2+U_x^2\delta_0(x)^2 .\eqno(3e.12)
$$
If we insert (3e.10) and (3e.11) into (3e.12)
we get
$$
	2c(-1)^xU_x^{-1}\delta_0(x)^{-1}(t_{xx}-\mu-U_xd_0(x))
	+c^2+|a_1|^2 =1.
$$
We therefore conclude that if $c\ne0$ then
$$
	t_{xx}-\mu-U_xd_0(x)=(-1)^xU_x\delta_0(x)C,\eqno(3e.13)
$$
where $C$ is a constant.
We can restate this as
$$
	T_\mu-UD_0=\pmatrix{t'&\cr\vrule height15ptwidth0pt
	&-t'}-U_x\delta_0(x)
	\pmatrix{(-1)^xC&1\cr\vrule height15ptwidth0pt
	1&-(-1)^xC}.\eqno(3e.14)
$$

We can choose a pseudo-spin rotation $W$ as in (3d.52) such that
$$
	W\pmatrix{(-1)^xC&1\cr\vrule height15ptwidth0pt
	1&-(-1)^xC}W^{\dagger}
	=\pmatrix{&1+C^2\cr\vrule height15ptwidth0pt
	1+C^2&}.\eqno(3e.15)
$$
Since $t'$ is assumed to be bipartite we see that the unitary
operator
$$
        V=\sum_x (-1)^x\pmatrix{&1\cr1&}\1_x
$$
has the effect
$VW(T_\mu-UD)W^\dagger V^{\dagger}=W(T_\mu-UD)W^\dagger$.

{F}rom (3b.31) and (3c.4)
we see that if $\Gamma'$ is the minimizer corresponding to $D_0$ then
$\Gamma'=F_\beta((T_\mu-UD_0))$  where $F_\beta$ is the function
$F_\beta(\lambda)=[1+\exp(\beta\lambda/2)]^{-1}$ (for $\beta=\infty$ we have
$F_\infty(\lambda)=\lim_{\beta\to\infty}F_\beta(\lambda)=\chi(\lambda)$).
Hence $VW\Gamma'W^\dagger V^{\dagger}=W{\Gamma'}W^{\dagger}$.
For the potential $D_0$ this implies that
$$
	\pmatrix{&1\cr1&}WD_0W^\dagger \pmatrix{&1\cr1&}=WD_0W^\dagger.
$$
Since $WD_0W^\dagger$ is traceless we see in particular
that the diagonal entries of $WD_0W^\dagger$ must be zero.
If we write
$$
	D_{0x}= \delta_0(x)\pmatrix{d_0(x)\delta_0(x)^{-1}&1\cr
	\vrule height15ptwidth0pt
        1&-d_0(x)\delta_0(x)^{-1}}
$$
we see by comparison with (3e.15) that we must have $d_0(x)\delta_0(x)^{-1}
=(-1)^xC$. It then follows from  (3e.13) that  $t_{xx}-\mu=0$.
Since we are in the case where $t'$ is bipartite this implies that
$t-\mu$ is bipartite contrary to our assumption in the lemma.
\lanbox

We can now state the main result of Chapter 3.

{\bf 3.12 MAIN THEOREM (Characterization of HF states):}  {\it The HF Gibbs
states or ground states are unique modulo global gauge transformations.
More precisely, all HF Gibbs states at the same inverse temperature
$\beta$ or all HF ground states (i.e., the minimizers of the functionals
(3b.23) and (3c.2) respectively) are related by global gauge
transformations.}

{\it Proof:}  By Theorems~3.3 and 3.4 there is a one to one correspondence
between the HF states described by $\Gamma^\prime$ minimizing $-\P_{\beta,
\mu}$ and the potentials $D$ minimizing $\sr_{\beta, \mu}$.  The uniqueness
of $D$ modulo gauge transformations follows immediately from the results
about $\sr_{\beta, \mu}$; Lemmas~3.6--3.8 and Lemmas~3.10 and 3.11. \lanbox

We remark that Theorem 3.12 allows us to characterize the HF Gibbs
states by the global gauge and the inverse temperature. Indeed, the
correspondence
between $\beta$ and the HF Gibbs state $\Gamma^\prime( \beta, w)$
is continuous in $\beta$ for every fixed global gauge $w$ ($w$
being a pseudo-spin rotation, or a phase change).
In particular, this mapping is continuous at $\beta = \infty$ and,
a posteriori, we could have spared the entire discussion
of the zero temperature state by simply appealing to this continuity.

We shall see in Sect.~III.g that if the temperature is high enough their
maybe no symmetry breaking and hence the Gibbs state is unique.  There is
no degeneracy caused by gauge transformations.  If however, the temperature
is small it follows from Lemma~3.5 that there is symmetry breaking for any
finite system.  In the translation invariant case studied in Sect.~III.g,
the symmetry breaking phase transition persists in the thermodynamic limit.
\bigskip
\bigskip\noindent
{\bf III.f  SPATIAL SYMMETRIES}

In Sect.~III.d we studied global gauge symmetries of the
Hamiltonian. In the present section we shall address the question of spatial
symmetries, i.e., symmetries of the lattice or, more precisely, of the $t$
matrix
and the coupling constants $U_x$.

As a special example, in the next section we shall
explicitly determine the minimizing matrix-valued
potential $D$ when the lattice is translation invariant and the coupling
constants
$U_x$ are independent of $x$.

By a {\bf spatial symmetry} we understand an invertible transformation
$\tau:\Lambda\mapsto\Lambda$  of the lattice $\Lambda$.
We say that $t$ is {\bf magnetically invariant} under $\tau$ if
there exists a map $\alpha_{\tau}:\Lambda\mapsto \R$ such that
$$t_{\tau(x)\tau(y)}=\exp[-i(\alpha_{\tau}(x)-\alpha_{\tau}(y))]t_{xy}.$$
Equivalently, if we realize $t$ as an operator on $\H$ magnetic invariance
means that
$t$ commutes with the unitary transformation $m(\tau)$ on $\H$ defined by
$$
        m(\tau)|x,\sigma\rangle=\exp(-i\alpha_\tau(x))|\tau(x),\sigma\rangle.
$$

We need the phase factor $\alpha$ in order to treat non-real $t$.
If $t$ is real we may of course choose $\alpha_{\tau}=0$.
As an example consider a two-dimensional torus, i.e., a
finite cube $\Lambda$ in $\Z^2$ with periodic boundary conditions.
The original (real) hopping matrix considered by Hubbard, Kanamori
and Gutzwiller is invariant
under pure translations ($m(\tau)$ with $\alpha_{\tau}=0$) on the torus.
Consider, however, the (complex)
hopping matrix $t$ which differs from the elements of the original matrix
by the multiplication of complex phases and which correspond to
having a fixed magnetic flux through each unit square. Then $t$ is not
invariant under pure translations  but rather under the {\it magnetic}
translations
$m(\tau)$ which are compositions of translations and gauge transformations
($\alpha_{\tau}\ne0$).

The family of all transformations $\tau$ for which $t$ is magnetically
invariant
and the coupling constants satisfy $U_{\tau(x)}=U_x$ for all $x$ in $\Lambda$
naturally forms a group $\G$ which we call the {\bf spatial symmetry group} (of
$t$ and $U$).
Notice that $\tau\mapsto m(\tau)$ need not be a unitary representation of
$\G$. In fact for the two-dimensional torus
the translations commute while the magnetic translations do not.
(If $\tau_1,\tau_2$ denote the translations of unit length along the
first and second
direction respectively we have $m(\tau_1)m(\tau_2)=\exp(i\phi)
m(\tau_2)m(\tau_1)$,
where $\phi$ is the flux through the unit squares.)
We emphasize that the group $\G$ itself need not be abelian, it could be
one of the crystallographic groups (e.g. if we study the Hubbard model
on the lattice formed by the carbon atoms in the
Buckminsterfullerene $C_{60}$ molecule).
We say that $t$ is {\bf translation invariant} if the spatial
symmetry group $\G$ acts transitively
on the lattice $\Lambda$, i.e., if for any two points $x,y\in\Lambda$ there is
a $\tau$ in $\G$ such that $\tau(x)=y$.
If $t$ is bipartite (and connected) it is easy to see that each element
$\tau$ of $\G$ must either map the $A$ and $B$ sublattices into themselves
($\tau(A)=A$ and $\tau(B)=B$) or map the $A$ sublattice to the $B$ sublattice
($\tau(A)=B$). It is clear that the latter type of transformations
exist only if $|A|=|B|$.

Corresponding to $ \tau\in \G$ we define a Bogoliubov transformation
$\W_{\tau}$
on the Fock space $\F$ by
$$
        \W_{\tau}c^{\dagger}_{x,\sigma}\W_{\tau}^{\dagger}
        =\exp(-i\alpha_{\tau}(x))c^{\dagger}_{\tau(x),\sigma}.
$$
The unitary matrix $W_{\tau}$ corresponding to $\W_{\tau}$ is
$$
        W_{\tau}=\pmatrix{m(\tau)&\cr&\overline{m(\tau)}}.
$$

If $\tau$ belongs to the spatial symmetry group then the Hubbard Hamiltonian
is invariant under $\W_{\tau}$.
We shall prove in the next lemma that if $t$ is also non-bipartite then the HF
Gibbs states are also invariant under $\W_{\tau}$. In the bipartite case
the states are also invariant under transformations such that $\tau(A)=A$ and
hence $\tau(B)=B$, they need, however, not be invariant under transformations
such that $\tau(A)=B$.

{\bf 3.13  THEOREM (Spatial invariance):}  {\it  Let $\G$ be the group of
spatial symmetries of $t$ and $U$.

(a) If $t$ is bipartite (but not necessarily real) and $\mu=0$
all minimizing potentials $D$ for $\sr_{\beta,0}$ satisfy
that $n(x)=\sqrt{\mfr1/2\Tr [D_x^2]}$ is invariant under $\G$,
i.e., $n(\tau(x))=n(x)$ for all $\tau\in\G$.
All minimizing
$\Gamma'$ for $-\P_{\beta,0}$  are invariant under $W_{\tau}$
unless $\tau$ maps the $A$ sublattice to the $B$ sublattice
($\tau(A)=B$). As a consequence the HF ground and Gibbs states
are invariant under $\W_{\tau}$ unless $\tau(A)=B$.

(b) Assume $t$ real and $t-\mu$ not bipartite
Then the minimizers $D$ for $\sr_{\beta, \mu}$, $\Gamma'$ for $-\P_{\beta,\mu}$
and the HF ground and Gibbs states are invariant under $\G$.}

{\it Proof:}
Part (a) follows from the strict convexity proved in Lemma~3.10
since it implies that $n$ is unique and hence invariant.
The second statement in part (a) is a consequence of Lemma~3.7.  where
all the possible minimizers $D$ are described. In fact, the
minimizers $D$ are invariant under $\tau$ unless $\tau(A)=B$ because the
$2\times2$-matrix valued function $w_x$ defined in Lemma~3.7
is constant on the $A$ sublattice and on the $B$ sublattice.

Part (b) follows from Lemma~3.11 and Lemma~3.6
\lanbox

\bigskip\noindent
{\bf III.g THE TRANSLATION INVARIANT CASE}

In this section we shall explicitly determine the minimizing matrix-valued
potential $D$ under the additional assumption that the lattice is translational
invariant, i.e., that the spatial isometry group $G$ of $t$ and $U$ acts
transitively. In particular this means that $U$ is constant independent
of $x$.

As shall be shown, there occurs a phase transition as the temperature
varies about a critical value $T_c$, or $\beta_c$,
respectively, provided $1 < {U \over \vert \Lambda \vert} \Tr_{\H_\Lambda}
[\vert t-\mu \vert^{-1}] < \infty$ holds
and assuming that the spectrum of $t-\mu$ is symmetric about $0$.
More precisely, we will show that $D_x
\equiv 0$, i.e. $n(x) = 0$ for all $x \in \Lambda$, if $\beta \leq \beta_c$
and $n(x)$ equal to a non-zero constant $n_0$
for all $x \in \Lambda$ in case of $\beta > \beta_c$.

Since, as we pointed out in section III.d, $n(x) = n_0 > 0$ goes along with
having non-vanishing aligned pseudo-spins everywhere in $\Lambda$, this
implies long range order which is off-diagonal in case that $\delta (x)
\not= 0$ for all $x \in \Lambda$.

Secondly, due to the analytical dependence of the pressure on $D$, the
transition from $n \equiv 0$ for $\beta \leq \beta_c$ to $n > 0$ for $\beta
> \beta_c$ also indicates that the pressure is nonanalytic at $\beta_c$.
Notice that in our model there is no thermodynamic limit $\vert \Lambda \vert
\longrightarrow \infty$ required to yield a nonanalytic thermodynamic
potential.

{\bf 3.14 THEOREM (Pressure in translation invariant case):}
{\it
If $t$ (with as usual $t-\mu$ either bipartite or real)  and $U$
are translation invariant the pressure is
$$\eqalignno{\P (\beta, \mu)
=&-\min_{d^2 \leq \eta}\Bigl\{-2\beta^{-1}
\Tr\left [\ln \cosh\mfr{\beta}/2
\sqrt{(t-\mu)^2-2Ud(t-\mu)+U^2\eta}\right]+U\eta|\Lambda|\Bigr\}\cr
&\phantom{\min_{d^2 \leq \eta}
\Bigl\{}+(2\beta^{-1}\ln2+\mu)|\Lambda|,\qquad
&(3g.1)}
$$
where the minimum is over real constants $d$ and $\eta$.
The minimium in (3g.1) occurs at unique values
$ d_0(\beta,\mu)$ and
$\eta_0(\beta,\mu)$ satisfying
$$
        d_0(\beta,\mu)^2\leq\eta_0(\beta,\mu)\leq\mfr1/4.
\eqno(3g.2)
$$
If $t-\mu$ is bipartite then $d_0(\beta,\mu)=0$.

Moreover, if $D$ is any minimizer for $\sr_{\beta,\mu}$ then
for all $x$ we have $n(x)^2=\mfr1/2\Tr[D_x^2]=\eta$. If
$t-\mu$ is not bipartite we also have that the
upper diagonal element of $D_x$ is $d(x)=d_0$ for all $x$.
}

{\it Proof:}
Since $x\mapsto-\ln\cosh\sqrt{x}$ is strictly convex and since we are
minimizing over the convex domain $\{d_0^2\leq\eta_0\}$ we conclude that
the minimum occurs at unique values. This is
just a special case of the more general statements in Lemmas~3.10 and 3.11.

If $t - \mu$ is bipartite it is unitarily equivalent to
$-(t-\mu)$ it then follows
from (3.113) that if $d_0$ and $\eta_0$ are minimizing values then so are
$-d_0$ and $\eta_0$. By uniqueness we therefore have
that $d_0=0$.

If $t-\mu$ is bipartite we may without loss of
generality assume $t$ bipartite and $\mu=0$. It then follows from
part (a) of Theorem~3.13 that all minimizers $D$ of $\sr_{\beta,\mu}$
satisfy that $n(x)^2=\mfr1/2\Tr[D_x^2]$ is a constant $\eta$.
We see from Lemma~3.7 that
$D_x$ is of a form such that  $(T_\mu-UD)^2=T^2+U^2D^2$. Since the minimum
in (3g.1) occurs for $d_0=0$ we see from (3b.30) that (3g.1) is
indeed a correct formula for the pressure.

If $t-\mu$ is not bipartite but real and if the functions $d$ and
$\eta$
define a minimizer for $\sr_{\beta,\mu}$ (in the sense of Lemma~3.11)
it follows from the uniqueness proved in Theorem~3.5bb that $d$ and $\eta$
are independent of $x$.
It is then clear from (3b.30) that (3g.1) is correct and that
the parameters $\eta=\eta_0$ and $d=d_0$ are the unique minimizers in (3g.1).
\lanbox

We now restrict to the case when $t-\mu$ is bipartite or more
generally to the case where
$t - \mu$ is unitarily equivalent to $-(t - \mu)$ and
$$1 < {U \over \vert \Lambda \vert} \Tr [\vert t - \mu \vert^{-1}] \leq \infty
\eqno(3g.3)$$
holds.  Then a critical inverse temperature $0 < \beta_c < \infty$ is
uniquely determined by
$$ 1 = {U \over \vert \Lambda \vert} \Tr \left[ \vert t- \mu \vert^{-1} \tanh
\biggl( \beta_c  \vert t - \mu \vert \biggr) \right] ,\eqno(3g.4)$$
since the right side in (3g.4) is continuous and grows monotonically with
$\beta_c$ from
0 to $U \vert \Lambda \vert^{-1} \Tr [\vert t- \mu \vert^{-1}]$.

The following theorem establishes
that $\beta_c$ is, indeed, critical, provided
the spectrum of $t-\mu$ is symmetric about $0$ because, then,
$d_0 (\beta, \mu)$ always vanishes and the
issue of determining the minimizer $D$ simplifies. Notice that in the case
where
$d_0=0$, the gap around zero in the spectrum of $T_\mu-UD$ is at least
$2U\eta_0^{1/2}$. If $t-\mu$ has a
zero eigenvalue then $2U\eta_0^{1/2}$ is precisely the value of gap.

{\bf 3.15  THEOREM (Gap equation):}
{\it Let $t$ be real, translation invariant and $U_x
= U > 0$ for all $x \in \Lambda$.  Assume that $t- \mu$ and $\mu -t$ are
unitarily equivalent as in the case of a bipartite lattice with $\mu=0$.
Let $\beta_c$ be given by (3g.4) in case (3g.3) holds and $\beta_c :=
\infty$ otherwise.
Then $\eta_0 = 0$ for $\beta \leq \beta_c$.
Moreover, $\eta_0 (\beta, \mu)$ is a strict monotonically
increasing function in $\beta_c < \beta$ given by the gap equation
$$1 = \mfr1/2{U \over \vert \Lambda \vert} \Tr \left[ \Bigl({(t - \mu)^2 + U^2
\eta_0}\Bigr)^{-1/2} \tanh \biggl(  \mfr{\beta}/2 \sqrt{(t - \mu)^2 + U^2
\eta_0} \biggr) \right], \eqno(3g.5)$$
provided $\beta_c < \infty$.}

{\it Proof:}
Equation (3g.5) follows by setting the derivative with respect to
$\eta$ of the expression on the right side of (3g.1) equal to zero.
Note that since $\tanh x<1$ (3g.5) implies
$$
        1\leq\mfr1/2{|U|\over|\Lambda|}\Tr\Bigl[(t-\mu)^2+U^2\eta_0]^{-1/2}
        \leq\mfr1/2 \eta_0^{-1/2},
$$
in agreement with our previous condition $\eta_0\leq\mfr1/4$.

The monotonicity of $\eta_0(\beta,\mu)$ in $\beta>\beta_c$ is
straightforward from (3g.5). \lanbox

On a translation invariant lattice we may represent $t-\mu$ by its
eigenvalues $\varepsilon_{\bf k}$ for ${\bf k}\in BZ$, the
Brillouin zone, whose volume we denote by $|BZ|$.
In the thermodynamic limit, $|\Lambda|\to\infty$, the gap equation
(3g.5) then takes the more familiar form
$$
	1=\mfr1/2 {|U|\over|BZ|} \int_{BZ}
	\Bigl(\varepsilon_{\bf k}^2+U^2\eta_0\Bigr)^{-1/2}
	\tanh \biggl(  \mfr{\beta}/2 \sqrt{\varepsilon_{\bf k}^2 + U^2
	\eta_0} \biggr) d{\bf k},
$$
which is the BCS gap equation.

\bigskip\bigskip
\noindent
{\bf IV.  THE GENERALIZED HF THEORY FOR THE \hfill\break
HUBBARD MODEL WITH REPULSIVE INTERACTION}
\bigskip\noindent
{\bf IV.a LINEARIZATION OF THE PRESSURE FUNCTIONAL  }

In this section the generalized HF theory will be applied to the Hubbard
model with repulsive interaction.
We continue to use the notation of Chapter III and consider the Hamiltonian
$$H_+ = \sum \limits_{\scriptstyle x,y \in \lambda \atop \scriptstyle
\sigma} t_{xy} c^\dagger_{x,\sigma} c_{y,\sigma} + \sum \limits_{x \in i}
U_x \big(c^\dagger_{x,\uparrow} c^{{\phantom{\dagger}}}_{x,\uparrow} -
\mfr1/2 \big) \big( c^\dagger_{x,\downarrow}
c^{{\phantom{\dagger}}}_{x,\downarrow} - \mfr1/2 \big).\eqno(4a.1) 
$$
which differs from $H_-$ in (3a.1) in the reversed sign of the interaction,
i.e., we again assume $U_x >0$. A close look at the pair interaction
reveals that
$c^\dagger_{x,\uparrow}c^{{\phantom{\dagger}}}_{x,\uparrow}
   c^\dagger_{x,\downarrow}c^{{\phantom{\dagger}}}_{x,\downarrow}
 = c^\dagger_{x,\uparrow}c^\dagger_{x,\downarrow}
    c^{{\phantom{\dagger}}}_{x,\downarrow}
    c^{{\phantom{\dagger}}}_{x,\uparrow} \geq 0$.
Thus the interaction (corresponding to the operator $V$ in Chapter II)
is repulsive and Theorem~2.11 applies. Hence, we may restrict
our attention to 1-$pdm$ of the form
$\Gamma = \pmatrix{\gamma & \cr &\1
             - \overline \gamma \cr}$
and the energy expectation reduces to
$$\E(\Gamma) = \Tr[ T^\prime_{0} \gamma] + \sum \limits_{x \in \Lambda}
U_x \big\{ [\gamma_{\uparrow}(x) - \mfr1/2 ]
                [ \gamma_\downarrow (x) - \mfr1/2 ]
         - \vert \gamma_* (x) \vert^2 \big\}, \eqno(4a.2) 
$$
where
$\gamma_\sigma(x) := \langle x, \sigma \vert \gamma \vert x, \sigma \rangle$,
$\gamma_*(x) := \langle x, \uparrow \vert \gamma \vert x, \downarrow \rangle$
and we denoted
$$T^\prime_\mu := \pmatrix{t-\mu & 0 \cr 0&t- \mu \cr} \eqno(4a.3) 
$$
on $\H$.
Obviously, $\E( \Gamma )$ depends on $\gamma$ only and we will write
$\E ( \gamma ) := \E ( \Gamma )$. In fact, dealing merely with $\gamma$
we will have to consider only operators on $\H$ rather than $\H \oplus \H$
throughout this chapter.

We denote $Q_x := \1_x Q \1_x$ for any operator
$Q$ on $\H$ where $\1_x$ is now the projection onto functions
in $\H$ vanishing everywhere except at $x$.

Equipped with this notation we may write
$$ \gamma_x = \pmatrix{ \gamma_\uparrow(x) & \overline{\gamma_*}(x) \cr
                \vrule height15ptwidth0pt
                     \gamma_*(x) & \gamma_\downarrow(x) \cr}\1_x
\eqno(4a.4) 
 $$
and one easily verifies that
$$ [\gamma_{\uparrow}(x) - \mfr1/2 ]
                [ \gamma_\downarrow (x) - \mfr1/2 ]
         - \vert \gamma_* (x) \vert^2
= \mfr1/2 \left( \Tr[ \gamma_x - \mfr1/2 \1_x ] \right)^2
- \mfr1/2 \Tr\left[ (\gamma_x - \mfr1/2 \1_x )^2 \right]. \eqno(4a.5)
$$
The entropy depends merely on $\gamma$, too, namely
$$\eqalignno{S(\gamma) & =
- \mfr1/2 \Tr
        [ \Gamma \ln \Gamma + (\1-\Gamma) \ln (\1-\Gamma) ] \cr & =
- \Tr[ \gamma \ln \gamma + (\1-\gamma) \ln (\1-\gamma) ] ,
\qquad&(4a.6) \cr } 
$$
and so does the pressure expectation
$$\eqalignno{-\P_{\beta, \mu} (\gamma) = &\Tr [T^\prime_\mu \gamma]
+ \mfr1/2 \sum
\limits_x U_x \left( \Tr [\gamma_x] -1 \right)^2
- \mfr1/2 \sum
\limits_x U_x \Tr \left[(\gamma_x - \mfr1/2 \1_x)^2 \right]  \cr
&+  \beta^{-1} \Tr [\gamma \ln \gamma
+ (\1 - \gamma) \ln (\1 -\gamma)],
\qquad&(4a.7) \cr} 
$$
where we again denoted $S(\gamma):=S(\Gamma)$ and
$\P_{\beta, \mu} (\gamma) := \P_{\beta, \mu}(\Gamma)$.
For the examination of the HF ground states we denote
$\P_{\infty, \mu} := \lim \nolimits_{\beta \rightarrow \infty}
\P_{\beta, \mu} = \E - \mu N$.
Let us introduce an auxiliary functional
$$\eqalignno{- \widehat{\P}_{\beta, \mu} (\gamma) = &\Tr [T^\prime_\mu \gamma]
- \mfr1/2 \sum
\limits_x U_x \Tr \left[(\gamma_x - \mfr1/2 \uprho(x)\1_x)^2 \right]  \cr
&+  \beta^{-1} \Tr [\gamma \ln \gamma
+ (\1 - \gamma) \ln (\1 -\gamma)],
\qquad&(4a.8) \cr} 
$$
denoting $\uprho(x) := \Tr[ \gamma_x ]$.
Of course, we write $\widehat\P_{\infty, \mu} :=
\lim \nolimits_{\beta \rightarrow \infty} \widehat\P_{\beta, \mu}$.
Notice the formal similarity between (4a.8) and (3b.23). Replacing $T_\mu$
by $T'_\mu$ and $\Gamma'$ by $\gamma$ in (3b.23) we arrive at (4a.8)
excpet that $\uprho(x)$ is missing (and that there is an extra unimportant term
$-\mu|\Lambda|$ in (3b.23)). The reason $\uprho(x)$ is missing in (3b.23) is
that $\Tr[\Gamma_x']=1$ is a consequence of the form (3b.13).
This is the crucial formal difference bewteen the attractive and repulsive
cases.
We claim that
$$ -\P_{\beta, \mu}(\gamma) = -\widehat{\P}_{\beta, \mu} (\gamma)
   + \mfr1/4 \sum \limits_x U_x [ \uprho(x) - 1 ]^2 . \eqno(4a.9)
$$
Indeed, since $\gamma_x - \mfr1/2 \uprho(x) \1_x$ has zero trace,
$$\eqalignno{- \P_{\beta, \mu} (\gamma) + \widehat{\P}_{\beta, \mu} (\gamma)
& =
\mfr1/2 \sum
\limits_x U_x \Bigl\{ [\uprho(x) -1]^2 -
\Tr \left[(\gamma_x - \mfr1/2 \1_x)^2 \right] +
\Tr \left[(\gamma_x - \mfr1/2 \uprho(x)\1_x)^2 \right] \Bigr\}  \cr
& =
\mfr1/2 \sum
\limits_x U_x \Bigl\{ [\uprho(x) -1]^2 -
\Tr \left[ \mfr1/2 [\uprho(x) -1]^2 \1_x \right] \Bigr\} \cr
& =
\mfr1/4 \sum \limits_x U_x [ \uprho(x) - 1 ]^2 .
\qquad&(4a.10) 
\cr}$$

Equation (4a.9) is an important observation.
With our machinery developed in Chapter III we can only determine
the minimizers for $-\widehat{\P}_{\beta, \mu}$ rather
than $- \P_{\beta, \mu}$.
This substitution is justified only if we can show that a 1-$pdm$ $\gamma$
minimizing $-\widehat{\P}_{\beta, \mu}$ also minimizes $- \P_{\beta, \mu}$.
We succeed in doing so {\it only} in case of bipartite hopping matrices $t$
with chemical potential $\mu = 0$. In fact, in this case we will prove that
$$ \uprho(x)=\Tr [ \gamma_{x} ] = \sum \limits_\sigma
 \langle x, \sigma \vert \gamma \vert x, \sigma \rangle =1 \eqno(4a.11)
$$
for all lattice points $x \in \Lambda$.
We shall refer to (4a.11) as the {\bf constant density lemma} because
of its similarity to the main theorem in [LLM].
Equation (4a.11) establishes the formal analogy between the attractive and
repulsive cases.

We start with the analysis of $-\widehat{\P}_{\beta, \mu}$ and its minimizer.
We will denote
$$ - \widehat{\P}(\beta, \mu) := \min \limits_{0 \leq \gamma \leq \1}
   - \widehat{\P}_{\beta, \mu} (\gamma) \eqno(4a.12) 
 $$
for positive or infinite $\beta$.
In analogy with (3b.25) we first observe that
$$ - \Tr\left[ ( \gamma_x - \mfr1/2 \uprho(x) \1_x)^2 \right]
= \min \limits_{d, \delta}
\left\{ -2 \Tr[ D_x \gamma_x] + \Tr[D_x^2] \right\}, \eqno(4a.13)
$$
where
$D_x := \pmatrix{ d(x) & \overline{\delta}(x) \cr \vrule height12ptwidth0pt
\delta(x) & -d(x)}$.
Because of (4a.13) the proof of the following lemma is a line-by-line
copy of the one for Theorems~3.3 and 3.4.

{\bf 4.1 LEMMA:}
{\it For all $0<\beta\leq\infty$
and all $\mu$ we can write the auxiliary functional
$\widehat\P (\beta, \mu)$ as the following
variation over the functions $d$ and $\delta$.
$$
        -\widehat\P (\beta, \mu)=\min_{d,\delta}
         \srr_{\beta,\mu}(d,\delta)
	-  2 \beta^{-1} |\Lambda|\ln 2,\eqno(4a.14) 
$$
where
$$\eqalignno{\srr_{\beta, \mu}(d,\delta): = \srr_{\beta, \mu}(D):=&
 -  \beta^{-1} \Tr [\ln \cosh  \mfr{\beta}/2 (T'_\mu - UD)] +
\mfr1/2 \sum \limits_x U_x \Tr [D^2_x],&(4a.15a)\cr 
\cr\noalign{\hbox{and}}
\srr_{\infty,\mu}(d,\delta):=\srr_{\infty,\mu}(D):=&
 -\mfr1/2\Tr |T'_\mu - UD| +
\mfr1/2 \sum \limits_x U_x \Tr [D^2_x].\qquad&(4.a15b)
}
$$
If a potential $D$ minimizes $\srr_{\beta, \mu}$
then the operator
$$\gamma = \left( \1 + \exp \biggl[ \beta  (T'_\mu -
UD) \biggr] \right)^{-1},\ \hbox{for $\beta<\infty$}\quad\hbox{or}
\quad \gamma=\chi(T'_\mu -UD),\ \hbox{for $\beta=\infty$}
\eqno(4a.16) 
$$
minimizes the auxiliary functional $-\widehat\P_{\beta,\mu}$ and
satisfies the consistency equation
$$\gamma_{x} = \left(D_{x} + \mfr1/2\Tr[\gamma_x]\right )\1_x.\eqno(4a.17)
$$
Conversely, if $\gamma$ is a minimizer for $-\widehat\P_{\beta,\mu}$
then the potential $D$ defined by (4a.17) minimizes $\srr_{\beta,\mu}$
and satisfies (4a.16).
}

As in the attractive case the minimum in (4a.14) occurs for functions
$d$ and $\delta$ satisfying $d(x)^2+|\delta(x)|^2\leq\mfr1/4$ for all
$x\in\Lambda$.

We prove an analog of Lemma~3.5 by merely replacing
$T_\mu$ by $T^\prime_\mu $ in the proof of that lemma.

{\bf 4.2 LEMMA (Gap estimate):}
{\it Let $D$ be a minimizing matrix-valued potential for $\srr_{\beta,\mu}$
or  $\srr_{\infty,\mu}$ and denote
the eigenvalues of $T_\mu - UD$ by $e_1, e_2, \ldots, e_{ 2
\vert \Lambda \vert
}$. Then, for any $j=1,2, \ldots, 2 \vert \Lambda \vert$, we have
$$ \vert e_j \vert \geq \mfr1/4 U_{\min} \vert \Lambda \vert^{-1} -
   \beta^{-1} 2 \ln 2 \qquad\hbox{or}\qquad
 \vert e_j \vert \geq \mfr1/4 U_{\min} \vert \Lambda \vert^{-1}
, \eqno(4a.18) 
$$
respectively, where $U_{\min} := \min \limits_{x \in \Lambda} \{ U_x \}$. }

Lemma~4.1 allows us to concentrate now on the determination of the functions
$d$ and $\delta$ that yield a minimizer $D$ for
$\sr_{\beta, \mu} $
or
$\sr_{\infty, \mu}$
respectively.
Note that $\sr_{\beta, \mu}$ in the repulsive case differs from
$\sr_{\beta, \mu}$ in the attractive case
only by the replacement of $T_\mu$ by $T_\mu^\prime$.

\bigskip
\vbox{\noindent{\bf IV.b CONSTANT DENSITY LEMMA FOR BIPARTITE LATTICES
\hfill \break AT HALF-FILLING  }

In this section we will assume that the hopping matrix $t$
is bipartite (possibly non-real) and the chemical potential $\mu$
equals zero. It will turn out that this choice of $t$ and $\mu$ allows
us to conclude that
$\Tr[ \gamma ] = \vert \Lambda \vert = \mfr1/2 \hbox{dim}\H$
which is the reason for calling this case {\it half-filling}.}

{\bf 4.3 LEMMA (Antiferromagnetic spin alignment):}
{\it Let $t$ be bipartite (but not necessarily real)
and $\mu = 0$. Then
$$\sr_{\beta,\mu} (d , \delta) \geq \sr_{\beta,\mu} \left( (-1)^x
\sqrt{d^2 + \vert \delta \vert^2}, 0 \right). \eqno(4b.1) 
$$
If $n(x) := \sqrt{d^2 (x) + \vert \delta (x)
\vert^2} > 0$ for all $x \in \Lambda$ equality holds in (4b.1) if and only if
for all $x \in \lambda$
$$\pmatrix{d(x) & \delta (x) \cr \vrule height15ptwidth0pt
\overline \delta (x) &-d(x)\cr} =
(-1)^x n(x) w \pmatrix{1 & \cr\vrule height15ptwidth0pt
 & -1} w^\dagger  \eqno(4b.2) 
$$
for some unitary $2 \times 2$-matrix $w$ independent of $x \in \Lambda$.}

The proof of Lemma~4.3 is in complete anology to the proof of Lemma~3.7.
We merely have to replace $T_0$ by $T_0^\prime$. Notice that (4b.2)
results from
$$t_{xy}\left[ {d(x) \over n(x)} - {d(y) \over n(y)} \right] =0, \ \ \ \ \ \
  t_{xy}\left[ {\delta(x) \over n(x)} - {\delta(y) \over n(y) }\right]
=0 , \eqno(4b.3) 
$$
which replaces the conditions (3d.25a) and (3d.25b) in the proof of
Lemma~3.7.  Note that (4b.2) shows that the potential $D$ has the
staggered order characteristic of antiferromagnetism.
The form (4b.2) is exactly what we need to prove the
constant density lemma.

{\bf 4.4  LEMMA (Constant density):}
{\it Let $t$ be bipartite (but not necessarily real)
and $\mu = 0$.
Define $\gamma$ by
$$ \gamma := \bigl( \1 + \exp[ \beta( T^\prime_0 - UD)] \bigr)^{-1},
\eqno(4b.4) 
$$
where $D_x$ is subject to condition (4b.2) for some unitary
$2 \times 2$-matrix $w$.
Then
$$ \Tr[\gamma_x] = \sum \limits_\sigma \langle x, \sigma \vert \gamma
\vert x, \sigma \rangle = 1. \eqno(4b.5) 
$$
In particular, $\Tr[\gamma]=\sum_x\Tr[\gamma_x]=|\Lambda|$.
}

{\it Proof:} We want to show that
$$0 =
\sum \limits_\sigma
\langle x, \sigma \vert \mfr1/2 - \gamma \vert x, \sigma \rangle
=
\mfr1/2
\sum \limits_\sigma
\langle x, \sigma \vert \tanh \mfr1/2 \beta ( T^\prime_0 -UD)
 \vert x, \sigma \rangle.
\eqno(4b.6) 
 $$
We define a unitary transformation
$$ V_w := w \pmatrix{ 0 & -\1 \cr \1 & 0 \cr} w^\dagger (-1)^x. \eqno(4b.7)
$$
One easily checks that
$$ V_w (T^\prime_0 -UD) V_w^\dagger = - (T^\prime_0 - U D).
\eqno(4b.8) 
$$
Thus,
$$\eqalignno{
\sum \limits_\sigma
\langle x, \sigma \vert \tanh \mfr1/2 \beta ( T^\prime_0 -UD)
 \vert x, \sigma \rangle &=
\sum \limits_\sigma
\langle x, \sigma \vert V_w \left[ \tanh \mfr1/2 \beta ( T^\prime_0 -UD)
 \right] V_w^\dagger \vert x, \sigma \rangle \cr &=
 - \sum \limits_\sigma
\langle x, \sigma \vert \tanh \mfr1/2 \beta ( T^\prime_0 -U D)
 \vert x, \sigma \rangle ,
\qquad&(4b.9) 
\cr}$$
which is equivalent to (4b.6). \lanbox

Now we are in a position to determine the actual pressure.

{\bf 4.5 MAIN THEOREM (Ground state and positive temperature pressure):}
{\it Let t be bipartite and $\mu = 0$. Then for all $0<\beta\leq\infty$
$$ - \P (\beta, 0) = -\widehat{\P} (\beta, 0) = \
\min_{d,\delta}\srr_{\beta,0}(d,\delta) -2\beta^{-1}|\Lambda|\ln2.
\eqno(4b.10) 
$$
If a potential $D$ minimizes $\srr_{\beta,0}$ for $0<\beta\leq\infty$
then $\gamma$
defined in (4a.16) is a HF Gibbs state ($\beta<\infty$) or
ground state ($\beta=\infty$).
This $\gamma$ satisfies the consistency equation
$$
        \gamma_x=(D_x+\mfr1/2)\1_x.\eqno(4b.11) 
$$
The function $n(x) = \sqrt{ d(x)^2 + \vert \delta(x) \vert^2} $
corresponding to a minimizer $D$ is unique and
vanishes either everywhere or else nowhere. The potential $D_x$ must
be of the form (4b.2) for some unitary matrix $w$.
As a consequence, the HF Gibbs states and ground states are unique
modulo gauge transformations.
}

{\it Proof:} The right side of (4b.10)
is clearly a lower bound to
$-\P(\beta, 0)$ since $-\P_{\beta, 0}( \gamma) \geq
-\widehat{\P}_{\beta, 0}( \gamma)$ for any $0 \leq \gamma \leq \1$.
On the other hand by Lemma~4.3 the minimum on the right side of
(4b.10) can be attained
by a potential $D$ satisfying (4b.2) (e.g. with $w$
equal to the identity matrix).
We then have a minimizer $\gamma$ for $\widehat\P$ defined in terms of
$D$ by (4a.16) (with $\mu=0$).
Note that because of the gap estimate we know that $\chi(T'_0-UD)=
\lim_{\beta\to\infty}(\1+\exp[\beta(T'_0-UD)])^{-1}$.
It therefore follows for both
finite and infinite $\beta$ that
$\gamma$ satisfies the constant density relation
(4b.5).
Thus by (4a.9),
$$ -\P(\beta, 0) \leq -\P_{\beta, 0}(\gamma) =
-\widehat{\P}_{\beta,0}(\gamma).  \eqno(4b.12) 
$$
This conludes the proof that $\gamma$ is a HF Gibbs or ground state.
The remaining part of the heorem follows by an analysis identical
to the one leading to Lemma~3.8, Lemma~3.10  and Theorem~3.12
(part of the conclusion is that
$\srr_{\beta,0}((-1)^xn,0)$ is convex as a function of $\eta=n^2$
and strictly convex at its minimum).\lanbox

Finally, we mention that a result on spatial symmetry analogous to
Theorem~3.13 (with the same proof) holds in the repulsive case.

{\bf  4.6 THEOREM (Spatial invariance):}  
{\it
Assume $t$ is bipartite (not necessarily real) and $\mu=0$
and let $\G$ be the group of spatial symmetries of $t$ and $U$.
For any minimizer $(d,\delta)$ of $\sr_{\beta,0}$  we have that
the function
$n(x) = \sqrt{ d(x)^2 + \vert \delta(x) \vert^2}$
is invariant under $\G$.
The HF ground and Gibbs states are invariant under $\W_{\tau}$ for
$\tau\in\G$ unless $\tau(A)=B$.}

\bigskip\noindent
{\bf IV.c PARTICLE-HOLE SYMMETRY}\par
In the special case of a {\it real} bipartite hopping matrix $t$ at
half-filling
$\mu=0$  there is a more elegant way of deducing Theorems~4.5 and 4.6
from Theorems~3.2, 3.5 and 3.8 by means of a {\bf partial particle-hole
transformation} only on the $\uparrow$-spins. More precisely, we
define the Bogoliubov transformation $
\W_{ph, \uparrow}
$ by
$$\eqalignno{
\W^{{\phantom{\dagger}}}_{ph, \uparrow} c^\dagger_{x, \uparrow}
 \W_{ph, \uparrow}^\dagger
&= (-1)^x c^{{\phantom{\dagger}}}_{x, \uparrow},
\cr
\W^{{\phantom{\dagger}}}_{ph, \uparrow} c^\dagger_{x, \downarrow}
\W_{ph, \uparrow}^\dagger
&= c^\dagger_{x, \downarrow}.
\qquad&(4c.1) 
\cr}$$
Then
$$\eqalignno{
\W_{ph, \uparrow} H_+ \W_{ph, \uparrow}^\dagger
=&
\sum \limits_{x,y} t_{xy}
( -c^{{\phantom{\dagger}}}_{x, \uparrow} c_{y, \uparrow}^\dagger
+ c^{\dagger}_{x,\downarrow} c^{{\phantom{\dagger}}}_{y, \downarrow} ) +
\sum \limits_x U_x ( c^{{\phantom{\dagger}}}_{x, \uparrow}
c^\dagger_{x, \uparrow} - \mfr1/2)
               ( c^\dagger_{x, \downarrow}
c^{{\phantom{\dagger}}}_{x, \downarrow} - \mfr1/2) \cr
=&
\sum \limits_{x,y} ( \overline{t}_{xy}
c^\dagger_{x, \uparrow} c^{{\phantom{\dagger}}}_{y, \uparrow}
+ t_{xy} c_{x,\downarrow}^\dagger c^{{\phantom{\dagger}}}_{y, \downarrow} )
- \sum \limits_x t_{xx} \cr & -
\sum \limits_x U_x ( c_{x, \uparrow}^\dagger
c^{{\phantom{\dagger}}}_{x, \uparrow} - \mfr1/2)
               ( c_{x, \downarrow}^\dagger
c^{{\phantom{\dagger}}}_{x, \downarrow} - \mfr1/2) \cr
=& H_-,
\qquad&(4c.2) 
\cr}$$
using $t_{xy}=\overline{t}_{xy}=t_{yx}$,
the canonical anticommutation relations and $\sum \nolimits_x t_{xx}=0$.
Since the Bogoliubov transformation leave the set of quasi-free states
invariant, it immediately follows that
$$\eqalignno{ E^{\HF}(H_-)
=&
\inf \left\{ \uprho(H_-) \left\vert \right. \uprho \ \hbox{is quasi-free
}\right\} \cr
=&
\inf \left\{ \uprho_{\W_{ph, \uparrow}}(H_-)
\left\vert \right. \uprho \ \hbox{is quasi-free }\right\} \cr
=&
\inf \left\{ \uprho(H_+) \left\vert \right. \uprho \ \hbox{is quasi-free }
\right\} \cr
=& E^{\HF}(H_+)
\qquad&(4c.3) 
\cr}$$
and a similar equality holds for the pressure at finite temperatures.
Thus, there is clearly a one-to-one correspondence between
the 1-$pdm$
$\Gamma_{+}$ minimizing the generalized HF pressure functional for $H_+$
and the  1-$pdm$
$\Gamma_{-}$ minimizing the generalized HF pressure functional for $H_-$
via
$$ \Gamma_{+} = \W^{{\phantom{\dagger}}}_{ph, \uparrow} \Gamma_{-}
\W_{ph, \uparrow}^\dagger .
\eqno(4c.4) 
$$

Note that if $t$ is {\it not} real the attractive and repulsive cases are
{\it not} unitarily equivalent. In fact, it follows from Theorems~3.5 and 4.5
that in the attractive case there may be no more than two minimizers, while
in the repulsive case there is a continuous family of minimizers related by
spin $SU(2)$ transformations.

\bigskip\noindent
{\bf IV.d FERROMAGNETISM AT INFINITE REPULSION}

In the preceding sections we found that for bipartite
lattices at {\it half-filling}, $N=|\Lambda|$,
the minimizing 1-$pdm$ always has
{\it antiferromagnetic order}, i.e., the sign of $d$ on the
A-sublattice is always opposite to the sign on the B-sublattice.
On the other hand Nagaoka's Theorem  states
that the true ground state (which happens to be the
free particle HF state) for
$N= \vert \Lambda \vert -1$ has maximal spin $S = N/2$,
provided one takes $U=\infty$ (see [NY], [TD], [TH]).
In our language this would mean that the ground state $\gamma$
would satisfy (after an $SU(2)$ rotation) $\gamma_{\downarrow}(x)=0$.
Setting $U=\infty$ mathematically, means, projecting out the vectors
with doubly occupied sites from the full Fock space.

In the context of the generalized HF approximation one can also make sense
of $U=\infty$ and we will derive an analog of
Nagaoka's Theorem  for the HF minimizer. The analogy goes too far,
however, because as we will prove for any
$1\leq N \leq \vert \Lambda \vert - 1$ the HF ground state $\gamma_0$
will be fully spin polarized.
In contrast, for the true Hubbard model
this does not hold, i.e. for
$N= \vert \Lambda \vert -2$
(see [DW], [DFR], [TB],[SA])  or for
$N\leq 0.51|\Lambda|$ [SKA],
the ground state does not have $S=N/2$.

Let us start by defining what is meant by infinite repulsion in the
context of generalized HF theory.
Recall from (4.2) that the energy functional for positive coupling
$U_x = U >0$ becomes
$$\eqalignno{ \E(\gamma) = &
\Tr[ T^\prime_{0} \gamma] + U \sum \limits_{x \in \Lambda}
 \big\{ [\gamma_{\uparrow}(x) - \mfr1/2 ]
                [ \gamma_\downarrow (x) - \mfr1/2 ]
         - \vert \gamma_* (x) \vert^2 \big\} \cr
=&
\Tr[ T^\prime_{0} \gamma] + U \sum \limits_{x \in \Lambda}
\left[\gamma_{\uparrow}(x) \gamma_\downarrow (x)
  - \vert \gamma_* (x) \vert^2 \right] + \mfr1/4 U \vert \Lambda \vert
-\mfr1/2 U\Tr[\gamma].
\qquad&(4d.1) 
 \cr}$$
We consider the particle number $N=\Tr[\gamma]$ fixed. The last two terms
in (4d.1) are therefore constants that we may ignore when determining the
minimizing $\gamma$.
The limit $U\to\infty$ yields the constraint
$$\gamma_{\uparrow}(x) \gamma_\downarrow (x)
  - \vert \gamma_* (x) \vert^2  = 0  \eqno(4d.2) 
$$
for any $x \in \Lambda$.
More precisely, if we define the Hartree-Fock energy of $N$ electrons
by
$$
        E_U^{\HF}(N):=\inf\Bigl\{\E(\gamma)+\mfr1/2UN-\mfr1/4U|\Lambda|\
\Bigl\vert\
        0 \leq \gamma \leq \1, \ \Tr[\gamma] = N
        \Bigr\} \eqno(4d.3)
$$
then $\lim_{U\to\infty}E_U^{\HF}(N)= E^{\HF}_{U=\infty}(N)$ where
$$ E^{\HF}_{U=\infty}(N) := \inf \Bigl\{  \Tr[T^\prime_0 \gamma]\
\Bigl\vert\  0 \leq \gamma \leq \1, \ \Tr[\gamma] = N, \gamma \ \hbox{fulfills
(4d.2) }\Bigr\}. \eqno(4d.4) 
$$
We remark that $N \leq \vert \Lambda \vert$ is automatic in (4d.4)
because the constraint
(4d.2) is equivalent to
Det$[\gamma_x]=0$, which together with $0\leq \gamma\leq\1$
implies that $\Tr[\gamma_x]\leq1$ and hence
$N=\sum_x\Tr[\gamma_x]\leq|\Lambda|$.

{\bf 4.7 THEOREM (Ferromagnetism at infinite $U$):} 
{\it
Let $e_1 \leq e_2 \leq \ldots \leq e_{\vert \Lambda \vert}$ denote the
eigenvalues of $t$. Then
$$ E^{\HF}_{U=\infty} (N) = \sum\limits_{i=1}^N e_i.  \eqno(4d.5)
$$
A family of minimizing $\gamma$ for the variation in (4d.4) corresponds to the
ferromagnetically saturated states
i.e., is given by
$$
        \gamma=w\pmatrix{P_N&\cr&0}w^\dagger,\eqno(4d.6) 
$$
where $w$ is any $SU(2)$ matrix and
$P_N$ is the spectral projection onto the eigenvectors
with eigenvalues $e_1, e_2,\ldots,e_N$.}

{\it Proof:}
It is clear that $ E^{\HF}_{U=\infty} (N)\leq\sum\limits_{i=1}^N e_i$.
Because any
matrix of the form (4d.6) is admissible for the variation in (4d.4).
On the other hand given any $\gamma$ satisfying (4d.2), $0\leq\gamma\leq\1$
and $\Tr[\gamma]=N$. We can write
$$ \gamma = \sum\limits_{i} \vert \varphi_i \rangle
\langle \varphi_i \vert = \sum\limits_{i} \vert f_i \oplus g_i \rangle
\langle f_i \oplus g_i \vert , \eqno(4d.7) 
$$
where $\langle \varphi_i\vert\varphi_j\rangle\leq\delta_{ij}$ for all $i,j$.
On the right side of (4d.7) we
regarded $\H$ as $\H_\uparrow \oplus \H_\downarrow$, and we denoted
$f_i(x) := \varphi_i(x, \uparrow)$ and
$g_i(x) := \varphi_i(x, \downarrow)$.
Now, define on $\H_\Lambda$
$$ \widetilde\gamma_{\uparrow} :=
\sum\limits_{i} \vert f_i \rangle
\langle f_i \vert
+ \vert g_i \rangle
\langle g_i \vert , \eqno(4d.8) 
$$
such that $\widetilde\gamma:=\pmatrix{\widetilde\gamma_{\uparrow}&\cr&0}$
is completely spin-up polarized. The 1-$pdm$
$\widetilde\gamma$ naturally fulfills (4d.2).
Our goal is to show that $\widetilde\gamma_{\uparrow} \leq \1$ and
hence $\widetilde\gamma\leq \1$ since this,
together with $0 \leq \widetilde\gamma$ and
$\Tr [\widetilde\gamma]=N$
implies (4d.5) because
$$ \Tr[ T^\prime_0 \gamma ] =
   \Tr\left[ T^\prime_0 \widetilde\gamma \right] = \Tr[ t
\widetilde\gamma_{\uparrow}]
\geq \sum\limits_{i=1}^N e_i. \eqno(4d.9) 
$$

In order to demonstrate $\widetilde\gamma_{\uparrow} \leq \1$ we need to take a
closer look at the constraint (4d.2) first. We may rewrite (4d.2) as
$$ \left( \sum\limits_{i} \vert f_i(x) \vert^2 \right)
 \left( \sum\limits_{i} \vert g_i(x) \vert^2 \right)
 - \left\vert \sum\limits_{i} \overline{f_i(x)} g_i(x) \right\vert^2
= 0 \eqno(4d.10) 
$$
for all $x \in \Lambda$.
Let us denote
$$  \Lambda_0 := \{ x \in \Lambda \ \vert \
  f_1(x) = f_2(x) = \cdots = f_N(x) =0 \}. \eqno(4d.11) 
$$
On $\Lambda_0$ (4d.10) holds trivially, but on the complement it yields
the existence of a complex number $\alpha(x)$ such that
$$ g_1(x) = \alpha(x) f_1(x), \ g_2(x) = \alpha(x) f_2(x), \ \ldots ,
   g_N(x) = \alpha(x) f_N(x)   \eqno(4d.12) 
$$
for any $x \in \Lambda \setminus \Lambda_0$ by Schwarz' ``equality''.
We define a {\it normal} operator $A$, $AA^\dagger = A^\dagger A$ on
$\H_\Lambda$
by
$$ A := \sum \limits_{x \in \Lambda \setminus \Lambda_0}
        \alpha(x)
\vert x \rangle \langle x \vert , \eqno(4d.13) 
$$
and denote the projection onto $\Lambda_0$ by
$B := \sum\nolimits_{x \in \Lambda_0}
\vert x \rangle \langle x \vert$. Hence, we may express (4d.12) as
$g_i = Af_i + Bg_i$.
By means of $A$ and $B$  we can write
$$\eqalignno{
\delta_{ij} \geq \langle\varphi_i\vert\varphi_j\rangle&=
\langle f_i \vert f_j \rangle +\langle g_i \vert g_j \rangle
 = \langle f_i \vert (\1 + A^\dagger A) f_j \rangle
    + \langle g_i \vert B g_j \rangle \cr
=&
\langle f_i \vert (\1 + AA^\dagger ) f_j \rangle
    + \langle g_i \vert B g_j \rangle .
\qquad&(4d.14) 
 \cr}$$
We define
$F_i := \vert f_i \rangle \langle f_i \vert$,
$G_i := \vert g_i \rangle \langle g_i \vert$,
$R_i := B \vert g_i \rangle \langle f_i \vert$.
Now, $\widetilde\gamma_{\uparrow}\leq \1$ and, therefore, (4d.5) are direct
consequences of the inequality
$$ \widetilde\gamma_{\uparrow} - \widetilde\gamma_{\uparrow}^2 \geq
\left( \sum\limits_{i} [A,F_i]+ R_i \right)
\left( \sum\limits_{i} [A,F_i]^\dagger + R_i^\dagger \right)
+ \left( \sum\limits_{i} R_i^\dagger \right)
\left( \sum\limits_{i} R_i \right).
\eqno(4d.15) 
$$
The inequality (4d.15) follows from (4d.14), $\widetilde\gamma_{\uparrow}
=\sum_i F_i+G_i$, and $AA^\dagger=A^{\dagger}A$ since
$$\eqalignno{
F_iF_j =&|f_i\rangle\langle f_i|f_j\rangle\langle f_j|\leq
        |f_i\rangle\Bigl(\delta_{ij}-\langle f_i|A^{\dagger}A|f_j\rangle
        -\langle g_i B g_j\rangle\Bigr)
        \langle f_j|\cr
        =&\delta_{ij}F_i -F_iA^{\dagger}AF_j-R_i^\dagger R_j,
        =\delta_{ij}F_i -F_iAA^\dagger F_j-R_i^\dagger R_j, \cr
G_iG_j=&\vert g_i \rangle \langle g_i|g_j\rangle\langle g_j|\leq
\vert g_i \rangle  \Bigl( \delta_{ij}
 - \langle f_i \vert f_j \rangle \Bigr)\langle g_j \vert \cr
=& \delta_{ij} G_i - AF_iF_jA^\dagger - R_i F_jA^\dagger
               - AF_iR_j^\dagger - R_i R_j^\dagger , \qquad&(4d.16) \cr
F_i G_j =&
F_i A F_j A^\dagger + F_i A R_j^\dagger .
&{\hbox{\lanbox}}
}$$

The previous theorem does not tell us when the ferromagnetically
saturated states are the {\it only} HF ground states.
The proof only gives the following criterion, which,
unfortunately may be very difficult to verify.
Assume that $e_N < e_{N+1}$ which ensures the uniqueness of $P_N$.
The criterion for uniqueness of the ferromagnetically
states given by 1-$pdm$ of the form (4d.6) is that
the spectral projection $P_N$ be connected in the sense that
any two points $x,y\in\Lambda$ can be connected by a path
$x=x_0,x_1,\ldots,x_n=y$ for which $\langle x_i|P_N| x_{i+1}\rangle\ne0$.
To prove this criterion suppose that $\gamma$ in (4d.4) is a minimizer, i.e.,
$$ \Tr[ T^\prime_0 \gamma ] =
   \Tr_\Lambda[ t \widetilde\gamma_{\uparrow}] =
 \sum\limits_{i=1}^N e_i, \eqno(4d.17) 
$$
where $\widetilde\gamma_{\uparrow}$ is defined as in (4d.7).
Hence, $\widetilde\gamma_{\uparrow} = P_N =
\widetilde\gamma_{\uparrow}^2$ and (4d.15) implies
$$ \sum\limits_{i} R_i = 0, \ \left[ A, \sum\limits_{i} F_i \right]=0,
\eqno(4d.18) 
$$
which, in turn, gives us
$$\eqalignno{ P_N = \widetilde\gamma_{\uparrow}&=
\sum\limits_{i} F_i + G_i =
\sum\limits_{i} F_i + A F_i A^\dagger + AR_i + R_i^\dagger A^\dagger
+ BG_iB \cr
& =
\sum\limits_{i} F_i + A F_i A^\dagger
+ BG_iB. \qquad&(4d.19) 
\cr}$$
The right side of (4d.19) is not connected between $\Lambda_0$ and
$\Lambda \setminus \Lambda_0$ unless $\Lambda_0 = \emptyset$ or
$\Lambda_0 = \Lambda$. The case $\Lambda_0 = \Lambda$ simply means
that $\gamma$ is completely spin-down polarized, in accordance with (4d.6).
Conversely assuming $\Lambda_0 = \emptyset$, we observe that
(4d.19) also implies $[ A, P_N ]=0$ and obtain
$$ \left[ \alpha(x) - \alpha(y) \right] \langle x \vert P_N \vert y \rangle =0
\eqno(4d.20) 
$$
for all $x,y \in \Lambda \setminus \Lambda_0 = \Lambda$.
But $P_N$ was assumed to be connected and, therefore,
$\alpha(x) =$const for all $x\in \Lambda$.
In other words, $f_i = \alpha g_i$ for all $i $
which implies (4d.6).

\bigskip\bigskip
\noindent{\bf V.  SUMMARY OF HF THEORY OF THE HUBBARD MODEL}

\bigskip\noindent
{\bf V.a INTRODUCTION}\par
Our aim here is to give
some perspective to the results in Sects.~III and IV by
summarizing them---with special emphasis on symmetries and
their breaking.  The first task is to define {\bf symmetry breaking}.
We start with a Hamiltonian $H$ that, in many cases, is invariant
under some symmetry group, $G$, each element of which is
represented by a unitary operator on our
Hilbert (Fock) space of dimension $4^{|\Lambda |}$.
The representation of the group $G$ might be a ray representation,
as in the case of ``magnetic translations''. The
unitary operator corresponding to an element $w\in G$ will be denoted by $W(w)$
on the one-particle space $\H\oplus\H$ and by $\W(w)$ on the Fock space $\F$.
Since the only unitaries that transform HF states into HF
states are Bogoliubov transformations,  we
restrict our attention to groups consisting of such transformations.
This restriction is not really a cause for disappointment
because all the symmetry groups that are usually considered, such as
rotations in real space, rotations in spin space, translations, etc.
are, in fact, represented by Bogoliubov unitaries.  The reason is
simply that most symmetry groups in physics are defined by their action on one
particle and are then extended to $N > 1$ particles by tensoring.
This is exactly what number conserving Bogoliubov transformations do.

We shall take the point of view {\it that a state breaks a
symmetry if the state is not invariant under the
action of the corresponding symmetry group}.  This is formalized as
follows.

{\bf 5.1 DEFINITION (Broken or unbroken symmetry):}  {\it
Let $G$ be a group represented by unitaries as described above.
Let $\rho$ be a state on the operators on Fock space $\F$ corresponding to the
Hilbert space $\H$.  We say that the $G$ symmetry is
{\bf unbroken} in the state $\rho$ if, for each $w\in G$, $\rho$ is
invariant under the Bogoliubov transformation $\W(w)$,
i.e., $\rho\, (\W(w) A\W(w)^\dagger) = \rho(A)$.  If the
state is not invariant for all $w\in G$ we say that the $G$ symmetry is
{\bf broken} in the state $\rho$.}
\smallskip

We are particularly interested in HF ground states or finite temperature
Gibbs states for a Hamiltonian $H$ invariant under the action of some group
$G$, i.e., $\W(w)H\W(w)^\dagger = H$ for all $w\in G$.
The interesting question is then whether or not the
$G$ symmetry is broken in these states.

The contrast between the usual theory and the HF theory should be kept in
mind.  While the original Schr\"odinger equation
$H\psi = E\psi$ defines a linear theory, the HF theory
that approximates it is intrinsically a {\bf nonlinear, one-particle
theory}.  Linear combinations of HF wave functions (which are Bogoliubov
transforms of the zero-particle vacuum) are {\bf not} HF
wave functions.  On the level of {\bf states}
(either pure states, $A\mapsto \langle\psi |A|\psi\rangle$, or
Gibbs states), the HF states do {\it not} form convex sets.  Usually,
if $\rho_1$ and $\rho_2$ are states (either ground states of
some Hamiltonian $H$ or Gibbs states of $H$ at some
temperature $T$---in the thermodynamic limit there can be more than
one) then $\rho = \lambda \rho_1 + (1 -  \lambda)\rho_2$ is an
admissible state (ground or Gibbs).  As discussed in the introduction
this is not true for HF
states because $\rho$ is not usually a HF state when $\rho_1$ and
$\rho_2$ are.  In the usual theory we can ask for the extremal states
(i.e. those states that are not convex combinations of other states
at the same temperature) and ask about their properties with
respect to symmetry operations.  An example to keep in mind here is
the Heisenberg Hamiltonian, for which there is a
magnetized ground state.  By taking a convex combination
of all ground states one can construct a ground state that is
$SU(2)$ invariant, but this state is not extremal in the
set of ground states.  From this example we learn that
in the usual theory symmetry breaking should be
sought only with extremal states---which correspond to pure phases.

The HF states, on the other hand, do not form convex sets
and therefore we cannot talk about extremal states.  We regard
each HF state, heuristically, as playing the role of an extremal state.
Indeed, in the usual theory
symmetry breaking in a finite system is infrequent; typically it is
necessary to pass to the thermodynamic limit in order to see
it.  In contrast, if symmetry breaking occurs in HF theory,
it is usually manifest for the finite system.
The following discussion refers to either the ground state or to
positive temperature states. It is to be understood that
the symmetry breaking displayed in our three tables may not actually
occur. In particular, they will usually not occur if the temperature
is high enough.
\bigskip

\noindent{\bf V.b SYMMETRIES OF THE HUBBARD HAMILTONIAN}

To begin our summary of symmetry breaking in the HF
theory of the Hubbard model we first list the symmetries of
the Hubbard Hamiltonian.
There are two types of possible symmetries of $H$, global gauge symmetries
and spatial symmetries. The spatial symmetries of $H$ depend on the spatial
symmetries
of $t$ and $U$ as explained in Sect.~III.f.
The gauge symmetries depend
on the presence or absence of the following three properties:
Bipartiteness of $t$, reality of $t$, and $\mu=0$.
Whether the gauge symmetries are broken may depend on the sign of the
interaction.

{\bf (i) Spatial symmetries:}
In Section III.f, Theorem~3.13,
we found that the spatial symmetries (if any are present)
are broken
only for a bipartite $t$ and then only when $\mu =0$.
Each spatial symmetry either maps $A$ to $A$ and $B$ to $B$ or maps
$A$ to $B$ and $B$ to $A$.
Even for bipartite $t$ and $\mu=0$, the $A$ to $A$ symmetries are never broken.
It is only the transformations
$\tau$ in the spatial symmetry group which take the $A$ sublattice into the
$B$ sublattice, i.e.,
$\tau(A)=B$ (hence of course $\tau(B)=A$) that are broken.
In fact, in this case we must require
that the two sublattices $A$ and $B$ have the same number of points
$(|A|=|B|)$.
Whenever we refer to a broken spatial symmetry
in the tables below, we always mean that it is
only the symmetry of maps from $A$ to $B$ that is broken. (Note that
there can be several maps from $A$ to $B$, but by modding out by
maps from $A$ to $A$ we are left with one map from $A$ to $B$.)

For the very special case of a {\it real} bipartite $t$, $\mu=0$
and attractive interaction there are states which do not even
break the $A$ to $B$ symmetry.  Indeed, according to (3d.19),
there are states for which $d(x)=0$ and $\delta(x)$ is constant
on the whole lattice.  These states are completely
translation invariant.

The gauge symmetries of the Hubbard Hamiltonian that we shall
consider are the following:

{\bf (ii)  Spin $ SU(2)$:}  The action of $SU(2)$ on the
Fock space was defined in (3d.40).  It is the
Bogoliubov transform corresponding to a global spin
rotation on the one-particle space $\H$.  Every
Hubbard Hamiltonian is invariant under this transformation.

{\bf (iii)  Phase $ U(1)$:}  The action of $U(1)$ is given by
the Bogoliubov transformation (3d.41). Again, every Hubbard
Hamiltonian is invariant under this $U(1)$ symmetry.  A
generator of the $U(1)$ transformation (3d.41) on Fock space is,
in fact, the number operator $\N$.  Thus, $U(1)$ invariance of a
state implies particle conservation, but not necessarily a definite
particle number. (See the remark after Theorem~2.3.)  A
normal state is precisely a state where the $U(1)$
symmetry is unbroken.

{\bf (iv)  Pseudo-spin $SU(2)$:}  If we are in the real,
bipartite case and $\mu=0$ we have a very
large symmetry group.  In fact the spin $SU(2)$ symmetry
is supplemented by the pseudo-spin $SU(2)$ symmetry defined
in (3d.50).

Although the pseudo-spin $SU(2)$ commutes with the spin $SU(2)$,
the pseudo-spin is not completely independent
of the spin. More precisely, the full symmetry group
generated by the spin $SU(2)$ and pseudo-spin $SU(2)$ transformations
is not isomorphic to the group $SU(2) \times SU(2)$.  The
full symmetry group is $SO(4)$, see [YZ].  This is so
because for the particular matrix $w = -\bf 1$ in $SU(2)$, the spin
transformation (i.e., the Bogoliubov transformation $\W$
on Fock space) corresponding to $(w,{\bf 1}) \in SU(2)\times SU(2)$ is
identical to the pseudo-spin transformation corresponding to
$({\bf 1},w) \in SU(2)\times SU(2)$.
This easily follows from (3d.51) with $w =- \bf 1$ since the
Bogoliubov transformation ${\W}_\s(w)$
commutes with the transformation ${\W}_\bp$. This also corresponds with
the observation that the representations we obtain for the two spins
are either both integer or both half-integer. For our
purposes it is more useful, however, to treat the spin
and pseudo-spin transformations independently.  Indeed, there
will be cases where the pseudo-spin $SU(2)$ is a
broken symmetry but the spin $SU(2)$ is not and vice versa.
Merely saying that the $SO(4)$ symmetry is broken would
convey much less information.  Likewise the $U(1)$ phase
symmetry is just a subgroup of the pseudo-spin
symmetry; the Bogoliubov transforamtion ${\W}_\theta$ in
(3d.41) is equal to the pseudo-spin transformation
${\W}_{\ps}$ corresponding to $w = \pmatrix {e^{i\theta} &0\cr
0 &e^{-i\theta}}$ in $SU(2)$.  We therefore emphasize
in Table 2 in the case of broken pseudo-spin ($t$ real and bipartite,
positive interaction and $\mu = 0$),
that not only is the pseudo-spin broken but there even exist HF
ground states or Gibbs states for which the subgroup
$U(1)$ is broken.  While the pseudo-spin
is always broken in this case, there are states for which $U(1)$ is
unbroken, i.e., there exist {\bf normal} ground states and
Gibbs states.

{\bf (v) Particle-hole ${\bf Z}_2$:}  The large symmetry group
consisting of the spin $SU(2)$ and pseudo-spin $SU(2)$
required reality, bipartiteness and half-filling.  If, however, we
give up the condition of reality we saw in Section III.d
that the particle-hole symmetry survives as the antiunitary
transformation given in (3d.45).

In case that we have a real $t$ we shall of course not
distinguish the unitary and antiunitary particle hole
transformations as different symmetries.  In this case,
breaking of pseudo-spin or of spin symmetry may or may not
imply breaking of particle-hole symmetry (See table 2).
According to (3d.47) (with an extra complex conjugation, because we are
in the real $t$ case)
the ${\bf Z}_2$ symmetry is unbroken if and only if

$$
\pmatrix{d(x) &\phantom{-} \delta(x)\cr
\vrule height15ptwidth0pt
\overline{\delta(x)} &-d(x)} \ =
- \pmatrix{d(x) &\phantom{-}\overline{\delta(x)}\cr
\vrule height15ptwidth0pt
\delta(x) &-d(x) },
 \eqno(5b.1) $$
i.e., $d{(x)} =0$ and $\delta (x)$ is purely imaginary.

Notice that in the case where the pseudo-spin is a broken symmetry
($t$ real bipartite, attractive interaction and $\mu =0$)
the $U(1)$ symmetry must
be broken when ${\bf Z}_2$ is unbroken and conversely
${\bf Z}_2$ must be broken when $U(1)$ is unbroken.  Thus,
there are two normal states in this case, related by a
particle-hole transformation, and there is one state
(which cannot be a normal state) that is
invariant under the particle-hole transformation.

Note that the spatial $A$--$B$ symmetry (if exists) is broken if
and only if the ${\bf Z}_2$ particle hole symmetry is broken. Although,
these are different symmetries, in the sense that they act as different
transformations on the Fock space, they are identical when restricted
to the HF ground  states and Gibbs states.

The following diagram illustrates the relationships among
the four gauge symetries: The particle-hole ${\bf Z}_2$ is a
combination of spin and pseudo-spin, while the phase $U(1)$ is really
a subgroup of the pseudo-spin $SU(2)$.
\bigskip
\bigskip
\vbox{\settabs 10\columns
\+&&&Spin $SU(2)$&&Pseudo-spin $SU(2)$&&&&\cr
\+&&&$\phantom{\hbox{Spin $SU$}}\Biggl\backslash$&&$\phantom{\hbox{Pse}}
\Biggr\slash\qquad\Biggl\backslash$&&&&\cr
\+&&&&Particle-hole ${\bf Z}_2$&& \qquad Phase $U(1)$&&&\cr
}
\bigskip\bigskip
\hoffset-0.5cm
\vbox{\tabskip=0pt  \offinterlineskip
\def\tablerule{\noalign{\hrule}}
\halign to 430pt{
\strut#& \vrule#\tabskip=1em plus 2em&
  \hfil#&\vrule#& \hfil#\hfil& \vrule#& \hfil#\hfil& \vrule#&
  \hfil#\hfil& \vrule#\tabskip=0pt\cr\tablerule
\omit&height 10pt&\multispan7&height 10pt\cr
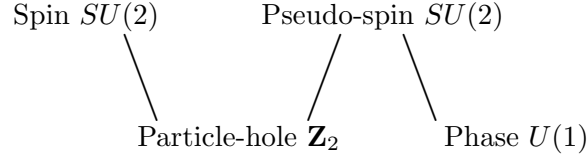
&&\multispan7\hfil {\bf TABLE NO. 1 Non-Real
$t$}\hfil&\cr
\omit&height 8pt&\multispan7&height 8pt\cr
\tablerule
\omit&height 4pt&\omit&&\omit&&\omit&&\omit&height 4pt\cr
&&\omit&&  {\it Symmetries of $H$}&&
{\it Broken Symmetries}&&
\hidewidth{\it Spatial invariance}\hidewidth&\cr
\omit&height 2pt&\omit&&\omit&&\omit&&\omit&height 2pt\cr
\tablerule
\omit&height 4pt&\omit&&\omit&&\multispan3&height 4pt\cr
&& \phantom{Xxxx}Non-bipartite\phantom{XX}\hfil  &&\omit\hidewidth
    &&\multispan3  No results except for $U=\infty$, where\hidewidth\hfil&\cr
&& and/or\hfil  &&Spin $SU(2),\, U(1)$\hfil &&\multispan2
      spin $SU(2)$ is broken but $U(1)$ and\hidewidth \hfil&\hfil&\cr
&& $\mu \neq 0$\hfil&&\omit &&\multispan2
     spatial invariance are not.\hidewidth\hfil &\hfil&\cr
\omit&height 2pt&\omit&&\omit&&\multispan3&height 2pt\cr
\tablerule
\omit&height 4pt&\omit&&\omit&&\omit&&\omit&height 4pt\cr
&& Bipartite\hfil &&Spin $SU(2),\, U(1)$,  &&\omit\hidewidth
    &&$A$-$B$ symmetry\phantom &\cr
&&  $\mu=0$\hfil        && \hfil${\bf Z}_2$\hfil
   &&\hfil ${\bf Z}_2$\hfil && \hfil  broken \hfil&\cr
&&\hidewidth Attractive interaction$^*$ &&\omit\hidewidth &&\omit\hidewidth
&&\omit\hidewidth
 &\cr
\omit&height 2pt&\omit&&\omit&&\omit&&\omit&height 2pt\cr
\tablerule
\omit&height 4pt&\omit&&\omit&&\omit&&\omit&height 4pt\cr
&& Bipartite\hfil &&Spin $SU(2),\, U(1)$,   &&\hskip.4truein Spin
$SU(2)$,\hskip.4truein &&$A$-$B$ symmetry&\cr
&& $\mu=0$\hfil && \hfil${\bf Z}_2$\hfil
   &&\hfil ${\bf Z}_2$\hfil && \hfil broken \hfil&\cr
&&\hidewidth Repulsive interaction$^*$ &&\omit\hidewidth &&\omit\hidewidth
&&\omit\hidewidth
 &\cr
\omit&height 2pt&\omit&&\omit&&\omit&&\omit&height 2pt\cr
\tablerule
&\multispan9*  Notice that even though  we are in the bipartite case and
$\mu=0$,
the attractive and repulsive \cr
&\multispan9 interaction Hamiltonians are not unitarily equivalent when $t$ is
not real
\hfil\cr}}\bigskip\bigskip\bigskip
\vbox{\tabskip=0pt  \offinterlineskip
\def\tablerule{\noalign{\hrule}}
\halign to 430pt{\strut#& \vrule#\tabskip=1em plus 2em&
  \hfil#&\vrule#& \hfil#\hfil& \vrule#& \hfil#\hfil& \vrule#&
  \hfil#& \vrule#\tabskip=0pt\cr\tablerule
\omit&height 10pt&\multispan7&height 10pt\cr
&&\multispan7 \hfill{\bf TABLE NO. 2 Real Bipartite
$t$}\hfill&\cr
\omit&height 8pt&\multispan7&height 8pt\cr\tablerule
\omit&height 4pt&\omit&&\omit&&\omit&&\omit&height 4pt\cr
&&\omit &&\hidewidth{\it Symmetries of $H$}\hidewidth&&
 \hidewidth{\it Broken Symmetries}\hidewidth&&
  \hidewidth{\it Spatial invariance}\hidewidth\hfil&\cr
\omit&height 2pt&\omit&&\omit&&\omit&&\omit&height 2pt\cr\tablerule
\omit&height 4pt&\omit&&\omit&&\omit&&\omit&height 4pt\cr
&& $\mu\ne0$\hfil  &&Spin $SU(2)$,
    &&$U(1) $&&\hfil Not broken\hfil &\cr
&&Attractive interaction &&$U(1)$ &&\omit\hidewidth&&\omit&\cr
\omit&height 2pt&\omit&&\omit&&\omit&&\omit&height 2pt\cr\tablerule
\omit&height 4pt&\omit&&\omit&&\omit&&\omit&height 4pt\cr
&&$\mu=0$\hfil &&Spin $SU(2)$, &&Pseudo-spin $SU(2)$\hfill
  &&\hfil $A$-$B$ symmetry\hfil &\cr
&& Attractive interaction$^*$&& Pseudo spin $SU(2)$ &&(in particular ${\bf
Z}_2$ and\hfill
  &&\hfil can be broken\hfill&\cr
 &&\omit\hidewidth &&\omit\hidewidth&& $U(1)$ can be broken)\hfill
  &&\omit&\cr
\omit&height 2pt&\omit&&\omit&&\omit&&\omit&height 2pt\cr
\tablerule
\omit&height 4pt&\omit&&\omit&&\multispan3&height 4pt\cr
 && $\mu\ne0$\hfil &&Spin $SU(2)$,
&&\multispan3  No results except for $U=\infty$, where\hidewidth&\cr
&& Repulsive interaction\phantom{$^*$}  &&$U(1)$ &&\multispan3
       spin $SU(2)$ is broken but $U(1)$ and \hidewidth&\cr
&& \omit\hidewidth &&\omit\hidewidth &&\multispan2
     spatial invariance are not. \hfil\hidewidth &\hfil&\cr
\omit&height 2pt&\omit&&\omit&&\multispan3&height 2pt\cr\tablerule
\omit&height 4pt&\omit&&\omit&&\omit&&\omit&height 4pt\cr
&& $\mu=0$\hfil &&Spin $SU(2)$,
   &&Spin $SU(2)$\hfill &&$A$-$B$ symmetry&\cr
&& Repulsive interaction$^*$ &&Pseudo-spin $SU(2)$ && (in particular ${\bf
Z}_2$ is
\hfill && broken\hfil &\cr
&& \omit\hidewidth &&\omit\hidewidth &&  broken)\hfill
  &&\omit\hidewidth
 &\cr\omit&height 2pt&\omit&&\omit&&\omit&&\omit&height 2pt\cr\tablerule
&\multispan9* For real, bipartite $t$ and $\mu=0$ the Hamiltonians with
attractive and repulsive interactions {\bf are}\cr
&\multispan9 unitarily equivalent. Because of the $(-1)^x$, a spatial symmetry
between the $A$ and $B$ sublattices\cr
&\multispan9 will not commute with this unitary transformation; hence Table 2
fails
 to be the same for the\cr
&\multispan9 repulsive and attractive cases.\hfill\cr
}}
\bigskip\bigskip
\vbox{\tabskip=0pt  \offinterlineskip
\def\tablerule{\noalign{\hrule}}
\halign to 430pt{\strut#& \vrule#\tabskip=1em plus 2em&
  \hfil#&\vrule#& \hfil#\hfil& \vrule#& \hfil#\hfil& \vrule#&
  \hfil#& \vrule#\tabskip=0pt\cr\tablerule
\omit&height 10pt&\multispan7&height 10pt\cr
&&\multispan7\hfil {\bf TABLE NO. 3 Real, Non-Bipartite
$t$}\hfil&\cr
\omit&height 8pt&\multispan7&height 8pt\cr
\tablerule
\omit&height 4pt&\omit&&\omit&&\omit&&\omit&height 4pt\cr
&&\omit\hidewidth&&{\it Symmetries of $H$}&&
\omit{\it Broken Symmetries }&&
\omit\hidewidth{\it Spatial invariance}\hfil\hidewidth&\cr
\omit&height 2pt&\omit&&\omit&&\omit&&\omit&height 2pt\cr
\tablerule
\omit&height 4pt&\omit&&\omit&&\omit&&\omit&height 4pt\cr
&&Attractive interaction &&Spin $SU(2)$, &&\hfil $U(1)$\hfil && \phantom{X}Not
broken
\phantom{X}\hfil&\cr
  &&Any $\mu$ \hfil&& \hfil $U(1)$
\hfil &&\omit &&\omit\hidewidth &\cr
\omit&height 2pt&\omit&&\omit&&\omit&&\omit&height 2pt\cr\tablerule
\omit&height 4pt&\omit&&\omit&&\multispan3&height 4pt\cr
&&Repulsive interaction &&Spin $SU(2)$, &&
    \multispan3  No results except for $U=\infty$, where \hfil &\cr
&& Any $\mu$\hfil &&\hfil$U(1)$\hfil &&\multispan3
       spin $SU(2)$ is broken but $U(1)$ and\hfil&\cr
&& \omit\hidewidth &&\omit\hidewidth &&\multispan3
 spatial invariance are not.\hfil&\cr
\omit&height 2pt&\omit&&\omit&&\multispan3&height 2pt\cr\tablerule}}

\bigskip
\bigskip
\def\newblock{}
\def\em{\it}
\centerline{{\bf REFERENCES}}
\medskip
\ninepoint
\item{[AH]}
H.~Araki.
\newblock On quasifree states of {CAR} and {B}ogoliubov automorphisms.
\newblock {\em Publ. RIMS Kyoto}, {\bf6}:385--442, 1970/71.

\item{[BCS]}J. Bardeen, L. N. Cooper and J. R. Schrieffer, Theory
of superconductivity, {\em Phys. Rev.}, {\bf108}:1175, 1947.

\item{[BLL]}
V.~Bach{,} E.H. Lieb{,} M. Loss{,} J. P. Solovej.
\newblock There are no unfilled shells in {H}artree-{F}ock theory.
\newblock {\em Preprint}, 1993.

\item{[BN]}
N.N. Bogoliubov, V. V. Tolmachev and D. V. Shirkov.
\newblock {\em A New Method in the Theory of Superconductivity},
Consultants Bureau, NY, 1959, Appendix 2.

\item{[BR]}
J.-P. Blaizot and G.~Ripka.
\newblock {\em Quantum Theory of finite Systems}.
\newblock MIT Press, Cambridge, Mass., 1986.

\item{[BV]}
V.~Bach.
\newblock Error bound for the {H}artree-{F}ock energy of atoms and molecules.
\newblock {\em Commun. Math. Phys.}, {\bf147}:527--548, 1992.

\item{[CM]}
M.~Cyrot.
\newblock Theory of {M}ott transition: Application to transient metal oxides.
\newblock {\em J. de Phys.}, {\bf33}:125--134, 1972.

\item{[dG]}
P.G. de~Gennes.
\newblock {\em Superconductivity of metals and alloys}.
\newblock Benjamin, New York, 1966.

\item{[DFR]}
E.~Dagatto Y.~Fand, A.E.~Ruckenstein and S.~Schmitt-Rink.
\newblock Holes in the infinite $U$ {H}ubbard model:
Instability of the {N}agaoka state.  \newblock {\em Phys. Rev. B},
{\bf40}:7406--7409, 1989.

\item{[DJK]} K. Dichtel, R.H. Jellito and H. Koppe. The ground state of
the neutral Hubbard model. {\em Z. Physik}, {\bf246}:248--260, 1971.
Thermodynamics of the Hubbard model. {\em Z. Physik}, {\bf251}:173--184,
1972.

\item{[DW]}
B.~Doucot and X.G. Wen.
\newblock Instability of the {N}agaoka state with more than one hole.
\newblock {\em Phys. Rev. B}, {\bf40}:2719--2722, 1989.

\item{[FE]}
E.~Fradkin.
\newblock {\em Field theories of condensed matter systems}.
\newblock Addison-Wesley, 1991.

\item{[GM]}
M.~Gaudin.
\newblock Une d{\'e}monstartion simplifi{\'e}e du th{\'e}or{\`e}me
de {W}ick en m{\'e}chanique statistique.
\newblock {\em Nucl. Phys.}, {\bf15}:89--91, 1960.

\item{[GMC]}
M.C. Gutzwiller, The effect of correlation on the
ferromagnetism of transition metals, {\em Phys. Rev. Lett.},
{\bf 10}:159-162 (1963).

\item{[HJ]}
J.~Hubbard.
\newblock Electron correlations in narrow energy bands.
\newblock {\em Proc. Roy. Soc. (London)}, {\bf A276}:238--257, 1963.

\item{[KJ]}
J. Kanamori, Electron correlation and ferromagnetism of
transition metals, {\em Prog. Theor. Phys.}, {\bf 30}:275-289 (1963).

\item{[KL]}
T.~Kennedy and E.H. Lieb.
\newblock An itinerant electron model with crystalline or
magnetic long range order.
\newblock {\em Physica}, {\bf138A}:320--358, 1986.

\item{[LE1]}
E.H. Lieb.
\newblock Variational principle for many-fermion systems.
\newblock {\em Phys. Rev. Lett.}, {\bf46}:457--459, 1981. Errata {\bf
47}, 69 (1981).

\item{[LE2]}
E.H. Lieb.
\newblock Two theorems on the {H}ubbard model.
\newblock {\em Phys. Rev. Lett.}, {\bf62}:1201--1204, 1989.

\item{[LE3]} E. H. Lieb. The Hubbard model: Some rigorous results and
open problems. In {\em Advances in Dynamical
Systems and Quantum Physics}, V. Figari et. al., eds., World Scinetific,
Singapore, in press.

\item{[LE4]} E. H. Lieb.
Thomas-Fermi and Hartree-Fock theory. In {\em Proc. Inter. Cong.
Math.}, Canad. Math. Soc., 1975, pp.383-386.

\item{[LL]}
E.H. Lieb and M.~Loss.
\newblock Fluxes, Laplacians and Kasteleyn's theorem.
\newblock {\em Duke Math. J.}, {\bf71}:337--363, 1993.

\item{[LLM]}
E.H. Lieb{,}~M. Loss and R. J. McCann.
\newblock Uniform density theorem for the {H}ubbard model.
\newblock {\em J. Math. Phys.}, {\bf34}:891--898, 1993.

\item{[LS]}
E.H. Lieb and B.~Simon.
\newblock The {H}artree-{F}ock theory for {C}oulomb systems.
\newblock {\em Commun. Math. Phys.}, {\bf53}:185--194, 1977.

\item{[NY]}
Y.~Nagaoka.
\newblock Ferromagnetism in a narrow, almost half-filled S-band.
\newblock {\em Phys. Rev.}, {\bf147}:392--405, 1966.

\item{[PD]} D. R. Penn. Stability theory of the magnetic phases for a
simple model of the transition metals. {\em Phys. Rev.},
{\bf142}:350--365, 1966.

\item{[SA]}
A.~S\"ut\H o.
\newblock Absence of highest spin ground states in the {H}ubbard model.
\newblock {\em Commun. Math. Phys.}, {\bf140}:43--62, 1991.

\item{[SKA]}
B.S. Shastry{,}~H.R. Krishnamurthy and P.W. Anderson.
\newblock Instability of the {N}agaoka ferromagnetic state of the
$U=\infty$ {H}ubbard model.
\newblock {\em Phys. Rev. B}, {\bf41}:275--2379, 1990.

\item{[TB]}
B.~T\'oth.
\newblock Failure of saturated ferromagnetism for the {H}ubbard model
with two holes.  \newblock {\em Lett. Math. Phys.}, {\bf22}:321--333, 1991.

\item{[TD]}
D.J. Thouless.
\newblock Exchange in solid $^3he$ and the {H}eisenberg {H}amiltonian.
\newblock {\em Proc. Phys. Soc.(London)}, {\bf86}:893--904, 1965.

\item{[TH]}
H.~Tasaki.
\newblock Extension of {N}agaoka's theorem on the large $U$ {H}ubbard model.
\newblock {\em Phys. Rev. B}, {\bf40}:9192--9193, 1989.

\item{[TW]} W. Thirring. {\it A course in Mathematical Physics: Vol.
4}, transl. E. M. Harrell, Springer, Vienna, 1980, p. 48.

\item{[VJ]} J. G. Valatin. Comments on the theory of superconductivity,
{\em Nuovo Cimento}, [X]{\bf 7}:843--857, 1958.

\item{[YZ]}
C. N. Yang and S. C. Zhang.
\newblock $SO_4$ symmetry in a {H}ubbard model.
\newblock {\em Mod. Phys. Lett.}, {\bf B4}:759--766, 1990.

\end